\documentclass[11pt]{article}
\usepackage{amscd,amsmath,amsxtra,amssymb,theorem,latexsym,amsfonts,graphics}
\usepackage[usenames,dvipsnames]{color}
\usepackage[all]{xy}
\usepackage{ulem}
\usepackage{bbm}
\usepackage{rotating}

\reversemarginpar
\renewcommand{\em}{\it}

\newtheorem{Thm}{Theorem}[section]
\newtheorem{Lemma}[Thm]{Lemma}
\newtheorem{Prop}[Thm]{Proposition}
\newtheorem{Proposition}[Thm]{Proposition}
\newtheorem{Remark}[Thm]{Remark}
\newtheorem{Cor}[Thm]{Corollary}

\newtheorem{Notation}[Thm]{Notation}

\newtheorem{Claim}[Thm]{Claim}

\DeclareMathOperator{\Coker}{Coker}
\DeclareMathOperator{\Ker}{Ker}
\newcommand\wa{\mathbf{W}^\alpha}
\newcommand\wma{\mathbf{W}^{-\alpha}}
\newcommand\wc{\mathbf{W}}
\newcommand\waprime{\wc'}

\newcommand{\bA}{\mathbf{A}}

\newcommand{\bD}{{\mathbf{D}}}
\newcommand{\bx}{{\mathbf{x}}}
\newcommand{\cD}{\mathcal{D}}
\newcommand{\cV}{{\mathcal{V}}}
\newcommand{\btN}{\tilde{\mathbf{N}}}
\newcommand{\bc}{\mathbf{c}}
\newcommand\TP{{\mathbf{TP}}}
\newcommand\XX{{\mathcal{X}}}
\newcommand\YY{{\mathcal{Y}}}
\renewcommand{\SS}{{\rm S.S.}}
\newcommand{\bN}{{\mathbf{N}}}
\newcommand{\bU}{{\mathbf{U}}}
\newcommand{\bV}{{\mathbf{V}}}
\newcommand{\cH}{\mathcal{H}}
\newcommand{\hatc}{{\hat c}}
\newcommand{\cO}{{\mathcal{O}}}
\newcommand{\mQ}{{\mathcal{Q}}}
\newcommand{\cR}{{\mathcal{R}}}
\newcommand{\cU}{{\mathcal{U}}}
\newcommand{\cF}{\mathcal{F}}
\newcommand{\zA}{{\mathcal{A}}}
\newcommand{\cJ}{{\mathcal{J}}}
\newcommand{\bS}{\mathbf{S}}
\newcommand{\bfq}{{\mathbf{q}}}
\newcommand{\be}{\mathbf{e}}
\newcommand{\tW}{{\tilde W}}
\newcommand{\tw}{{\tilde w}}
\newcommand{\tpsi}{{\tilde \psi}}
\newcommand{\bw}{{\mathbf{w}}}
\newcommand{\tnu}{\tilde \nu}
\newcommand{\bfun}{{\mathbf 1}}
\newcommand{\tcS}{{\tilde S}}
\newcommand{\tN}{{\tilde N}}
\newcommand{\bB}{{\mathbf B}}
\newcommand{\bC}{{\mathbf C}}
\newcommand{\bPi}{{\mathbf\Pi}}
\newcommand{\bff}{\mathbf{f}}
\newcommand{\sU}{\mathbf{U}}
\newcommand{\sV}{\mathbf{V}}
\newcommand{\tGamma}{{\tilde \Gamma}}
\newcommand{\tS}{{\tilde S}}
\newcommand{\tA}{{\tilde A}}
\newcommand{\tI}{{\tilde I}}
\newcommand{\tiota}{{\tilde\iota}}

\newcommand{\bP}{\mathbf P}
\newcommand{\vt}{\vartheta}
\newcommand\ihom{{\mathcal H}om}
\DeclareMathOperator{\End}{End}
\newcommand{\R}{{\mathbb R}}

\newcommand\ba{{\mathbf{a}}}

\newcommand{\C}{{\mathbb C}}
\newcommand{\Z}{{\mathbb Z}}

\newcommand{\cK}{{\mathbb K}}
\newcommand{\ext}{\text{ext}}
\newcommand{\elr}{\mathbf{r}}
\newcommand{\into}{\hookrightarrow}

\newcommand{\zq}{{{q_0}}}
\newcommand{\Gr}{{\mathbf{Gr}}}

\newcommand{\Id}{{\rm Id }}

\renewcommand{\Im}{{\rm Im \ }}
\renewcommand{\Re}{{\rm Re \ }}

\newcommand{\supp}{{\rm supp \ }}
\newcommand{\Int}{{\rm Int}}

\newcommand{\xu}{{\underline{x}}}

\newcommand{\id}{{\rm id}}

\renewcommand{\Im}{{\rm Im \ }}

\newcommand{\Sol}{{Sol}}
\newcommand{\rest}{\mathbf{res}}
\newcommand{\un}{\underline}

\newcommand{\DerCat}{{\mathbf D}}
\newcommand{\trim}{{\rm trim}}
\newcommand{\levo}{{\rm left}}
\newcommand{\pravo}{{\rm right}}
\newcommand{\Lotimes}{{\otimes^{\mathbb L}}}
\newcommand{\vs}{\varsigma}

\newcommand{\Co}{\mathbb{C}}

\newcommand{\iP}{{\stackrel\circ{P}}}

\newcommand{\ve}{{\varepsilon}}
\newcommand{\bF}{\mathbf{F}}
\usepackage[small]{caption} 

\newcommand{\inthesetbelow}{{\text{\rotatebox{-90}{$\in$}}}}
\newcommand{\inthesetabove}{{\text{\rotatebox{90}{$\in$}}}}

\makeatletter
\newif\if@caption@empty
\newcommand{\captionempty}
{%
  \@caption@emptytrue
  \caption{}%
}
\renewcommand{\@makecaption}[2]
{%
  \centering
  \itshape
  \rm #1%
  \if@caption@empty
    \global\@caption@emptyfalse
  \else
    \textbf{:} #2%
  \fi
}
\makeatother

\def\bL{\mathbb{L}}

\def\cC{\mathcal{C}}
\def\cS{\mathcal{S}}
\def\perpC{{{}^\perp \cC}}
\def\bZ{\mathbb{Z}}
\def\lang{\langle}
\def\rang{\rangle}
\def\fs{{\mathbb{F}}}
\def\fS{\fs}
\DeclareMathOperator\Hom{Hom}
\def\cI{{\mathbf{I}}}
\def\pt{\mathbf{pt}}
\DeclareMathOperator{\Cone}{Cone}
\def\cL{{\mathcal{L}}}
\def\cM{\mathcal{M}}
\def\cG{{\mathcal{G}}}

\begin{document}

\title{Microlocal properties of sheaves and complex WKB.} \author{Alexander GETMANENKO \footnote{Kavli IPMU, University of Tokyo, Japan; {\tt alexander.getmanenko@ipmu.jp} }, Dmitry TAMARKIN \footnote{Mathematics Department, Northwestern University, U.S.A.; {\tt tamarkin@math.northwestern.edu}} }
\maketitle

\begin{abstract} Kashiwara-Schapira style sheaf theory is used to justify analytic continuability of solutions of the Laplace transformed Schr\"odinger equation with a small parameter. This partially proves the description of the Stokes phenomenon for WKB asymptotics predicted by Voros in 1983.
\end{abstract}

\section{Introduction} 

In this paper we are going to study the following PDE on one unknown function $\Psi$ in two complex 
variables $x,s$:
\begin{equation}
-\Psi_{xx}+V(x)\Psi_{ss}=0, \label{Laplace}
\end{equation}
where $V(x)$ is a given polynomial; the weakest possible assumptions on $V(x)$ will be formulated in Sec.\ref{WeakestPossible}. 

This equation is related to the Schr\"odinger 
equation \begin{equation} -h^2\partial_x^2 \psi(x,h) + V(x) \psi(x,h) = 0 \label{SchE} \end{equation}
by means of the Laplace transform $1/h\mapsto \partial_s$.
According to resurgent analysis, the analytic behavior of $\Psi(x,s)$ determines 
quasi-classical asymptotics of  solutions of  \eqref{SchE}. 

 A multivalued solution $\Psi$ of (\ref{Laplace}) can be specified by means of prescribing
 its initial values.
Our problem is now as follows.  Consider  a class of initial value problems for (\ref{Laplace}) with a fixed
type of the analytic behavior of the initial data; we are to find a manifold
 where solutions of these problems are defined.

\subsection{Cauchy problem}\label{cauchyy}
We study the Cauchy problem for \eqref{Laplace} of the following type. 
We fix a point $x_0\in \C$ and prescribe 
{$\Psi(x_0,s)=\psi_0(s)$}  and  {$\frac{\partial \Psi(x,s)}{\partial x}|_{x=x_0}=\psi_1(s)$}
as multivalued analytic functions of $s$.  Let us now  give a more precise account.

\subsubsection{Initial data} \label{InDataS}
Fix an acute angle $\alpha\in (0,\pi/2)$. Let 
$S_\alpha:=(0,\infty)\times (-\alpha,\alpha+2\pi)$ be an open sector of aperture $2\pi+2\alpha$.
Let $\pi_{S_\alpha}:S_\alpha\to \Co$ be the covering map $\pi_{S_\alpha}(r,\phi):=re^{i\phi}$.
The map $\pi_{S_\alpha}$ induces a complex structure on $S_\alpha$ so that $\pi_{S_\alpha}$
is a local biholomorphism. The initial conditions  are given by  two  holomorphic
functions
\begin{equation}\label{initialdata}
\text{$\psi_0$ and $\psi_1$ on $S_\alpha$.}
\end{equation}
\subsection{Multi-valued solution to a multi-valued Cauchy problem}\label{init}
We first fix a complex surface $\cS$ along  with a local biholomorphism
$p_\cS:\cS\to \Co\times \Co$. Let us also fix a map 
\begin{equation}\label{isa}
h:S_\alpha\to \cS
\end{equation}
 fitting into
the following commutative diagram
$$
\xymatrix{ \Co\ar[r]^{i_{x_0}}& \Co\times \Co\\
S_\alpha\ar[r]^{h}\ar[u]^{\pi_{S_\alpha}}& \cS\ar[u]^{p_{\cS}}}
$$
where $i_{x_0}:\Co\to \Co\times\Co$ is given by the formula $i_{x_0}(s)=(x_0,s)$.

The equation (\ref{Laplace}) gets transferred onto $\cS$ by means of a local biholomorphism $p_\cS$.
Call this equation "the transferred equation".

The coordinates $(x,s)$ on $\Co\times \Co$ give rise to local coordinates on $\cS$. Given a function
$\Psi$ on $\cS$,  we then have a well defined 
derivative $\displaystyle\frac{\partial \Psi}{\partial x}$ as a holomorphic function on $\cS$.

We say that a solution $\Psi$ of the transferred equation {\em is a solution  of the Cauchy problem
with initial data (\ref{initialdata})} on $\cS$, if
$\Psi\circ h=\psi_0$; $\displaystyle\frac{\partial \Psi}{\partial x}\circ h=\psi_1$.

\subsection{Formulation of the result} Our main result is a construction of a complex
surface $\cS$ and a map $h$ as in (\ref{isa}), such that for every choice of  the initial data,
 there exists  a unique solution $\Psi$ of the Cauchy problem on $\cS$.

We  prove (Sec. \ref{beskon}) that the surface $\cS$ is ``extends infinitely in the direction of  $K$",
where $K\in \Co$ is the following  cone: 
\begin{equation}\label{coneK}
K:=\{re^{i\phi};r\geq 0; -\alpha\leq \phi\leq \alpha\}.
\end{equation}
Let us give a more precise formulation.  Fix a point $x\in \Co$ such that $V(x)\neq 0$.
Consider a one-dimensional
complex manifold $\cS^x:=p_{\cS}^{-1}(x\times \Co)$, where the projection onto $x\times \Co$
gives a local biholomorphism $P^x:\cS^x\to \Co$. Let $\bU\subset \Co$ be an open  parallelogram 
whose sides are parallel to vectors $e^{i\alpha}$ and $e^{-i\alpha}$.
Let $\sigma:\bU\to \cS^x$ be a section of $P^x$.  Let also $\elr_{-\alpha}\subset K$
be the ray $[0,\infty).e^{-i\alpha}$.

We prove that
\begin{Thm}\label{prodolzhenie}
 There exists a set $\Gamma\subset \Co$ satisfying:

1) for every point $s\in \Co$, the intersection $(s-K)\cap \Gamma$ is at most finite, 

2) $\bU\subset  (\bU+K)\backslash (\Gamma+\elr_{-\alpha})$;

3) $\sigma$ extends uniquely onto $ (\bU+K)\backslash (\Gamma+\elr_{-\alpha})$.

\end{Thm}
This theorem is proved in Sec.\ref{beskon}: it easily follows from Theorem \ref{prodolzhenie1}, as explained
after its formulation.

Theorem \ref{prodolzhenie} assumes existence of a nonempty set $\bU$ and a section $\sigma$; this fact is the content of the theorem \ref{prodolzhenie1}.

Our construction of $\cS$, as well as the  proof of the above Theorem \ref{prodolzhenie},
 are based on sheaf-theoretical methods
\cite{KS}.  The relation between linear  PDEs and sheaves is  well known and consitutes the subject
 of Algebraic Analysis. {Our paper is also motivated by the classical work of Voros ~\cite[Sec.6]{V83} where an explicit description  singularities of solutions of  \eqref{Laplace} was derived heuristically, see ~\cite{V83}, p.213, line 15 from the bottom; additional insights came from ~\cite{ShSt} and ~\cite{G09}. 
Important works on this problem using methods of hard analysis include ~\cite{AKT91} and ~\cite{KK11}.}

In the next subsection, we will briefly describe the idea of our sheaf-theoretic approach.
\subsection{Introducing sheaves}
We start with introducing a covering space  $X$  of $\Co$, and defining the so-called action function on $X$.
\subsubsection{A covering space $X$}
 Let $\TP$ 
be the set of zeros of $V(x)$  -- ``turning points" of $V(x)$. We assume throughout the
paper that $\TP$ is finite.
 We  also
assume $x_0\not\in \TP$. Let $X$ be the universal covering of $\C\backslash TP$. We can choose 
a determination of $\sqrt{V(x)}$ and
 its primitive $S(x)=\int^x \sqrt{V(\xi)}d\xi$ on $X$.
It will be more convenient for us to use the notation $z:=S(x)$.
 Since $dS(x)$ is nowhere 
vanishing on $X$, we can use $z$ as a local coordinate on $X$. As above, we denote by $s$ the coordinate
on $\Co$, so that $(z,s)$ are local coordinates on $X\times \Co$.

Equation \eqref{Laplace} gets transfered onto $X\times \C$ and in the 
coordinates $(z,s)$ it looks as follows:
\begin{equation} -\Psi_{zz} + \Psi_{ss} + \text{l.o.t.} \ = \ 0 \label{ActionSch} \end{equation}
where l.o.t. stands for a differential operator of order $\le 1$ applied to $\Psi$. 
We now pass to a  sheaf-theoretical consideration.
\subsubsection{Solution sheaf and its singular support}
Let $\Sol$ be the solution sheaf of \eqref{ActionSch}.  
According to ~\cite[Th.11.3.3]{KS}, the singular
 support of $\Sol$ is of a very special form
 which is determined by the highest order term of \eqref{ActionSch} (see Sec. \ref{sisu} for more details).
More specifically,
let $(z,s,\zeta dz + \sigma ds)$ be local coordinates on $T^*(X\times \Co)$. Then 
\begin{equation} S.S. \, \Sol \ \subset \ \Omega_X \ : = \{ (z,s,\zeta dz + \sigma ds) \ : \ \zeta = \sigma \ \text{or} \ \zeta=-\sigma \}. \label{DefOmega} \end{equation}  
It turns out that this condition  contains enough information on $\Sol$ in order to deal with solving
the Cauchy problem. In fact, at this stage, we abstact from our PDE, and only remember that its
solution sheaf has its singular support as specified.

\subsubsection{Initial value problem in sheaf-theoretical terms}

Choose and fix a preimage $\bx_0\in X$ of $x_0$.  
 Define a map  $g: S_\alpha \to X\times\C$  by setting 
$g(\tilde s):= (\bx_0,\pi_{S_\alpha}(\tilde s))$. Cauchy-Kowalewski theorem implies  that
the initial conditions (\ref{initialdata})
are in 1-to-1 correspondence with elements of $\Gamma(S_\alpha, g^{-1}\Sol)$,
see Sec. \ref{inico} for more detail.

As explained in the same Sec., the latter group 
can be identified with $R^0\Hom_{X\times \Co}(Rg_! \Z_{S_\alpha}[-2], \Sol)$.
Therefore, the initial data (\ref{initialdata}) can be interpreted as a map 
\begin{equation} m_\psi:Rg_! \Z_{S_\alpha}[-2]\to \Sol, \label{mpsidefd} \end{equation}  see (\ref{otozha}).

\subsubsection{Semi-orhogonal decomposition of $Rg_!\Z_{S_\alpha}[-2]$.}
Let ${\DerCat}(X\times \C)$ be the bounded derived category of sheaves of abelian
 groups on $X\times \C$.  Let ${\cal C}\subset  {\DerCat}(X\times \C)$ be the full triangulated 
subcategory consisting of all objects whose singular
 support is contained in $\Omega_X$ as in \eqref{DefOmega}.  
Let ${}^\perp \cC\subset {\DerCat}(X\times \C)$
be the so-called left semi-orthogonal complement to $\cC$,
i.e. a full subcategory consisting of all objects $Y$ such that $R\hom(Y,X)=0$ for all $X\in \cC$.
We prove
\begin{Thm} \label{ThOctOpus}
1) There exists the following distinguished triangle in $\bD(X\times \C)$:
$$
\to Rg_!\Z_{S_\alpha}[-2]\stackrel{i_{\Phi}}{\to} \Phi\to\delta\stackrel{+1}{\to}
$$
where $\Phi\in \cC$, $\delta\in {}^\perp\cC$ (``semi-orthogonal decomposition");

2) Stalks of $\Phi$ at every point  of $X\times \Co$ have no negative cohomology. 
\end{Thm}
This theorem coincides (up-to slight reformulations) with Theorem \ref{thsemiort}.
The object $\Phi$ and the map $i_{\Phi}:Rg_!\bZ_{S_\alpha}[-2]\to \Phi$ are constructed 
in Sec \ref{StrucObPhi}-\ref{phikonec}. The bulk of the paper (Sec. \ref{simplified}--Sec. \ref{PfSemior})
is devoted to showing that  the constructed  object $\Phi$ and  a map $i_\Phi$ satisfy the above theorem.

It is well known that the distinguished triangle in  part 1 of Th.\ref{ThOctOpus} , if exists, is unique up to a unique
 isomorphism, meaning that $\Phi$ is defined uniquely. It also follows that the precomposition with $i_\Phi$:
$$ i_\Phi:\circ - \ : \  R^0\Hom_{X\times \Co}( \Phi, \Sol) \ \to \ R^0\Hom(Rg_! \Z_{S_\alpha}[-2], \Sol) $$
is an isomorphism of groups. This implies that the map $m_\psi$, {cf. \eqref{mpsidefd}},  uniquely factors as follows:
$$  Rg_! \Z_{S_\alpha}[-2] \ \to  \Phi\ \stackrel{m_\psi}{\to} \ \Sol. $$ 

Let $\Phi_0:= \tau_{\le 0} \Phi$. 
Condition 2) of Theorem {\ref{ThOctOpus}} implies that $\Phi_0$ is a sheaf of abelian groups. 
We have a composition 
$$ (m_\psi)_{0} \ : \  \Phi_0 \to \Phi \to \Sol. $$

\subsubsection{\'Etale space of $\Phi_0$ and solving the initial data problem} \label{IntroEtale}

Let  ${\cal S'}$ be the \'etale space of $\Phi_0$.  We have a local homeomorphism
$p_{\cS'}:\cS'\to X\times \Co$ so that we have a unique complex structure on $\cS'$
making $p_{\cS'}$ into a local biholomorphism.
It turns out, that the map $(m_\psi)_0$ gives rise to a solution of the transferred equation on $\cal S'$.
Indeed, every such a solution can be equivalently described as an element in 
$\Psi\in \Gamma(\cS';p_{\cS'}^{-1}\Sol)$.   We also have a canonical section
$\rho\in \Gamma(\cS';p_{\cS'}^{-1}\Phi_0)$ (by the construction of the \'etale space); the map
$(m_\psi)_0$ induces a map $\nu:p_{\cS'}^{-1}\Phi_0\to p_{\cS'}^{-1}\Sol$, and we set
$\Psi:=\nu(\rho)$.

It is now straigtforward (Sec. \ref{SolInVP}) to  prove that thus constructed  solution
$\Psi$ is a solution on $\cS'$ of the Cauchy problem with the initial data (\ref{initialdata}).

By choosing an appropriate connected component $\cS$ of $\cS'$ we finish the construction.


\section{Conventions and Notations}

Throughout the paper, we fix an acute angle $\alpha\in (0,\pi/2)$.
\subsection{Various subsets of $\Co$}\label{varioussubsets}
We introduce the following subsets of $\Co$:

--- \ \ $K$  is the closed  cone consisting of all complex numbers whose argument belongs
 to $[-\alpha,\alpha]$, including 0;

--- \ \ $\elr_\alpha:=e^{i\alpha}.[0,\infty)$; $\elr_{-\alpha}:=e^{-i\alpha}.[0,\infty)$;

\subsection{Sector $S_\alpha$}\label{sectorsalpha} We set $S_\alpha:=
{\{ \tau\in \C \, :\, -\alpha< \Im \tau < 2\pi+\alpha \} } $.
Let $\pi_{S_\alpha}:S_\alpha\to\Co$ be the map given by $\pi_{S_\alpha}({\tau}):={e^{\tau}}$.  
{Some c}omplex analysts call $S_\alpha$ an {\'etale} open sector with
aperture $2\pi+2\alpha$.

\subsection{Potential $V(x)$.  Stokes {curves}. Assumptions}
Throughout the paper,
  we fix an entire function $V(x)$ on $\Co$.
We assume that $V(x)$ has only finitely many zeros which are traditionally
called 'turning points'.

The conditions in Sec \ref{fas} below will be also assumed throughout the paper.

\subsubsection{Stokes {curves} and further assumptions}

Let $w\in \C$, $V(w)=0$ be a $k$-fold zero of $V(x)$.  We define
 {\em an $\alpha$-Stokes curve  $z(t)$, $0\leq t<C$,
 emanating from $w$} as follows:

 ---$z(t)$ is a smooth curve with $z(0)=w$ and $-V(z)(dz/dt)^2\in e^{2 i\alpha}\R_{>0}$.

The following facts are well known, {~\cite{EvFe}.}

1) There are exactly  $k+2$  $\alpha$-Stokes curves 
emanating from $w$.

2)  One can choose $C$ (to be a positive real number or $+\infty$) in such a way  that 
either $ z(C):=\lim\limits_{t\to C}$ coincides with another turning point of $V(x)$, or
$z(C)=\infty$. In the latter case we say that {\em the Stokes {curve} terminates at infinity}.

\subsubsection{Further assumptions}\label{fas}
We will  assume the following properties of $V(z)$.

a) All $\alpha$- and $(-\alpha)$-Stokes curves terminate at infinity.

b) Every $\alpha$-Stokes curve intersects only finitely many $-\alpha$-Stokes curves, and 
every $(-\alpha)$-Stokes curve intersects only finitely many $\alpha$-Stokes curves.

\vskip2.5pc

It is well known in the complex WKB theory that
 for every {polynomial} $V(x)$ one can find an $\alpha$ satisfying these assumptions.

\subsection{Universal cover $X$} \label{UniCovSubs}\label{stripsregions}
Let ${\cal U}$  be the complement in $\Co$ to the (finite)  set of  turning points of the potential $V(x)$.
$\alpha$-Stokes curves split ${\cal U}$ into regions called {\em $\alpha$-Stokes regions}; similarly, one can define
$-\alpha$-regions. 
Throughout the paper, we denote by  $X$ the universal cover of ${\cal U}$, and by 
$p_X:X\to \cU\to \Co$ the covering map. 

\subsection{Initial point $x_0$}\label{initialpoint} We fix a point $x_0\in X$. We assume that {\em $p_X(x_0)$ does not
belong to any of $\alpha$- or $-\alpha$-Stokes lines.}


\subsection{Action function on $X$}
 Fix a choice of $\sqrt{V(x)}$ on ${\cal U}$ and a function 
\begin{equation}\label{ac}
z:X\to \C \ \ \ : \ \  dz(x)=\sqrt{V(x)}dx.
\end{equation} 
It follows that  $dz$ is nowhere vanishing, i.e. $z$ is a local coordinate near every point of $X$.
The function $z$ has the meaning of the action function.  We use the notation $z$  because
$z$ will play the role of a local coordinate on $X$. The function $z$ should not be confused with the map
map $p_X:X\to \Co$.

\subsection{Subdivision of $X$ into $\alpha$-strips}\label{poloski}
 Let $\bP\subset {\cal U}$ be a closed $\alpha$-Stokes region on ${\cal U}$, that is, $\bP$ is one of the regions
 into which the complex plane $\Co$ is subdivided by $\alpha$-Stokes curves.

\begin{figure} \includegraphics{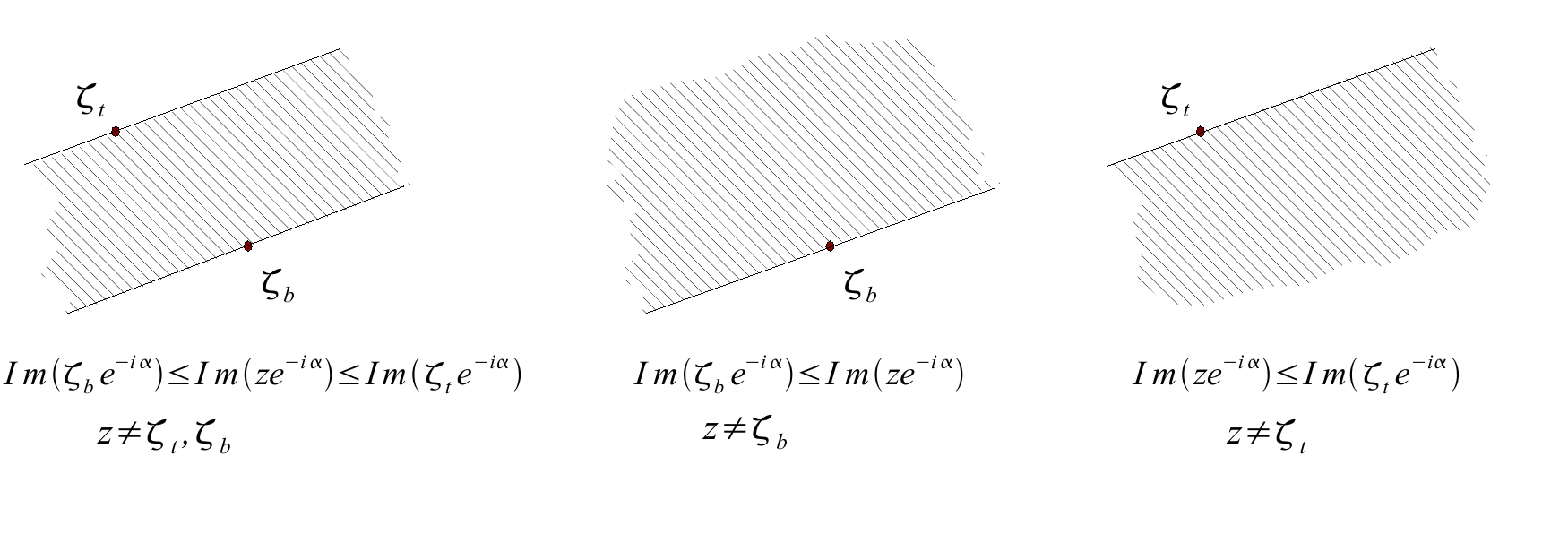} \caption{Three types of $\alpha$-strips}
 \label{CoShWKBp11} \end{figure}

Let us now switch to the universal cover $p:X\to {\cal U}$.
 It follows that $p^{-1}\bP$ splits into a disjoint union of its  connected components
$\bP=\coprod_{\gamma\in \Gamma_\bP} P_\gamma$, where 
$p:P_\gamma \stackrel{\sim}{\to} \bP$. Call each  such $P_\gamma$  (for every $\alpha$-Stokes region $\bP$)
{\em an  $\alpha$-strip}. {By ~\cite[\S 2.2]{EvFe},} 
 the function $z$ maps each $\alpha$-strip homeomorphically  into {\em a generalized strip on $\C$}, i.e.  
a subset 
of $\C$ of one of the following types, fig. \ref{CoShWKBp11}. Here the removed points
 $\zeta_t,\zeta_b$ correspond to the turning points of $V(x)$.

Throughout the paper $\alpha$-strips will be denoted by means of the letter $P$ with different subscripts.
We will often identify $\alpha$ strips with their images in  $\Co$ under $z$.

\subsubsection{{Weakest Possible Assumptions on $V(x)$} } \label{WeakestPossible} 

{The results and proofs of our paper also hold true for any entire function $V(x)$ with finitely many zeros, satisfying the following condition that corresponds to Condition A of ~\cite[\S 2.2]{EvFe}:
$$ \lim_{x\to \infty; \, x\in C} |S(x)| = \infty $$
for any curve $C$ in $\C$ satisfying $\arg S(x) = \pm \alpha$. 
}

\subsubsection{Boundary rays}\label{ray}
Let $P_1,P_2$ be $\alpha$-strips and $P_1\cap P_2\ne \emptyset$. 
Then $\ell=P_1\cap P_2$ is a ray on $X$ which is identified by means of $z$ with either
  $\hat c(\ell) + e^{i\alpha}.(0,\infty)\subset\Co$ or  $\hat c(\ell) - e^{i\alpha}.(0,\infty)\subset\Co$,
where $\hatc(\ell)$ is a complex number.
We denote by $\cL^\alpha$ the set of all such  rays, to be called {\em boundary $\alpha$-rays}.
  Every
boundary $\alpha$- ray belongs to the  boundaries
of exactly two $\alpha$-strips; the boundary of every $\alpha$-strip is  a disjoint union
of boundary $\alpha$-rays. Boundary $\alpha$-rays will be often denoted by the letter $\ell$ with 
different subscripts.

We say that a  boundary $\alpha$-ray $\ell$ {\em goes to the left} if its 
image under $z$ is $\hat c(\ell) - e^{i\alpha}.(0,\infty)$.
Otherwise we say that a boundary $\alpha$-ray {\em $\ell$ goes to the right}. Accordingly, we get a splitting
 $\cL^\alpha=\cL^\alpha_\levo\sqcup\cL^\alpha_\pravo$.
\subsubsection{Strips form a tree}
Consider a graph whose vertices are $\alpha$-strips and we join two distinct vertices with an edge
if the corresponding strips intersect (along some boundary $\alpha$-ray). 
Since $X$ is simply connected, it follows that
this graph is a tree.

\subsection{$(-\alpha)$-Strips}

One has a  similar decomposition of $X$ into $(-\alpha)$-strips which are defined based on $-\alpha$-Stokes 
regions of $X$. Throughout the paper, $-\alpha$-strips will be denoted by means of the letter $\Pi$
with different subscripts. Similar to above,  every $-\alpha$-strip is homeomorphically
mapped under $z$ into a generalized strip whose  each boundary ray  is parallel to the line $e^{-i\alpha}.\R$.
We define  boundary $-\alpha$ rays in a similar way (as intersection rays of two $-\alpha$-strips).
The function $z$ identifies each boundary ray $\ell$ with either
$\hat c(\ell)+e^{-i\alpha}.(0,\infty)$ (we then say $\ell$ goes to the right), or
$\hat c(\ell)-e^{-i\alpha}.(0,\infty)$ ($\ell$ goes to the left). We denote the set of 
all boundary $-\alpha$-rays by $\cL^{-\alpha}$. We have a splitting
$\cL^{-\alpha}=\cL^{-\alpha}_\levo\sqcup \cL^{-\alpha}_\pravo$.
Bounday $-\alpha$-rays will be denoted by the letter $\ell$ with various subscripts.

\subsection{ Interaction of $\alpha$ and $-\alpha$-strips}

\begin{figure} \begin{center} \includegraphics{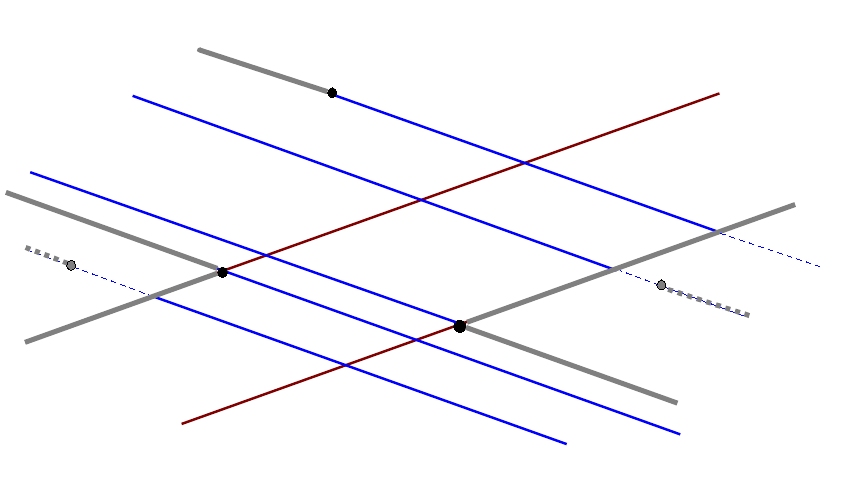} \end{center} \caption{Intersection of an $\alpha$-strip with 
several $(-\alpha$)-strips. Thick gray lines indicate branch cuts arising from
 the many sheets of the projection $X\to \C_x$.} \label{CoShWKBp15} 
\end{figure}

Choose a (red) $\alpha$-strip and look at all $(-\alpha)$-strips (blue) that intersect it. 
These $(-\alpha)$-strips cut the $\alpha$-strips into parallelograms and two semi-infinite
 parallelograms, e.g., fig. \ref{CoShWKBp15}.

\subsection{Categories} For a topological space $M$, we denote by $\bD({M})$ the bounded derived
category of sheaves of abelian groups on $M$. 
\subsubsection{Sub-categories $\cC^Y$; ${}^\perp\cC^Y$}\label{singul}  Let $Y$ be a one dimensional complex
manifold equipped with a local biholomorphism $z:Y\to \Co$.  For example, $Y=X$.

 We then refer to points of $T^*(Y\times \Co)$
as follows $(y,s,\zeta dz,\sigma ds)$, where $y\in Y$, $s\in \Co$ and  $\zeta,\sigma\in \Co$, 
so that $(y,s)\in Y\times \Co$
and $(\zeta,\sigma)$ define the following  real 1-form on $Y\times \Co$:
$$
(\zeta dz+\overline \zeta d\overline z +\sigma ds +\overline \sigma d\overline s)/2.
$$

Let us fix a  closed subset $\Omega_Y\subset T^*(Y\times \Co)$ to consist of all points
$(y,s,\zeta,\sigma)$, where $\zeta=\pm \sigma$.

We denote by $\cC^Y\subset \bD(Y\times \Co)$ the full triangulated subcategory consisting of
all objects $F$ with $\SS(F)\subset \cC^Y$.  We denote by ${}^\perp\cC^Y\subset \bD(Y\times \Co)$
the full subcategory consisting of all objects $G$ such that $R\hom(G,F)=0$ for all $F\in \cC^Y$.

\subsection{Sheaves}\label{LambdaU} Let $Y$ be a topological space endowed with a continuous
map $z:Y\to \Co$.  If $Y\subset X$, then we always assume that $z:Y\to \Co$ is the restriction
of the action function $z:X\to\Co$. We define the following sheaves on $Y\times \Co$:
$$
\Lambda_Y^{K+}:=\bZ_{\{(y,s)|s+z(y)\in K};\quad \Lambda^{K-}_Y:=\bZ_{\{(y,s)|s-z(y)\in K\}}.
$$

\section{ Statement of the problem and Main resuts}
We start this section with  giving a precise formulation for the problem of analytic continuation of solutions
to (\ref{Laplace}).  It turns out to  be  more convenient  to transfer this PDE to $X\times \Co$ by
means of the covering map $p_X:X\to \Co$.

Next, we  give a sheaf-theoretical reformulation of  the probem, and explain how the solution (i.e.
a complex surface $\cS$ along with a local biholomorphism $p_\cS:\cS\to X\times \Co$)
can be deduced from of a certain semi-orthogonal decomposition Theorem \ref{thsemiort}.
The rest of this section is devoted
to proving basic properties of $\cS$ modulo Theorem \ref{thsemiort}, namely
 Hausdorffness and infinite continuabilty in the direction of $K$, which are the main results of this paper.
To this end we need an explicit construction of the distinguished triangle of the semi-orthogonal decomposition
in Theorem \ref{thsemiort}. This triangle is obtained via combining four other distinguished triangles.

It now remains to prove Theorem \ref{thsemiort}, which is now reduced
 to showing that each of the above mentioned four triangles (and hence the combined triangle)
 gives a semi-orthogonal decomposition.   This is done in the rest of the paper.

\subsection{Transfer of the equation ${-}\Psi_{xx}+V(x)\Psi_{ss}=0$ to $X\times \Co$}

\begin{figure} \begin{center} \includegraphics{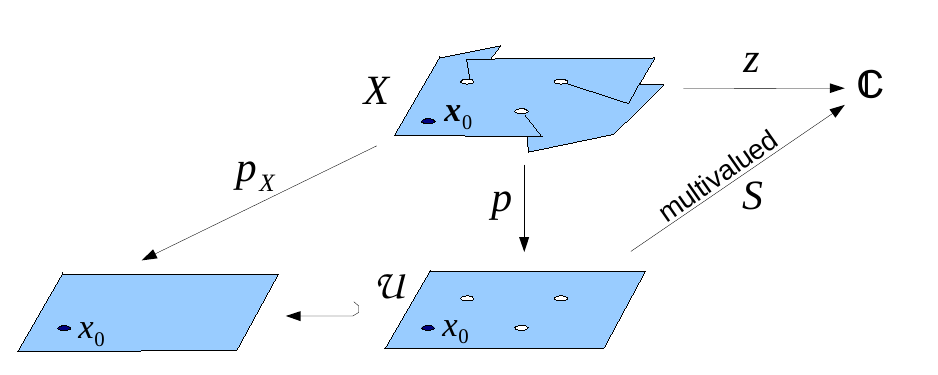} \end{center} \captionempty \label{CoShWKBp82x}
\end{figure}

Our main equation (\ref{Laplace}) can be transferred to $X\times \Co$ 
via the covering map $p\times \Id_\Co:X\times \C \to {\cal U}\times \C$.
 We will use the  action function $z$ on $X$ as in (\ref{ac}). 
 Recall that $z$ is a local coordinate near every point of $X$. Our notation is summarized in fig.\ref{CoShWKBp82x}.

 It is easy to see that the transferred equation has the following form
\begin{equation}\label{trans}
-\Psi_{zz}+ \Psi_{ss}+l.o.t=0,
\end{equation}
where   l.o.t stands for the 
differential operator of order $\leq 1$ applied to  $\Psi$.

Let $\Sol$ be the sheaf of solutions of our transferred equation: ${\Sol}$ is a sheaf of abelian groups
 on $X\times \C$.

\subsection{Singular support of the solution sheaf $\Sol$}\label{sisu}
It is well known that  to every linear PDE on a  manifold $M$ one can put into correspondence a 
 $\cD_M$-module,
where $\cD_M$ is the sheaf of differential operators on $M$; the solution sheaf of the  PDE will then match
with the solution sheaf of the $\cD_M$ module.

In our situation, let us rewrite the equation (\ref{trans}) in the form $L\Psi=0$ for an appropriate linear
differential operator $L$ on $X\times \Co$. Define a $\cD_{X\times\Co}$-module
$\cM$ as follows
$$
\cM=\cD_{X\times \Co}/\cD_{X\times \Co}L.
$$

We then have an obvious isomorphism
\begin{equation}\label{dmod}
\Sol\to \ihom_{\cD_{X\times \Co}}(\cM;\cO_{X\times \Co}).
\end{equation}
Indeed, every solution $\Psi$ of (\ref{trans}) on an open subset $U\subset X\times \Co$
 gives rise to a $\cD_{X\times \Co}$-module
map
$$
l_\Psi:\cD_{X\times \Co}|_U\to  \cO_{X\times\Co}|_U
$$
where $l_\Psi(T):=T\Psi$.  {Then, for any $T'\in \cD_{X\times\Co}(U)$, $l_\Psi(T'L)=T'L\Psi=0$}.
Hence, $l_\Psi$ descends to a map
$$
l_\Psi:\cM|_U\to \cO_{X\times \Co}|_U,
$$
which determines the map  (\ref{dmod}). It is straightforward to see that thus constructed  map (\ref{dmod})
 is in fact
an isomorphism of sheaves.

The usefulness of this fact comes from a  Kashiwara-Schapira's theorem on singular support
of the object
\begin{equation}\label{rsol}
R\ihom_{\cD_{X\times \Co}}(\cM;\cO_{X\times \Co})\in \DerCat(X\times \Co)
\end{equation}
(derived solution sheaf of $\cM$).  Let us now  prove that this object is quasi-isomorphic to $\Sol$.
 
The object (\ref{rsol}) can be conveniently computed by means of the following free resolution $\cR$ of $\cM$:
$$
(\cR) \ \ : \ \ 0\to \cD_{X\times \Co}\stackrel \lambda\to \cD_{X\times \Co}\to  0,
$$
where the map $\lambda$ is as follows: $\lambda(T)=TL$.
 We obtain that the object 
$\ihom_{\cD_{X\times \Co}}(\cM;\cO_{X\times \Co})$ is represented in $\DerCat^b(X\times \Co)$
by the two term complex
$$
\ihom_{\cD_{X\times \Co}}(\cR;\cO_{X\times \Co})
$$
which is  the same as
\begin{equation}\label{cmp} 
0\to \cO_{X\times \Co}\stackrel{L}\to \cO_{X\times \Co}\to 0.
\end{equation}
It is classically known, e.g. ~\cite[Th.3.1.1]{Sch}, that the action of the operator $L$ is locally surjective, meaning that we have
a short exact sequence of sheaves
$$
0\to \Sol\to\cO_{X\times \Co}\stackrel{L}\to \cO_{X\times \Co}\to 0.
$$
This means that the complex of sheaves (\ref{cmp}) is quasi-isomorphic to $\Sol$
so that finally
$$
\Sol\cong R\hom_{\cD_{X\times\Co}}(\cM;\cO_{X\times \Co}).
$$

Kashiwara-Schapira's theorem ~\cite[Th.11.3.3]{KS} says that the singular support of the object (\ref{rsol})
equals the characteristic variety of the $\cD_{X\times\Co}$-module $\cM$. In our situation, this
characteristic variety is well-known to be equal to the zero set  of the principal symbol of the operator $L$.
This set  is
\begin{equation}\label{redbox}
 \{ (z,s,\zeta dz+\sigma ds) \ : \ \zeta=\pm \sigma \} \ \subset \ T^*(X\times \Co),
\end{equation} 
which is the same  as  $\Omega_X$ from Sec. \ref{singul}.
Thus, by Kashiwara-Schapira's theorem, \cite[Th 11.3.3]{KS},  we conclude
that
$$
S.S. \Sol \ = \  \Omega_X, \quad \Sol\in \cC^X,
$$
where $\cC^X$ is defined in Sec. \ref{singul}.

\subsection{Initial conditions}\label{inico} Let $x_0\in X$ be an initial point satifying the  assumptions
from Sec \ref{initialpoint}.
  Let us pose a Cauchy problem for the equation (\ref{trans}) similar
to  Sec. \ref{init}.

Let 
$ S_\alpha$ 
and  $\pi_{S_\alpha}:S_\alpha\to \Co$ be the same as in Sec \ref{sectorsalpha}.
Set $q:=\Id_X\times \pi_{S_\alpha}:X\times S_\alpha\to X\times \Co$.
The equation (\ref{trans}) gets transfered to $X\times S_\alpha$ by means of the map $q$.
The  transfered equation  is of the form
\begin{equation}\label{trans1}
L'\Psi=0,
\end{equation}
where $\Psi$ is an unknown function on $X\times S_\alpha$ and $L'$ is a linear differential operator
$$ {L'=-\Psi_{zz} + e^{-2\tau}\Psi_{\tau \tau} + l.o.t,}$$
and all coefficients of $L'$ are holomorphic on $X\times S_\alpha$ because $\partial_s=e^{-\tau}\partial_\tau$. 
The solution sheaf of this equation is canonically isomorphic to  $q^{-1}\Sol$.

Let us fix two holomorphic functions $\psi_0,\psi_1$ on $S_\alpha$ and pose the initial conditions by requiring 
$$
\Psi(\bx_0,s)=\psi^0(s) \text{ and }\partial_z\Psi(\bx_0,s) = \psi^1(s), \ \ \ s\in S_\alpha.
$$

Cauchy-Kowalewski theorem implies that there exists
a neighborhood 
\begin{equation}\label{kowalewski}
U\subset X\times S_\alpha
\end{equation}
on which there exists a unique solution $\Psi\in \Gamma(U,q^{-1}\Sol)$ of our Cauchy problem.
We have a natural map
$$
\Gamma(U,q^{-1}\Sol)\to \Gamma(\bx_0\times S_\alpha,q^{-1}\Sol|_{\bx_0\times S_\alpha})=
\Gamma(S_\alpha;g^{-1}\Sol),
$$
where 
\begin{equation} g: S_\alpha\to X\times \Co \ \ : \ \ g(s)=(\bx_0,\pi_{S_\alpha}(s)). \label{se8e12a} \end{equation}

Thus, our initial data give rise to an element  
\begin{equation}\label{initcond}
\psi\in \Gamma(S_\alpha;g^{-1}\Sol).
\end{equation}
\subsubsection{Definition of a solution} \label{defmult}
Let us formulate the definition of a multivalued
 solution of the initial value problem in the sheaf-theoretical language.

Suppose we are given  a complex surface $\Sigma$ endowed with
 a local biholomorphism
$p_\Sigma:\Sigma\to X\times \Co$.
We  can  now transfer our differential equation from $X\times \Co$ to $\Sigma$. The solution sheaf
of the transferred equation is then $\Sol_{\Sigma}:=p_\Sigma^{-1}\Sol$.

In order to transfer the initial condition (\ref{initcond}), let us fix 
a factorization  $h$ of the map $g$:
\begin{equation}
S_\alpha\stackrel{h}\to \Sigma\stackrel {p_\Sigma}\to X\times \Co, \label{se26e131}
\end{equation}
where $h$ is a complex-analytic map. 
 We then have
$$
\Gamma(S_\alpha;g^{-1}\Sol)=\Gamma(S_\alpha;h^{-1}p_\Sigma^{-1}\Sol)=\Gamma(S_\alpha;h^{-1}\Sol_\Sigma).
$$
The initial condition  $\psi$ now gives rise to an element
$\psi_\Sigma\in \Gamma(S_\alpha;h^{-1}\Sol_\Sigma)$.

Let us now formulate the notion of a solution to this problem.

We have a restriction map $\rest:\Gamma(\Sigma;\Sol_\Sigma)\to \Gamma(S_\alpha;h^{-1}\Sol_\Sigma)$,
which is defined as follows:
$$
\rest:\Gamma(\Sigma;\Sol_\Sigma)=\hom(\bZ_{\Sigma};\Sol_\Sigma)\to \hom(h^{-1}\bZ_{\Sigma};
h^{-1}\Sol_\Sigma)=
\hom(\bZ_{S_\alpha};h^{-1}\Sol_\Sigma)=\Gamma(S_\alpha;h^{-1}\Sol_\Sigma).
$$

We call an element $\Psi\in \Gamma(\Sigma;\Sol_\Sigma)$ {\em a solution of the initial value problem with the initial
data $\psi$},
if $\rest(\Psi)=\psi_\Sigma$.  Since $\Sol_\Sigma$ is a sub-sheaf of $\cO_\Sigma$ ( the sheaf of analytic functions),
the unicity of analytic continuation implies:  
\begin{Claim} \label{OneAtMost} Suppose $\Sigma$ is connected. For every initial condition $\psi$, the initial 
value problem
has at most a unique solution.
\end{Claim}
\subsubsection{Equivalent formulation} One can define a notion of a solution to the initial value 
problem directly in terms of the initial data $\psi^0,\psi^1$: we can require that a solution
$\Psi$ should satisfy: $\Psi\circ h=\psi^0$; $\displaystyle\frac{ \partial\Psi}{\partial z}{\circ h}=\psi^1$.
It is clear that this new notion of a solution coincides with the one from the previous subsection.
Indeed, the restriction of $\Psi$ onto the neighborhood $U$ as in (\ref{kowalewski}) must coincide
with the solution provided by the Cauchy-Kowalewski theorem.

The notion of solution from this (or previous) subsection is related to the notion of solution from
Sec  \ref{cauchyy} as follows. First of all we have $dz=\sqrt{V(x)}dx$, where $\sqrt{V(x)}$ is a nowhere
vanishing holomorphic function on $X$.  
Set $\psi_0=\psi^0$ and $\psi_1(s)= {\sqrt{V(x_0)}}\psi^1(s)$. We then see that the
 notion of solution of the Cauchy problem
with the initial data $\psi_0,\psi_1$, as in Sec \ref{cauchyy}, coincides with the current notion of solution of
 the initial value problem
given by the initial data $\psi^0,\psi^1$.

\subsubsection{Formulation of the analytic continuation problem}\label{analcont} Our analytic continuation problem is now as follows.
Find a  connected complex surface $\cS$ along with a complex analytic local diffeomorphism 
$p_\cS:\cS\to X\times \Co$ and a factorization $g=hp_\cS$, where $h:S_\alpha\to \cS$ is as in the previous
subsection,  satisfying: given any initial condition $\psi$ as in (\ref{initcond}), there should exist a solution
to the 
initial value problem with the initial data $\psi$. By Claim \ref{OneAtMost}, this solution 
is then unique.

\subsection{Semi-orthogonal decomposition of $\cF_0$} Our main tool in solving 
the analytic continuation problem is  a  certain  semi-orthogonal decomposition theorem, to be 
now stated.

Let ${\cal F}_0=Rg_!\Z_{S_\alpha}[-2]$; let $\cC^X,{}^\perp\cC^X$ be the same as in Sec. \ref{singul}.

\begin{Thm} \label{thsemiort}1) There exists a distinguished triangle
\begin{equation}\label{semiort}
\to{\cal F}_0 \stackrel{i_{\Phi}}{\to} \Phi \to \delta \stackrel{+1}{\longrightarrow}
\end{equation}
where  $\Phi\in \cC^X$ and $\delta\in {}^\perp{\cal C}^X$.

2) The object $\Phi$ belongs to $\bD_{\geq 0}(X\times \Co)$ (that is: the stalks of $\Phi$ at every point
of $X\times \Co$ have no negative cohomology).
\end{Thm} 

{\bf Remark.} The distinguished rectangle (\ref{semiort}) is called ``left {semi-}orthogonal decomposition of $\cF_0$''.
It is  well known that such a triangle, if exists, is unique up-to a unique isomorphism.

We will devote the rest of this section by deducing a solution to the analytic continuation problem from 
this theorem.

\subsubsection{Factorization of the initial condition} \label{OkSec341}
Since $g:S_\alpha\to X\times \Co$ is locally a closed embedding of codimension 2, whose normal bundle
is canonically trivialized, we have a natural transformation of functors 
\begin{equation} \kappa \ : \ g^{-1}\to g^![2]. \label{se26e12} \end{equation}
Since $\Sol$ is microsupported on $\Omega_X$,  one can easily check that 
$\Sol$ is non-characteristic with respect to $g$. Accoriding to \cite[Prop.5.4.13]{KS},  
$\kappa$ induces an isomorphism $g^{-1}\Sol\to g^!\Sol[2]$.
We now have an isomorphism
\begin{equation}\label{otozha}
\Gamma(S_\alpha;g^{-1}\Sol)=R^0\hom(\bZ_{S_\alpha};g^{-1}\Sol)
=R^0\hom(\bZ_{S_\alpha};g^{!}\Sol[2])=R^0\hom(Rg_!\bZ_{S_\alpha}[-2];\Sol).
\end{equation}
Let us denote the images of $\psi$ under these identifications as follows:
$$
\nu_\psi:\bZ_{S_\alpha}\to g^{-1}\Sol;
$$
$$
m'_\psi:\bZ_{S_\alpha}\to g^!\Sol[2];
$$
$$
m_\psi:g_!\bZ_{S_\alpha}[-2]\to \Sol.
$$

Since $\Sol\in \cC$, the semi-orthogonal decomposition theorem {\ref{semiort}} implies that $m_\psi$ uniquely factors as
\begin{equation}\label{mpsi}
m_\psi:Rg_!\bZ_{S_\alpha}[-2]\stackrel{i_{\Phi}}\to \Phi\stackrel{\psi'}\to \Sol.
\end{equation}

The map $i_{\Phi}$ defines, by the conjugacy,  a map
$
\cI':\bZ_{S_\alpha}\to g^!\Phi[2].
$
Let also $\psi_1:g^!\Phi[2]\to g^!\Sol[2]$ be the map induced by $\psi'$.
The equation (\ref{mpsi}) now implies the following factorization (by the conjugacy between $Rg_!$ and $g^!$):
\begin{equation}\label{mpsi1}
m'_\psi:\bZ_{S_\alpha}\stackrel{\cI'}\to g^!\Phi[2]\stackrel{{\psi_1}}\to g^!\Sol[2].
\end{equation}

Since $\Phi[2]$ is microsupported within 
$\Omega_X$,  the transformation $\kappa$, cf. \eqref{se26e12}, induces an isomorphism $\kappa_\Phi:g^{-1}\Phi\to g^!\Phi[2]$
so that we have a unique map $\cI:\bZ_{S_\alpha}\to g^{-1}\Phi$ such that $\cI'=\kappa_\Phi \cI$.
Let $\tpsi:g^{-1}\Phi\to g^{-1}\Sol$ be the map induced by $\psi'$. We can now rewrite  (\ref{mpsi1})
as follows:
\begin{equation}
\nu_\psi:\bZ_{S_\alpha}\stackrel{\cI}{\to} g^{-1}\Phi\stackrel{\tpsi}\to g^{-1}\Sol.
\end{equation}

\subsubsection{Truncation}  The second statement of the theorem implies that $\Phi_0:=\tau_{\leq 0}\Phi$
is a sheaf of abelian groups.  The canonical
map $c:\tau_{\leq 0}\Phi\to \Phi$ induces a map
$c':g^{-1}\Phi_0\to g^{-1}\Phi$.

Let us show that 
\begin{Proposition} The map 
$\cI$ factorizes uniquely through $c'$.
\end{Proposition}
\textsc{Proof.}

We have a distinguished triangle 
$$\stackrel{+1}\to g^{-1}\Phi_0\stackrel{c'}\to g^{-1}\Phi\to g^{-1}\tau_{>0}\Phi\stackrel{+1}\to,
$$
which induces a long exact sequence
$$
\cdots R^{-1}\hom(\bZ_{S_\alpha};g^{-1}\tau_{>0}\Phi)\to R^0\hom(\bZ_{S_\alpha};g^{-1}\Phi_0)
\stackrel{*}\to  R^0\hom(\bZ_{S_\alpha};g^{-1}\Phi)\to 
R^0\hom(\bZ_{S_\alpha};g^{-1}\tau_{>0}\Phi)\cdots.
$$
where the arrow $*$ is given by the composition with $c'$.
Since the functor $g^{-1}$ is exact, $g^{-1}\tau_{>0}\Phi\in \bD_{>0}(S_\alpha)$ so that
$R^{\leq 0}\hom(\bZ_{S_\alpha};g^{-1}\tau_{>0}\Phi)=0$, meaning that the map $*$ is an isomorhism.
 This implies the statement.
$\Box$

Denote by 
\begin{equation}\label{cio}
\cI_0:\bZ_{S_\alpha}\to g^{-1}\Phi_0
\end{equation}
the factorization map (unique by the above Proposition):
$$
\cI:\bZ_{S_\alpha}\stackrel{\cI_0}\to g^{-1}\Phi_0\stackrel {c'}\to g^{-1}\Phi.
$$
We can also factorize:
$$
\nu_\psi:\bZ_{S_\alpha}\stackrel{\cI_0}\to g^{-1}\Phi_0\stackrel{\tpsi\circ c'}\to g^{-1}\Sol.
$$

\subsection{Etale space of $\Phi_0$}\label{etalespacecS}

\subsubsection{Choice of a covering space $\Sigma$} \label{pSigma}

Set $p_{\Sigma}:\Sigma\to X\times \Co$ to be the etale space of $\Phi_0$.
 Observe that the etale space of $g^{-1}\Phi_0$ is $S_\alpha\times_{X\times \Co} \Sigma$.
The etale space of $\bZ_{S_\alpha}$ is $S_\alpha\times \bZ$, 
so that we have a map 
$$S_\alpha\times\bZ\to S_\alpha\times_{X\times \Co}\Sigma$$ over $S_\alpha$, induced by the map $\cI_0$.
Let us restrict this map to $S_\alpha=S_\alpha\times 1$ and denote by $h$ the through map
\begin{equation}\label{opredh}
h:S_\alpha=S_\alpha\times 1\to S_\alpha\times\bZ\to S_\alpha\times_{X\times \Co}\Sigma\to \Sigma.
\end{equation}
By the definition of fibered product, we have $p_{\Sigma} h=g$.

Thus, $p_{\Sigma}:\Sigma\to X\times \C$ and $h:S_\alpha\to \Sigma$ yield a factorization 
of the map \eqref{se8e12a},
 as required by \eqref{se26e131}.

\subsubsection{Solving the initial value problem} \label{SolInVP} 

Let us show that the  initial value problem $\psi\in \Gamma(S_\alpha;g^{-1}\Sol)$ has a solution on $\Sigma$,
 in the sense of Sec. \ref{defmult},  where $\Sigma$ is as in {Sec.\ref{pSigma}}.

We have a canonical map $\lambda:\bZ_\Sigma\to p_{\Sigma}^{-1}\Phi_0$ which comes
 from the canonical section of $p_{\Sigma}^{-1}\Phi_0$: over a point of $\Sigma$ corresponding to
 $((x,s),\varphi_{(x,s)}\in (\Phi_0)_{(x,s)})$, the stalk of this canonical section equals $\varphi_{(x,s)}$.
Let us   apply the functor $h^{-1}$ {and obtain a map}
$$
\cI':\bZ_{S_\alpha}=h^{-1}\Z_\Sigma\to h^{-1}p_\Sigma^{-1}\Phi_0=g^{-1}\Phi_0.
$$
\begin{Lemma} We have $\cI'=\cI$.
\end{Lemma}
\textsc{Proof}
It is easy to see that for each $s\in S_\alpha$, the map $\cI'$ induces the same map
on stalks as $\cI$.
$\Box$

We have a composition $F_\psi:\bZ_\Sigma\stackrel{\lambda}{\to} p_\Sigma^{-1}\Phi_0
\stackrel{\tpsi c'}\to p_\Sigma^{-1}\Sol$.
Let us prove that $F_\psi$ is a solution to the initial value problem.
Indeed, applying $h^{-1}$ induces a map
$\bZ_{S_\alpha}\to g^{-1}\Sol$ which, by virtue of Lemma,  coincides with $\nu_\psi$,
which means that $F_\psi$ is a solution.

\subsubsection{Solving the analytic continuation problem} \label{Sec9o}

We replace $\Sigma$ with its connected component $\cS$ containing the image of $h$.  It is clear
that $\cS$ is a solution to the analytic continuation problem as in Sec. \ref{analcont}.

\subsection{Structure of the object $\Phi$.} \label{StrucObPhi}

We construct the semi-orthogonal  decomposition of $g_!\bZ_{S_\alpha}[-2]$ via  representing 
$g_!\bZ_{S_\alpha}[-2]$ as a cone of some arrow $A\to B$,  and then constructing the semi-orthogonal
decompositions for $A$ and $B$; these decompositions are then glued into the desired
 decomposition of $g_!\bZ_{S_\alpha}[-2]$.

\subsubsection{Decomposition of $\pi_!\Z_{S_\alpha}\in \bD(\Co)$}
\def\restr{\rho}
Let $\pi_{S_\alpha}:{S_\alpha}\to \Co$ be the projection. We are going to represent
 $\pi_{{S_\alpha}!}\bZ_{S_\alpha}$
as a cone of a certain map. 
To this end let us introduce the following subsets of $\Co$ (same as in Sec \ref{varioussubsets})
$$ K \ = \ \{ re^{i\varphi} \ : \ r\ge 0; \ -\alpha\le \varphi \le \alpha  \};   $$
$$ \elr_\alpha \ = \ \{ re^{i\varphi} \ : \ r\ge 0; \ \varphi = \alpha \}; $$
$$ \elr_{-\alpha} \ = \ \{ re^{i\varphi} \ : \ r\ge 0; \ \varphi = -\alpha \}. $$
We have natural restriction maps
$$ \Z_\C \ \stackrel{\restr_{\C K}}\to \ \Z_{K}  \ \stackrel{\restr_{K\elr_{\pm \alpha}}}\to \ \Z_{\elr_{\pm \alpha}}$$
in $\bD(\Co).$ 

The identification $\bZ_{S_\alpha}=\pi_{S_\alpha}^!\bZ_{\Co}$ induces, by conjugacy, a map
$$
p_\Co:\pi_{S_\alpha!}\bZ_{S_\alpha}\to \bZ_\Co.
$$

We are now up to defining a map
$p_K:\pi_{S_\alpha!}\bZ_{S_\alpha}\to \bZ_K.$  We have 
$$\pi^{-1}_{S_\alpha}K=(0,\infty)\times (-\alpha;\alpha]\sqcup
(0,\infty)\times [2\pi-\alpha;2\pi+\alpha)=:K_1\sqcup K_2.
$$
Denote by $i_1:K_1\to S_\alpha$ the closed embedding.
We have natural surjections of sheaves on $S_\alpha$:\\
$\iota_1:\bZ_{S_\alpha}\to i_{1{!}}\bZ_{K_1}$ and $\iota_2:\bZ_{S_\alpha}\to i_{2{!}}\bZ_{K_2}$.

The map $\pi_{S_\alpha}$ induces open embeddings $\pi_{S_\alpha}i_1:K_1\to K$  and 
$\pi_{S_\alpha}i_2:K_2\to K$.
We have $\pi_{S_\alpha}(K_1)=K\backslash \elr_\alpha$;  
$\pi_{S_\alpha}K_2=K\backslash \elr_{-\alpha}$. These open embeddings induce the following embeddings
of sheaves on $\Co$: $\pi_{S_\alpha!}i_{1!}\bZ_{K_1}\to  \bZ_K$; $\pi_{S_\alpha!}i_{2!}\bZ_{K_2}\to 
\bZ_K$. Combining these maps with $\iota_1,\iota_2$,  we get
the following through map
$$
p_K:\pi_{S_\alpha!}\bZ_{S_\alpha}\stackrel{\iota_1}\to \pi_{S_\alpha!}i_{1!}\bZ_{K_1}\to \bZ_K.
$$

One checks that $\restr_{K\elr_\alpha}p_K=\restr_{C \elr_\alpha}p_C$.
Let us now construct the following sequence of maps
\begin{equation} \xymatrix{  && \Z_\C \ar[rr]^{\restr_{C\elr_\alpha}} && \Z_{\elr_\alpha}& \\
0 \ar[r] & \pi_{S_\alpha!}\Z_{S_\alpha} \ar[ur]^{p_C} \ar[rd]^{p_K} &\oplus &&\oplus  \ar[r] & 0 \\
&& \Z_{K} \ar[rr]^{\restr_{K\elr_{-\alpha}}} \ar[uurr]^{-\restr_{K\elr_\alpha}} && \Z_{\elr_{-\alpha}}& } 
\label{eq145} 
\end{equation}
It is clear that the composition of every  two consecutive maps is zero. In fact, {\em this sequence is exact}, which can
be shown by proving exactness of the induced sequences on stalks for every point $z\in \Co$.

Let $g':\Co\to X\times \Co$ be given by $g'(s)=(\bx_0,s)$ so that $g=g'\pi_{{S_\alpha}}$.  Applying $g'_!$ to the exact sequence above
yields the following exact sequence of sheaves:
\begin{equation} \xymatrix{  && \Z_{\bx_0\times \C} \ar[rr] && \Z_{\bx_0\times \elr_\alpha}& \\
0 \ar[r] & g_! \Z_{S_\alpha} \ar[ur]^{g'_!(p_C)} \ar[rd]^{g'_!(p_K)} &\oplus &&\oplus  \ar[r] & 0 \\
&& \Z_{\bx_0\times K} \ar[rr]^{g'_!(\restr_{K\elr_{-\alpha}})} \ar[uurr]^{-g'_!(\restr_{K\elr_\alpha})} && 
\Z_{\bx_0\times \elr_{-\alpha}}& } 
\label{eq144} 
\end{equation}

\subsubsection{Semi-orthogonal
 decomposition for $\Z_{\bx_0\times \Co},\Z_{\bx_0\times K},\Z_{\bx_0\times \elr_{\pm \alpha}}$} 
\label{FourOrtho}

\begin{Thm} \label{CKl} There are objects $\Phi^\C$, $\Phi^{K}$, $\Phi^{\elr_\alpha}$, 
$\Phi^{\elr_{-\alpha}}$ in the category of sheaves of abelian groups and maps in $\DerCat^b(X\times\C)$:
$$ \begin{array}{ccc}  i_{\Phi^\C}:\Z_{\bx_0\times \C}[-2] \to \Phi^\C && 
i_{\Phi^K}:\Z_{\bx_0\times K}[-2] \to \Phi^K  \\ i_{\Phi^{\elr_\alpha}}:\Z_{\bx_0\times \elr_\alpha}[-2] \to \Phi^{\elr_\alpha} &&
i_{\Phi^{\elr_{-\alpha}}}: \Z_{\bx_0\times \elr_{-\alpha}}[-2] \to \Phi^{\elr_{-\alpha}} \end{array}
$$
whose cones are in ${}^\perp {\cal C}$ and 
$\Phi^{\C}, \Phi^K, \Phi^{\elr_\alpha}, \Phi^{\elr_{-\alpha}} \in {\cal C}$.
\end{Thm}

 Based on this theorem, let us construct a semi-orthogonal decomposition of $g_!\bZ_{S_\alpha}$.
Let us rewrite the sequence (\ref{eq144}) as 
$$
0\to g_!\bZ_{S_\alpha}\stackrel\iota\to \XX\stackrel q\to \YY\to 0,
$$
where
$\XX=\bZ_{\bx_0\times \Co}\oplus \bZ_{\bx_0\times K}$ and $\YY=\bZ_{\bx_0\times \elr_\alpha}\oplus
 \bZ_{\bx_0\times \elr_{-\alpha}}$.
By virtue of Theorem \ref{CKl} we have semi-orthogonal decompositions of $\XX$ and $\YY$
$$
\to\xi\to \XX[-2]\stackrel{P_\XX}\to \XX'\stackrel{+1}{\to};\quad \eta\to \YY[-2]\stackrel{P_\YY}\to 
\YY'\stackrel{+1}{\to},
$$
where $\XX'=\Phi^\Co\oplus \Phi^K\in \cC$; $\YY'=\Phi^{\elr_\alpha}\oplus \Phi^{\elr_{-\alpha}}\in\cC$;
$\xi,\eta\in {}^ \perp\cC$.
The map $P_\YY q:\XX[-2]\to \YY'$, by the univerality of $\XX'$, uniquely factors as 
\begin{equation}\label{zvyazok}
P_\YY q=q'P_\XX
\end{equation}
 for some $q':\XX'\to \YY'$
so that  we have a commutative diagram
$$
\xymatrix {\XX[-2]\ar[r]^{q}\ar[d]^{P_\XX} & \YY[-2]\ar[d]^{P_\YY}\\
\XX'\ar[r]^{q'}&\YY'.}
$$
We have $g_!\bZ_{S_\alpha}[-2]\cong \Cone q[-1]$.
Set $\Phi:=\Cone q'[-1]$.
It is well known that the commutative diagram above implies existence of a map 
\begin{equation} i_\Phi:g_!\bZ_{S_\alpha}[-2]\to {\Phi} \label{iPhiConstrd} \end{equation}
fitting into the following commutative diagram whose {rows} are distinguished triangles:
$$\xymatrix{
\ar[r]& g_!\bZ_{S_\alpha}[-2]\ar[r]\ar[d]^{i_\Phi}& \XX[-2]\ar[r]^q\ar[d]^{P_\XX}&\YY[-2]\ar[r]^{+1}
\ar[d]^{P_\YY}&\\
\ar[r]& \Phi\ar[r]& \XX'\ar[r]^{q'}&\YY'\ar[r]^{+1}&}
$$

Furthermore, we have a distinguished triangle
$$
\to \Cone(i_\Phi)\to \Cone P_\XX\to \Cone P_\YY\stackrel{+1}\to,
$$
which implies that $\delta:=\Cone(i_\Phi)\in {}^\perp \cC$ satisfies all the conditions of Theorem
 \ref{thsemiort}.

We will now give an explicit description of the sheaves $\Phi^\Co,\Phi^K,\Phi^{\elr_{\pm \alpha}}$, as 
well as
the maps $i_{\Phi^\Co},i_{\Phi^K},i_{\Phi^{\elr_{\pm \alpha}}}$ from Theorem \ref{CKl}.
This theorem will be proven below.

\subsubsection{$\Phi^\Co$}
We set $\Phi^\Co=\bZ_{\XX\times \Co}$. We have a codimension 2 embedding
$$
i_{\Co,\bx_0}:\Co\to X\times \Co,
$$
so that we have a  natural map
$$
\bZ_{\bx_0\times \Co}[-2]\to \bZ_{X\times \Co},
$$
and we assign $i_{\Phi^\Co}$ to be this map.
\subsection{Notation: convolution functor $\bD(X\times \Co)\times \bD(\Co)\to \bD(X\times \Co)$}
 Define a convolution functor 
\begin{equation} * \ : \ \bD(X\times \C) \times \bD(\C) \ \to \ \bD(X\times\C) \label{jn8eq1}\end{equation}
as follows. Let ${\cal F}\in\bD(X\times \C)$, $\Sigma\in\bD(\C)$. Let 
$$ a: X\times \C \times \C \to X\times \C \ : \ a(x,s_1,s_2) = (x,s_1+s_2) $$
Set 
$$ {\cal F} * \Sigma \ = \ Ra_!({\cal F} \boxtimes \Sigma ). $$

\subsection{Construction of $\Phi^K$}
\label{fik}
\subsubsection{Subdivision into $\alpha$-strips}
 Let us split $X$ into $\alpha$-strips as in Sec. \ref{poloski}. We will freely use the notation from this
section below.

We  will define a sheaf $\Phi^K$ on $X\times \Co$ via prescribing the following data.

1) For each $\alpha$-strip $P$ we  will  define a sheaf
$\Phi^K_{P}$ on $P\times \Co$. Recall that  by  $\alpha$-strip we always mean a closed $\alpha$-strip.
 
2) Let $P_1$, $P_2$ be intersecting closed $\alpha$-strips so that $P_1\cap P_2=\ell\in \cL^\alpha$.
We will construct an isomorphism
$$
\Gamma^{P_1P_2}_{\Phi^K}:\Phi^K_{P_1}|_{\ell\times \Co}\stackrel\sim\to \Phi^K_{P_2}|_{\ell\times \Co},
$$
 where we assume $\Gamma^{P_2P_1}_{\Phi^K}=(\Gamma^{P_1P_2}_{\Phi^K})^{-1}$.


Since every triple of distinct closed $\alpha$-strips has an empty intersection,  the data 1),2) define
a sheaf $\Phi^K$ unambiguously.
 More precisely,  there exists
 a 
sheaf $\Phi^K$  endowed with the following structure:

 --- isomorphisms $j_P:\Phi^K|_{P\times \Co}\stackrel\sim\to \Phi^K_P$ for every $\alpha$-strip $P$ satisfying:
for every pair of intersecting strips $P_1$ and $P_2$, $P_1\cap P_2=\ell$, the following maps must  conicide:
$$
\Phi^K|_{\ell\times \Co}\stackrel{j_{P_1}|_{\ell}}\longrightarrow
  \Phi^K_{P_1}|_{\ell\times \Co}\stackrel{\Gamma^{P_1P_2}_{\Phi^K}}\longrightarrow
 \Phi^K_{P_2}|_{\ell\times \Co}
$$
and 
$$
\Phi^K|_{\ell\times \Co}\stackrel{j_{P_2}|_{\ell\times \Co}}\longrightarrow \Phi^K_{P_2}|_{\ell\times \Co}.
$$
The sheaf $\Phi^K$ is unique up-to a unique isomorphism compatible with  all the structure maps $j_P$.

\subsubsection{Words} We will use the notation from Sec. \ref{ray}.
Let $\wa$ be the set of words from the alphabet ${\cal L}^\alpha \cup \{ L, R\} $ such that: 

1) each word is non-empty and its rightmost letter in $L$ or $R$

2) every word is either of the form
\begin{equation}  (\ell_n ... \ell_3 \ell_2 \ell_1 L) \label{Lword} \end{equation}
where $$ \ell_1, \ell_3, \ell_5, ... \in {\cal L}^\alpha_\pravo, \ \ \ \ell_2, \ell_4, \ell_6, ... \in {\cal L}^\alpha_\levo $$
or
\begin{equation} (\ell_n ... \ell_1 R) \label{Rword} \end{equation}
where
$$ \ell_1, \ell_3, ... \in {\cal L}^\alpha_\levo; \ \ \ \ell_2, \ell_4, \ell_6, .. \in {\cal L}^\alpha_\pravo $$
(alternating pattern).

Let $\wa= \wa_\levo \cup \wa_\pravo $, where
$$ \wa_\levo = \{ (\ell_n  ... )\ : \ \ell_n \in {\cal L}^\alpha_\levo \} \cup \{ L \}; \ \ \ \ \ 
\wa_\pravo = \{ (\ell_n ... )\ : \ \ell_n \in {\cal L}_\pravo^\alpha \} \cup \{ R\} .$$
Let us stress
that $\wa_\levo$ contains words both ending with $L$ and words ending with $R$, and the same 
is true for $\wa_\pravo$.

\subsubsection{Sheaves $S_\ell, S_w$ on $\Co$}
Given a ray $\ell\in \cL_\levo^\alpha$, let is define the following sheaf on $\Co$: 
\begin{equation} S_\ell := \Z_{\{ s\in +2\hat c(\ell) +K \} }, 
 \label{jn10eq81} \end{equation}
Given a ray $\ell\in \cL_\pravo^\alpha$, we set
 $$S_\ell := \Z_{\{ s\in -2\hat c(\ell) +K \} }. $$

 Set
\begin{equation} S_L := \Z_{\{ s\in z(\bx_0)+K \} }; \ \ \ S_R := \Z_{\{ s\in -z(\bx_0)+K \} }. \label{jn8eq340} 
\end{equation}



%

Let 
$$ S_w := S_{\ell_1}*S_{\ell_2} * ... * S_{\ell_n} * S_L,  \ \ \ \text{if} \ w:=\ell_1 .. \ell_n L \in \wa,$$
$$ S_w := S_{\ell_1}*S_{\ell_2} * ... * S_{\ell_n} * S_R,  \ \ \ \text{if} \  
w:=\ell_1 ... \ell_n R\in \wa,$$
where $*$ denotes the convolution functor $\bD(\Co)\times \bD(\Co)\to \bD(\Co)$
 in the sense of \eqref{jn8eq1}.
It is clear that
$S_w=\bZ_{\hatc(w)+K}$, where we set:
\begin{equation}
\hatc{(}w{)}=z(\bx_0)-2\hatc(\ell_n)+2\hatc(\ell_{n-1})-\cdots+(-1)^n\hatc(\ell_1) \ \ \ \text{if} \ w:=\ell_1 .. \ell_n L; \label{oc21e7}
\end{equation}
\begin{equation}
\hatc{(}w{)}=-z(\bx_0)+2\hatc(\ell_n)-2\hatc(\ell_{n-1})+\cdots-(-1)^n\hatc(\ell_1) \ \ \ \text{if} \ w:=\ell_1 .. \ell_n R. \label{oc21e8}
\end{equation}
Let us further set
\begin{equation} S_- := \oplus_{w\in \wa_\pravo} S_w; \ \
 \ \ S_+ := \oplus_{w\in \wa_\levo} S_w. \label{SpmDefd} \end{equation}

\subsubsection{Definition of $\Phi^K_P$} \label{definitionphikp}
For any subset $U\subset X$, we define the following sheaf on $U\times \Co$:
\begin{equation} \Phi_U^K  \ := \ \Lambda^{K-}_U * S_- \ \oplus \ \Lambda^{K+}_U * S_+, \label{jn8eq415} \end{equation}
where 
$\Lambda_U^{K\pm}:=\bZ_{\{(x,s)|s\pm z(x)\in K\}}
$
are the same as in Sec \ref{LambdaU}.

 Set $\Phi^{K\pm}_U = \Lambda^{K\pm}_U * S_\pm $. In particular, we have defined
sheaves $\Phi^{K\pm}_P$ for every $\alpha$-strip $P$.

\subsubsection{Constructuion of the identification $\Gamma_{\Phi^K}^{P_1P_2}$}\label{oboznvt}

We have identifications:
$$
\Phi^K_{P_1}|_{\ell\times \Co}=\Phi^K_{P_2}|_{\ell\times \Co} =
\Lambda^{K+}_\ell*S_+\oplus \Lambda^{K-}_\ell*S_-.
$$

Let us now construct the gluing maps 
$$
\Gamma^{P_1P_2}_{\Phi^K}:\Lambda^{K+}_\ell*S_+\oplus \Lambda^{K-}_\ell*S_-\to 
 \Lambda^{K+}_\ell*S_+\oplus \Lambda^{K-}_\ell*S_-.
$$
There are two cases. 

\begin{figure} \includegraphics{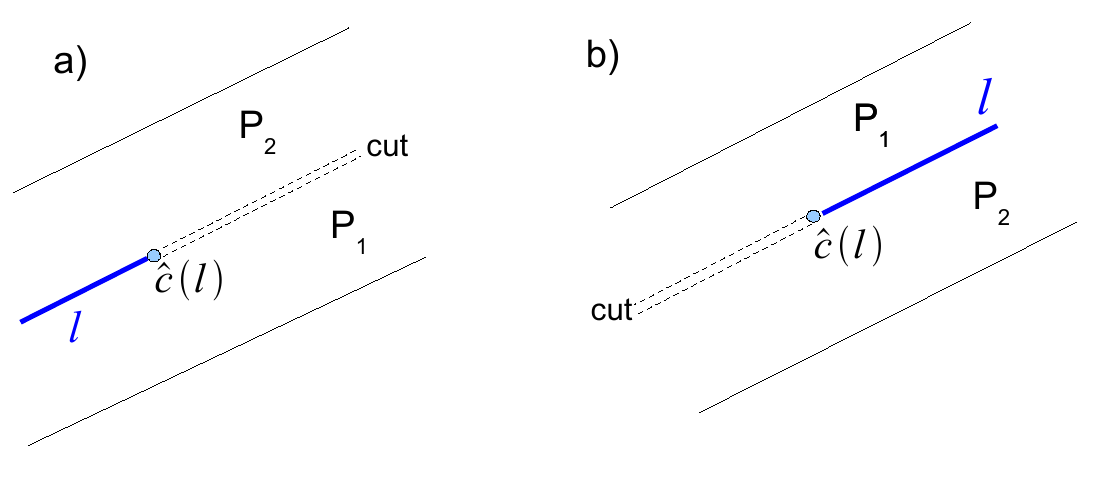} \caption{Notations in the construction of the sheaf $\Phi^K$: a) 
$\ell\in {\cal L}_\levo^{\alpha}$, b) $\ell\in {\cal L}_\pravo^{\alpha}$ } \label{CoShWKBp80} \end{figure}

\underline{Case A).} Let $\ell \in {\cal L}_\levo^\alpha$.\\
 Assume that the  $z$-image  of $P_2$ is above the $z$-image of $P_1$ in the complex plane, fig. \ref{CoShWKBp80}, a).

Let us define the following morphism of sheaves on $\ell\times \C$ 
\begin{equation}
\nu^K_\ell: \Lambda^{K-}_{\ell}\ \to S_{\ell}* \Lambda^{K+}_{\ell}, \label{jn9eq605}
\end{equation}
or, more explicitly, 
\begin{equation} \nu^K_\ell:\Z_{\{z\in \hat c(\ell) -e^{i\alpha}.[0,\infty), \ s-z\in K\}}
 \to  \Z_{\{s\in  2\hat c(\ell)+K\}} *\Z_{\{z\in \hat c(\ell)-e^{i\alpha}.[0,\infty), s+z\in K \}}. \label{jn10eq8}
 \end{equation}
 We have
 $\Z_{\{s\in  2\hat c(\ell)+K\}} *\Z_{\{z\in \hat c(\ell)-e^{i\alpha}.[0,\infty), s+z\in K \}}=
\Z_{\{z\in \hat c(\ell)-e^{i\alpha}.[0,\infty); s\in -z+2\hat c(\ell)+K \}}$.
The map $\nu^K_\ell$  is thus determined by a closed embedding
$$
\{z\in \hat c(\ell)-e^{i\alpha}.[0,\infty); s\in -z+2\hat c(\ell)+K \}
\subset\{z\in \hat c(\ell) -e^{i\alpha}.[0,\infty), \ s-z\in K\}.
$$
Let us now define a map
$$ N^K_\ell: \Lambda^{K-}_\ell* S_- \to \Lambda^{K+}_\ell* S_+. $$
as follows. We have
$\Lambda^{K-}_\ell*S_-=\bigoplus_{w\in \wa_\pravo} \Lambda^{K-}_{\ell}*S_w$.

We denote
\begin{equation}\label{nellw}
N_\ell^w:\Lambda^{K-}_\ell*S_w\stackrel {\nu^K_\ell}\to
 \Lambda^{K+}_\ell*S_\ell*S_w=\Lambda^{K+}_\ell*S_{\ell w}.
\end{equation}
Observe that $\ell w\in \wa_\levo$, so that $\Lambda^{K+}_\ell*S_{\ell w}$ is a direct summand
of $\Lambda^{K+}_\ell*S_+$. 
We therefore can define $N^K_\ell$ as the direct sum of all $N_\ell^w$, $w\in \wa_\pravo$.

Let $$\bN^K_\ell:\Lambda^{K-}_\ell*S_-\oplus \Lambda^{K+}_\ell*S_+\to
 \Lambda^{K-}_\ell*S_-\oplus \Lambda^{K+}_\ell*S_+$$
be the extension of $N^K_\ell$ whose all components  are zero, except for 
$\Lambda^{K-}_\ell*S_-\to \Lambda^{K+}_\ell*S_+$ which equals $N^K_\ell$.

We set \begin{equation}\Gamma^{P_1P_2}_{\Phi^K}:=\Id+\bN^K_\ell. \label{oct3e3a}
\end{equation}
Finally, we set 
$$
\Gamma^{P_2P_1}_{\Phi^K}:=(\Gamma^{P_1P_2}_{\Phi^K})^{-1}=\Id-\bN^K_\ell.
$$

Let us now rewrite the definition for the gluing maps in a more  uniform way.
Let $P$ and $P'$ be two neighboring strips such that 
$P\cap P'$ goes to the left.  Let us define the sign 
\begin{equation}\label{vtleft1}
\text{$\vt(P,P')=1$ if $P'$ is above $P$, and $\vt(P,P')=-1$ if $P'$ is below $P$. }
\end{equation}
 We now  have
\begin{equation}\label{gammappleft}
\Gamma^{PP'}_{\Phi^K}:=\Id+\vt(P,P')\bN^K_\ell .
\end{equation}

\underline{Case B).} Let $\ell\in {\cal L}_\pravo$,  fig. \ref{CoShWKBp80},b). 
Assume first that $P_2$ is below $P_1$.

The formulas are similar to the case A but $+$ and $-$ get exchanged. We have a map
\begin{equation}\nu^K_\ell: \Lambda^{K+}_{\ell} \to \Lambda^{K-}_{\ell} * S_\ell \label{jn29eq142}
 \end{equation}
which gives rise to a map
\begin{equation}\label{nell}
 N_\ell^K \ : \ \Lambda^{K+}_\ell * S_+ \ {\stackrel{\nu^K_\ell}\to} \  
\Lambda^{K-}_\ell  * S_\ell *S_+ \  \to \  \Lambda^{K-}_\ell *S_- . 
\end{equation}

Similar to above, we define a map
 $$
\bN^K_\ell:\Lambda^{K+}_\ell*S_+\oplus \Lambda^{K-}_\ell*S_-\to 
\Lambda^{K+}_\ell*S_+\oplus \Lambda^{K-}_\ell*S_-
$$
as the extension of $N^K_\ell$ whose all components are zero except for
$\Lambda^{K+}_\ell*S_+\to \Lambda^{K-}_\ell*S_-$ which is $N^K_\ell$. 

We set \begin{equation}\Gamma^{P_1P_2}_{\Phi^K}:=\Id+\bN^K_\ell;\label{oct3e3b}
\end{equation}
$$
\Gamma^{P_2P_1}_{\Phi^K}:=(\Gamma^{P_1P_2}_{\Phi^K})^{-1}=\Id-\bN^K_\ell.
$$

Similarly to above, let us rewrite the definition as follows. Let $P$ and $P'$ be two neighboring strips such that 
$P\cap P'$ goes to the right.  Let us define the sign 
\begin{equation}\label{vtright}
\text{$\vt(P,P')=1$ if $P'$ is below $P$; $\vt(P,P')=-1$ if $P$ is below $P'$.}
\end{equation}

We now have
\begin{equation}\label{gammappright}
\Gamma^{PP'}_{\Phi^K}:=\Id+\vt(P,P')\bN^K_\ell.
\end{equation}

\subsubsection{Description of the map $i_{\Phi^K}:\bZ_{\bx_0\times K}[-2]\to \Phi^K$}\label{defiphik}

 Let $P_0$ be the strip such that $\bx_0\in \Int P_0$. 


By construction, 
$$ \Phi^K|_{\Int P_0\times \C} = \Lambda^{K+}_{\Int P_0} * S_+ \oplus \Lambda^{K-}_{\Int P_0} * S_-. $$
The direct summand inclusions
$$ S_L\to S_+ \ ; \ \ \ S_R \to S_- $$
induce   maps
$ \Lambda^{K+}_{\Int P_0} * S_L 
\to 
\Lambda^{K+}_{\Int P_0}* S_+ ,$
$  \Lambda^{K- }_{\Int P_0}* S_R \to 
\Lambda^{K- }_{\Int P_0}* S_- .$

We have the following closed embedding of codimension 2: 
$$ \left\{ \begin{array}{c} x=\bx_0 \\ s\in  
K \end{array} \right\} \hookrightarrow  \left\{ \begin{array}{c} x\in \Int P_0\\ s\pm z(x)\in \pm z(\bx_0)+K \end{array}
 \right\}. $$


We have the following maps in $\bD(\Int P_0\times \Co)$:
\begin{equation} \xymatrix{ & \Z_{ \left\{ \begin{array}{c} x\in \Int P_0\\ s+z(x)\in z(\bx_0) +K \end{array}
 \right\} }  \ar[r] &  \Lambda^{K+}_{\Int P_0} *S_L \\
\Z_{ \left\{ \begin{array}{c} x=\bx_0\\ s\in K \end{array} \right\} } [-2] \ar[ur] \ar[dr] & \oplus & \oplus 
\ar@{}[r]|{\to} & \Phi^K|_{\Int P_0 \times \C} \\
& \Z_{ \left\{ \begin{array}{c} x\in \Int P_0\\ s-z(x)\in  -z(\bx_0)+K \end{array} \right\} }  \ar[r] &  
\Lambda^{K-}_{\Int P_0} *S_R
} \label{i11eq128} \end{equation}

We  thus have constructed a map
\begin{equation}\label{intophik}
 \Z_{ \left\{ \begin{array}{c} x=\bx_0\\ s\in K \end{array} \right\} } [-2]  \
 = \bZ_{\bx_0\times K}[-2] \to \Phi^K|_{\Int P_0\times \Co} \end{equation}

As $\bZ_{\bx_0\times K}[-2]$ is supported on $\Int P_0$, our map extends canonically to  a map
$ i_{\Phi^K}:\Z_{\bx_0\times K}[-2] \to \Phi^K $ in $\bD(X\times \Co)$.

\subsection{Alternative construction of $\Phi^K$ via $-\alpha$-strips}
It is clear that one can repeat all the steps of the previous section using $-\alpha$-strips instead of
$\alpha$ strips. We denote the resulting sheaf $\Psi^K$; we also get an  analogue of the map $i_{\Phi^K}$,
to be denoted by 
\begin{equation}\label{intopsik1}
i_{\Psi^K}:\bZ_{\bx_0\times K}[-2]\to \Psi^K.
\end{equation}
 By means of $\Psi^K$, we also
get a semiorthogonal decomposition of $\bZ_{\bx_0\times K}[-2]$. This implies the existence of a unique isomorphism
\begin{equation}\label{opredipsiphi}
I_{\Psi\Phi}:\Psi^K\to\Phi^K
\end{equation}
 satisfying $i_{\Phi^K}=I_{\Psi\Phi}i_{\Psi^K}$ (because of the unicity
of semiorthogonal decomposition).
We will now briefly go over the construction of $\Psi^K$.
\subsubsection{Notation for $-\alpha$-strips}
Let ${\cal L}^{-\alpha} = {\cal L}_\levo^{-\alpha} \cup {\cal L}_\pravo^{-\alpha}$ be the set of all
 intersection rays of
 $-\alpha$-strips. 
$ {\cal L}^{-\alpha}_\levo$ consists of the rays
 going to the left, ${\cal L}^{-\alpha}_\pravo$ consists of the
 rays going to the right.  Every ray $\ell\in \cL_\levo^{-\alpha}$ (resp. $\ell\in  \cL_\pravo^{-\alpha}$)  is of the form
$p_z(\ell)=\hat c(\ell)-(0,\infty)e^{-i\alpha}$;  (resp. $p_z(\ell)=\hat c(\ell)+(0,\infty)e^{-i\alpha}$)
for some $\hat c(\ell)\in \Co$.


Let $\wma, \wma_\levo, \wma_\pravo$ be defined in the same way as $\wa,\wa_\levo, \wa_\pravo$. 
($\wma_\levo$ consists of
 words of the form $w=\ell_n \ell_{n-1}... \ell_2 \ell_1 L$ or
 $w=\ell_n ... \ell_1 R$ where $\ell_n\in{\cal L}^{-\alpha}_\levo$ and we have an alternating 
pattern $\ell_{n-1}\in  {\cal L}^{-\alpha}_\pravo$, $\ell_{n-1}\in  {\cal L}^{-\alpha}_\levo$,... ; 
if $\ell_1\in {\cal L}^{-\alpha}_\pravo$, then the right-most letter of $w$ is $L$; if $\ell_1\in  {\cal L}^{-\alpha}_\levo$ then
 the right-most letter of $w$ is $R$; we also add  a one letter word $L$ to $\wma_\levo$. )
Similarly to the previous section, we set
$$\tilde S_\ell:= \Z_{\{s : s \in 2\hat c(\ell)+K\} }\in \DerCat(\C), \ \ \ \ell\in {{\cal L}^{-\alpha}_\levo};$$
$$\tilde S_\ell:= \Z_{\{s : s \in -2\hat c(\ell)+K\} }\in \DerCat(\C), \ \ \ \ell\in{{\cal L}^{-\alpha}_\pravo};$$
$$\tilde S_L := \Z_{\{s : s \in z(\bx_0)+K\} }\in \DerCat(\C); $$
$$\tilde S_R := \Z_{\{s : s \in -z(\bx_0)+K\} }\in \DerCat(\C) $$
 For $w\in \wma$, $w=\ell_n...\ell_1 (L \ \text{or} \ R)$ set 
$$ \tilde S_w = \tilde S_{\ell_n} * \tilde S_{\ell_{n-1}} * ... * \tilde S_{\ell_1} * ( \tilde S_L \ \text{or} \ \tilde S_R). $$

Set 
$$ \tilde S_{-} := \oplus_{w\in \wma_\pravo} \tilde S_w; \ \ \ \ \tilde S_{+} := \oplus_{w\in \wma_\levo} \tilde S_w . $$

\subsubsection{Sheaves  $\Psi^K_\Pi$}
Let $\Lambda_U^{K{\pm}}$ mean the same thing as in {Sec.\ref{LambdaU}}.
On every $(-\alpha)$-strip $\Pi$ consider the sheaf on $\Pi$
$$ \Psi_\Pi^K\ := \  \Lambda^{K+}_\Pi *\tilde S_+ \oplus \Lambda^{K-}_\Pi * \tilde S_-.  $$

\subsubsection{Gluing maps}
Let $\Pi_1$, $\Pi_2$ be neighboring strips, $\Pi_1\cap \Pi_2 = \ell$.

\begin{figure} \includegraphics{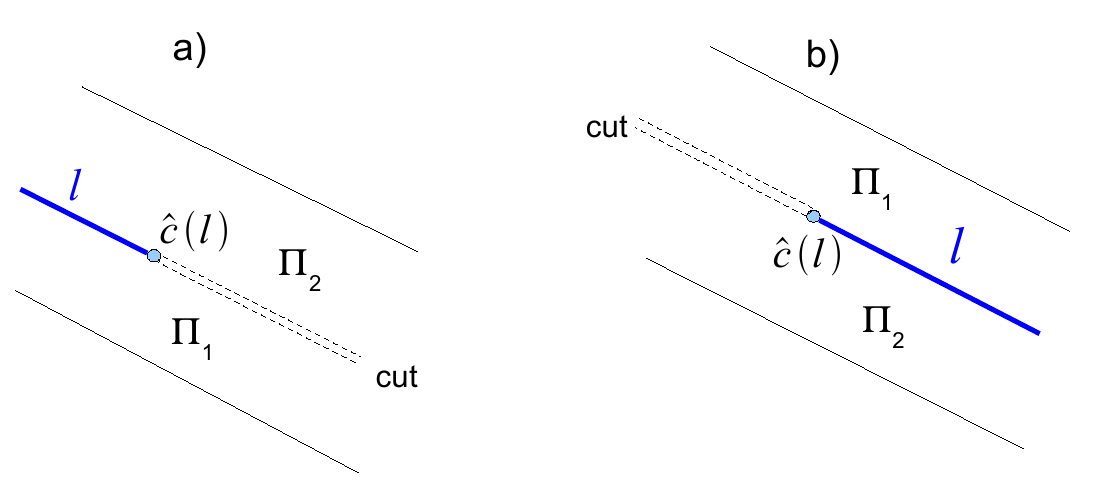} \caption{Notations in the construction of the sheaf $\Psi^K$: a) $\ell\in {\cal L}_\levo$, b) $\ell\in {\cal L}_\pravo$ } \label{CoShWKBp81} \end{figure}

\underline{Case A.} 
If $\ell$ goes to the left, we denote by $\Pi_1$ the bottom strip, fig. \ref{CoShWKBp81}, a). \label{au19pc626} \\
We then define a map
$$
\tnu_\ell^K:\Lambda^{K-}_\ell\to \Lambda^{K+}_\ell*\tcS_\ell
$$
similar to $\nu^K_\ell$ from the previous subsection. The  maps $\tnu^K_\ell$
induce  maps
$$
\tN^K_\ell:\Lambda^{K-}_\ell*\tS_+\to \Lambda^{K+}_\ell*\tS_-
$$
and
$$
\btN^K_\ell:\Lambda^{K+}_\ell*\tS_+\oplus \Lambda^{K-}_\ell*\tS_-\to
\Lambda^{K+}_\ell*\tS_+\to \Lambda^{K-}_\ell*\tS_-,
$$
in the same way as in Sec \ref{oboznvt}.

We now set

\begin{equation}
\Gamma^{\Pi_1\Pi_2}_{\Psi^K}:=\Id+\btN^K_\ell.
\label{se21e843}
\end{equation}

We set $\Gamma^{\Pi_2\Pi_1}_{\Psi^K}:=(\Gamma^{\Pi_1\Pi_2}_{\Psi_K})^{-1}=
\Id-\btN^K_\ell$. 

Similarly to the previous subsection, we can combine the definitions as follows.
Let $\Pi$  and $\Pi'$ be intersecting $-\alpha$-strips whose intersection ray $\ell:=\Pi\cap\Pi'$
goes to the left. Define a number $\vt(\Pi,\Pi')=1$ if $\Pi$ is below $\Pi'$ and $\vt(\Pi,\Pi')=-1$
otherwise. We then have $\Gamma^{\Pi \Pi'}_{\Psi^K}=\Id+\vt(\Pi,\Pi')\bN_\ell^K$.


\underline{Case B.} Analogously,  assume that
 $\ell=\Pi_1\cap \Pi_2$ goes to the right
 and that $\Pi_2$ is below $\Pi_2$, fig. \ref{CoShWKBp81}, b).
Similar to above, we have a map
\begin{equation}  \tnu^K_\ell:\Lambda^{K+}_\ell   \to
 \Lambda^{K-}_\ell * \tilde S_\ell   \label{au22e3}, \end{equation}
which enables us to define maps
$$ \tN_\ell^K  : \Lambda^{K+}_\ell * \tilde S_+ \to 
\Lambda^{K-}_\ell * \tilde{S}_-
 ;$$
$$
\btN_\ell^K:\Lambda^{K+}_\ell * \tilde S_+ \oplus
\Lambda^{K-}_\ell * \tilde S_-\to \Lambda^{K+}_\ell * \tilde S_+ \oplus
\Lambda^{K-}_\ell * \tilde S_-
$$
in the same way as above. 
We set 
\begin{equation} \Gamma^{\Pi_1\Pi_2}_{\Psi^K}:=\Id+\btN_\ell^K;\label{se21e844}
\end{equation}
\begin{equation} 
\Gamma^{\Pi_2\Pi_1}_{\Psi^K}:=(\Gamma^{\Pi_1\Pi_2}_{\Psi^K})^{-1}=\Id-\btN_\ell^K.\label{se21e1004} 
\end{equation}
Finally, given two  intersecting $-\alpha$-strips $\Pi$ and $\Pi'$ whose intersection ray $\ell$
goes to the right, we set $\vt(\Pi, \Pi')=1$ if $\Pi'$ is below $\Pi$ and $\vt(\Pi,\Pi')=-1$ otherwise so
that
$\Gamma^{\Pi\Pi'}_{\Psi^K}=\Id+\vt(\Pi,\Pi')\btN_\ell^K$.

The sheaf $\Psi^K$ is obtained by gluing of the sheaves $\Psi_\Pi$ along the boundary rays by means of the
 maps $\Gamma^{\Pi \Pi'}_{\Psi^K}$, similarly to $\Phi^K$.

The map 
\begin{equation}\label{intopsik}
i_{\Psi^K}:\bZ_{\bx_0\times K}[-2]\to \Psi^K,
\end{equation}
same as in (\ref{intopsik1}),
 is constructed similarly to $i_{\Phi^K}$.

\subsection{The map $I_{\Psi\Phi}$} 
We now pass to discussing the  identification $I_{\Psi\Phi}:\Psi^K\to \Phi^K$ as in (\ref{opredipsiphi}).
Explicit formulas for the map $I_{\Psi\Phi}$ 
are complicated,
see Sec.  \ref{psiphi}. Let us, however, formulate a result on this map, to be proven in Sec . \ref{psiphi}.

Let $P$ be  an $\alpha$-strip and $\Pi$ be a $-\alpha$-strip. Suppose $P\cap \Pi\neq \emptyset$.
We have identifications 
$$
\Phi^K|_{P\cap \Pi}=\Phi_P|_{P\cap \Pi}=
\Lambda^{K+}_{P\cap \Pi}*S_+\oplus \Lambda^{K-}_{P\cap \Pi}*S_-;
$$
$$
\Psi^K|_{P\cap \Pi}=\Psi_\Pi|_{P\cap \Pi}=
\Lambda^{K+}_{P\cap \Pi}*\tS_+\oplus \Lambda^{K-}_{P\cap \Pi}*\tS_-.
$$
Set $i_{\Pi P}:=I_{\Psi\Phi}|_{P\cap \Pi}$. In view of the above identifications, we can rewrite:
$$
i_{\Pi P}:\Lambda^{K+}_{P\cap \Pi}*\tS_+\oplus \Lambda^{K-}_{P\cap \Pi}*\tS_-
\to \Lambda^{K+}_{P\cap \Pi}*S_+\oplus \Lambda^{K-}_{P\cap \Pi}*S_-.
$$
We  are now going to take advantage of  direct sum decompositions of both parts of this map.

\subsubsection{Decomposing  $i_{\Pi P}$ into components}\label{akaw}
Let us now rewrite both sides of this map as follows. 

For a $w\in \wa_\levo$ or $w\in \wma_\levo$, we define ${\zA}(K,w)\subset (P\cap \Pi)\times \Co$: 
$$
{\zA}(K,w):=\{(x,s)| s+p_z(x)\in \hat c(w)+K\},
$$
where $\hatc(w)$ is as in {\eqref{oc21e7}, \eqref{oc21e8}}.
 

We then have
$$
\Lambda^{K+}_{P\cap \Pi}*S_+\oplus\Lambda^{K-}_{P\cap \Pi}*S_-=\bigoplus\limits_{w\in \wa} \bZ_{{\zA}(K,w)};
$$
$$
\Lambda^{K+}_{P\cap \Pi}*\tS_+\oplus\Lambda^{K-}_{P\cap \Pi}*\tS_-=\bigoplus\limits_{\tw\in \wma} \bZ_{{\zA}(K,\tw)}.
$$
Next,
$$
\Hom(\bigoplus\limits_{\tw\in \wma} \bZ_{{\zA}(K,\tw)};\bigoplus\limits_{w\in \wa} \bZ_{{\zA}(K,w)})=
\prod\limits_{\tw\in \wma}\Hom(\bZ_{{\zA}(K,\tw)};\bigoplus\limits_{w\in \wa} \bZ_{{\zA}(K,w)})
$$
\begin{equation}\label{vkl}
\into\prod\limits_{\tw\in \wma;w\in \wa}\Hom(\bZ_{{\zA}(K,\tw)};\bZ_{{\zA}(K,w)}).
\end{equation}

In Sec \ref{pipa}   we prove that  
$\Hom(\bZ_{{\zA}(K,\tw)};\bZ_{{\zA}(K,w)})=0$ unless ${\zA}(K,w)\subset {\zA}(K,\tw)$, in which case 
$\Hom(\bZ_{{\zA}(K,\tw)};\bZ_{{\zA}(K,w)})=\bZ.e_{\tw,w}$, where $e_{\tw,w}$ is the homomorphism
induced by the embedding ${\zA}(K,w)\subset {\zA}(K,\tw)$.  Elements 
of $\prod\limits_{\tw\in \wma;w\in \wa}\Hom(\bZ_{{\zA}(K,\tw)};\bZ_{{\zA}(K,w)})$ are thus identified with infinite
sums of the form
\begin{equation}\label{sumat}
\sum\limits_{\tw,w} n_{\tw w}e_{\tw w}, 
\end{equation}
where $n_{\tw w}\in \bZ$, and  ${\zA}(K,w)\subset {\zA}(K,\tw)$.
By Prop.\ref{PiPHom}, under the inclusion (\ref{vkl})
the set  $\hom(\bigoplus\limits_{\tw\in \wma} \bZ_{{\zA}(K,\tw)};\bigoplus\limits_{w\in \wa} \bZ_{{\zA}(K,w)})$
is identified with the set of all  sums as in  (\ref{sumat}), satisfying 

{\it for every point $y\in (P\cap \Pi)\times \Co$ and every $\tw\in \wma$,
 there are only finitely many $w\in \wa$ such 
that $n_{\tw w}\neq 0$ and $y\in {\zA}(K,w)$.}

\subsubsection{Identification $\wma\to \wa$.} \label{Afat}

 Let us first define an identification $\bA:\cL^{-\alpha}\to\cL^{\alpha}$.
Let $\ell\in \cL^{-\alpha}$. Suppose $\ell$ goes to the right. Let $P$ be the leftmost strip
among all $\alpha$-strips that intersect $\ell$.   There are exactly two boundary rays of $P$,
$\ell_l$ and $\ell_r$ such that  $\hat c(\ell_l)=\hat c(\ell_r)=\hat c(\ell)$, 
$\ell_l$  goes to the left, and $\ell_r$  goes  to the right. Let us assign $\bA(\ell)=\ell_r$.

Similarly, if $\ell\in\cL^{-\alpha}$, $\ell$ goes to the left, we consider the leftmost strip $P$ 
among all $\alpha$-strips that intersect $\ell$.   There are exactly two boundary rays of $P$,
$\ell_l$ and $\ell_r$ such that  
\begin{equation}\label{sovpad}
\hat c(\ell_l)=\hat c(\ell_r)=\hat c(\ell).
\end{equation}
$\ell_l$  goes to the left, and $\ell_r$  goes  to the right. Let us assign $\bA(\ell)=\ell_l$.
The map $\bA$ extends in the obvious way to a map
$\bA:\wma\to \wa$: a word $\ell_n\cdots \ell_1 L\in \wma$ (resp. $\ell_n\cdots \ell_1 R\in \wma$) is mapped into
$\bA(\ell_n)\cdots \bA(\ell_1)L$ (resp. $\bA(\ell_n)\cdots \bA(\ell_1)R$).  Because of (\ref{sovpad}),
we have ${\zA}(K,\tw)={\zA}(K,\bA(\tw))$ for all $\tw\in \wma$.
\subsubsection{Formulation of the result}
Let us write $i_{\Pi P}$ in the form (\ref{sumat}):
\begin{equation}
i_{\Pi P}=\sum \limits_{\tw\in \wma;w\in \wa} n_{\tw w}e_{\tw w}.
\end{equation}
In order to formulate the result, let us introduce some notation. 
For $\tw\in \wma$,
$\tw=\ell_n\cdots \ell_1 L\in \wma$ (resp. $\tw=\ell_n\cdots \ell_1 R\in \wma$), set $|\tw|:=n$, to be the length
of $\tw$ ( in particular $|L|=|R|=0$).

\begin{Proposition}\label{ipip}
1) We have $n_{\tw \bA(\tw)}=(-1)^{|\tw|}$;

2) If $n_{\tw w}\neq 0$ {and $w\ne {\zA}(\tilde w)$}, then ${\zA}(K,w)\neq {\zA}(K,\tw)$ (we have a strict embedding ${\zA}(K,w)\subset {\zA}(K,\tw)$).
\end{Proposition}
This proposition is proven in Sec \ref{proofipip}.

\subsection{Description of $\Phi^{\elr_\alpha}$} We construct the  sheaf $\Phi^{\elr_\alpha}$ 
and a map $i_{\Phi^{\elr_\alpha}}$
in a way very similar to the construction $\Phi^K$, using the decomposition of $X$ into $\alpha$-strips
and replacing $K$ with $\elr_\alpha$ everywhere.
We then get sheaves
$$
\Lambda^{\elr_\alpha\pm}_U:=\bZ_{\{(x,s)| x\in U,s\in \Co;s\pm x\in \elr_\alpha\}}.
$$
$$
\Phi^{\elr_\alpha}_P:=\Lambda^{\elr_\alpha+}_P*S_+\oplus \Lambda^{\elr_\alpha-}_P*S_-.
$$
 If $\ell$ goes to the left (resp. to the right) we still have a map
$$
\nu_{\ell}^{\elr_\alpha}:\Lambda^{\elr_\alpha-}_\ell \to \Lambda^{\elr_\alpha+}_\ell *S_\ell \text{ resp. }
\nu_{\ell}^{\elr_\alpha}:\Lambda^{\elr_\alpha+}_\ell \to \Lambda^{\elr_\alpha-}_\ell*S_\ell,
$$
so that we can define the  gluing maps $\Gamma^{P_1P_2}_{\Phi^{\elr_\alpha}}$ similarly
to $\Gamma^{P_1P_2}_{\Phi^K}$.
\subsection{Description of $\Phi^{\elr_{-\alpha}}$} In order to construct $\Phi^{\elr_{-\alpha}}$ 
and $i_{\Phi^{\elr_{-\alpha}}}$  we switch to $-\alpha$-strips ( sticking to  $\alpha$-strips leads
to a failure to define the maps $\nu_\ell^{\elr_{-\alpha}}$).
The construction is then  similar to the construction of $\Psi^K$ (just replace $K$ with $\elr_{-\alpha}$
everywhere).

\subsection{Constructing the map (\ref{zvyazok})}\label{phikonec}

Let us construct a  map $\mQ$,
satisfying (\ref{zvyazok}).  
It will be convenient for us to replace $\Phi^K$ with the isomorphic sheaf $\Psi^K$. 
 
First, we will  construct maps
$
q_{C\elr_\alpha}:\Phi^C\to \Phi^{\elr_\alpha};
$
$q_{K\elr_{\pm\alpha}}:\Psi^K\to \Psi^{\elr_{\pm \alpha}}$
satisfying
$i_{\Psi^K}=q_{C\elr_\alpha}i_{\Phi^C}$; $i_{\Phi^{\elr_{\pm \alpha}}}=q_{K\elr_{\pm \alpha}}i_{\Psi^K}$.

We   define   $\mQ$  as follows:
\begin{equation} \mQ \ : \  \raisebox{2pc}{\xymatrix{ 
\Phi^C\ar@{}[d]|{\oplus}\ar[rr]^{q_{C\elr_\alpha}} && \Phi^{\elr_\alpha} \ar@{}[d]|{\oplus}  \\
 \Psi^K \ar[rru]^{q_{K\elr_\alpha}} \ar[rr]^{q_{K\elr_{-\alpha}}}&&
\Phi^{\elr_{-\alpha}}.}}  \label{zvyazok2} \end{equation}
The categorical definition of the maps in this diagram was discussed in section \ref{StrucObPhi}. 

Let us now pass to constructing the above mentioned maps $q_{C\elr_\alpha}$ and $q_{K\elr_{\pm \alpha}}$.

\subsubsection{The map $q_{C\elr_\alpha}$} \label{Celra}

We have $\Phi^C=\bZ_{X\times \Co}$ so that 
$$
\hom(\Phi^C;\Phi^{\elr_\alpha})=\Gamma(X\times \Co;\Phi^{\elr_\alpha})
$$
so that a map $q_{C\elr_\alpha}$ can be defined by means of specifying a section
$\bfq\in\Gamma(X\times \Co;\Phi^{\elr_\alpha})$. This can be done strip-wise: we can instead specify, for every closed strip $P$, 
sections $\bfq_P\in \Gamma(P\times \Co;\Phi^{\elr_\alpha}_P)$  which agree on intresections as follows. 
Let $P_1\cap P_2= \ell $.
We then have restriction maps
$$
|_{\ell\times \Co}: \Gamma(P_i\times \Co;\Phi^{\elr_\alpha}_{P_i})\to 
\Gamma(\ell\times \Co;\Phi^{\elr_\alpha}_\ell),\quad i=1,2.
$$
We  then should have 
\begin{equation}\label{secklei}
\bfq_{P_1}|_{\ell\times \Co}=\bfq_{P_2}|_{\ell\times \Co}.
\end{equation}
It is clear that any collection of data $\bfq_{P}$, satisfying (\ref{secklei}) for all pairs of neighboring
strips,
 determines
a section $\bfq\in \Gamma(X\times\Co;\Phi^{\elr_\alpha})$ in a unique way.

We have $\bZ=\Gamma(P \times \Co;\Lambda^{\elr_\alpha \pm}_P*S_w)$ for all $w\in \wa$.

Let us take the direct sum
of these identifications over all $w\in \wa$ so as to get a map
$$
s_P:\bZ[\wa]\to \Gamma(P\times \Co;\Phi^{\elr_\alpha}_P),
$$
where $\bZ[\wa]$ is the $\bZ$-span of the set $\wa$.
Similarly, we define $$ s_\ell:\bZ[\wa]\to \Gamma(\ell\times\Co;\Phi^{\elr_\alpha}_\ell), $$
 where $\ell$ is the  intersection ray
of a pair of neighboring $\alpha$-strips .
The maps $s_P,s_\ell$ are inclusions; denote by
 $\Gamma'(P\times \Co;\Phi^{\elr_\alpha}_P),\Gamma'(\ell\times \Co;\Phi^{\elr_\alpha}_\ell)$ the images
 of these inclusions.  As easily follows from the definition of the gluing maps 
$ \Gamma^{P_1 P_2}_{\Phi^{\elr_\alpha}}$,
the restriction maps induce isomorphisms
$$
|_{\ell\times \Co}:\Gamma'(P\times \Co;\Phi^{\elr_\alpha}_P)\to \Gamma'(\ell\times \Co;\Phi^{\elr_\alpha}_\ell),
$$
where $\ell$ is a boundary ray of $P$.

Since the graph formed by $\alpha$-strips and their intersection rays is a tree, it follows that
given an element $\bfq_{P_0}\in \Gamma'(P_0\times \Co;\Phi^{\elr_\alpha}_{P_0})$, we  have unique elements
$$
\bfq_{P}\in \Gamma'(P\times \Co;\Phi^{\elr_\alpha}_P)
$$
satisfying (\ref{secklei}).   We set $\bfq_{P_0}:=s_{P_0}(L+R)$, where $L,R$ are words of of length 1 in $\wa$
viewed as elements in $\bZ[\wa]$. 
This way we get a section $\bfq$ and a map $q_{C\elr_\alpha}$. It is clear that
 Condition $i_{\Phi^{\elr_\alpha}}=q_{C\elr_\alpha}i_{\Phi^C}$
is satisfied.

Denote by $\be_P\in \bZ[\wa]$  a unique element such that $s_P(\be_P)=\bfq_P$.  
Denote by $W_P\subset \wa$ a finite subset such that 
$$
\be_P=\sum\limits_{w\in W_P}\be_{Pw}w,
$$
where $\be_{Pw}\in \bZ\backslash 0$.

\subsubsection{ Map $q_{K\elr_{-\alpha}}:\Psi^K\to \Phi^{\elr_{-\alpha}}$} 
Let us define this map stripwise. For every $-\alpha$-strip $\Pi$ we have
a map $\Lambda^{K\pm}_\Pi\to \Lambda^{\elr_{-\alpha}\pm}_\Pi$  induced by the embedding
of the corresponding closed subsets of $\Pi\times \Co$. Whence induced maps
$
\Lambda^{K\pm}_\Pi*\tS_w\to \Lambda^{\elr_{-\alpha}\pm}_\Pi*\tS_w.
$
Taking a direct sum over  all $w\in \wa$ yields a map
$$
\Lambda^{K+}_\Pi*\tS_+\oplus \Lambda^{K-}_\Pi*\tS_-\to\Lambda^{\elr_{-\alpha}+}_\Pi*\tS_+\oplus
 \Lambda^{\elr_{-\alpha}-}_\Pi*\tS_-,
$$
and we assign $q_{K\elr_{-\alpha},\Pi}:\Psi^K_\Pi\to \Phi^{\elr_{-\alpha}}_\Pi$ to be this map.
It is clear that thus defined maps agree on all intersection rays, thereby defining the desired map 
$q_{K\elr_{-\alpha}}$. The condition $i_{\Phi^{\elr_{-\alpha}}}=q_{K\elr_{-\alpha}}i_{\Psi^K}$ is clearly satisfied.

\subsubsection{Map $q_{K\elr_{\alpha}}:\Psi^K\to \Phi^{\elr_{\alpha}}$} \label{qstr}

We first construct  a map  $q'_{K\elr_{\alpha}}:\Phi^K\to \Phi^{\elr_{\alpha}}$
using $\alpha$ strip in the same way as we constructed $q_{K\elr_{-\alpha}}$.

We set
$$
q_{K\elr_{\alpha}}:=q'_{K\elr_{\alpha}}I_{\Psi\Phi}.
$$
The condition $i_{\Phi^{\elr_{\alpha}}}=q_{K\elr_{\alpha}}i_{\Psi^K}$ is clearly satisfied.

\subsubsection{Restriction of ${\cal Q}$ to a parallelogram}

Let $P$ and $\Pi$ be a pair of intersecting $\alpha$- and $(-\alpha)$-strips. 

First, in view of identification $\bA$, let us write $w$ instead of $\bA^{-1}w\in \wma$.
Next, for a $w\in \wa$  and a subset $\Delta\in \Co$, let us define a subset
${\zA}(\Delta,w)\subset (P\cap \Pi)\times \Co$ as follows. If $w\in \wa_\levo$ (resp., $w\in \wa_\pravo$), we set
$
{\zA}(\Delta,w)=\{(x,s)|s+z(x)\in \hatc(w)+\Delta\}
$
(resp., ${\zA}(\Delta,w)=\{(x,s)|s-z(x)\in \hatc(w)+\Delta\}
$; these notations are compatible with those of section \ref{akaw}.
Set $A_0:=(\Pi\cap P)\times \C$.
We then have  identifications
$$
\Phi^C_{\Pi\cap P}=\bZ_{A_0};
$$
$$
\Psi^K_{\Pi\cap P}=\bigoplus\limits_{w\in \wma} \bZ_{{\zA}(K,w)};
$$
$$
\Phi^{\elr_{\alpha}}_{\Pi\cap P}=\bigoplus\limits_{w\in \wa} \bZ_{{\zA}(\elr_{ \alpha};w)};
$$
$$
\Phi^{\elr_{-\alpha}}_{\Pi\cap P}=\bigoplus\limits_{w\in \wma} \bZ_{{\zA}(\elr_{- \alpha};w)}.
$$
Let us now rewrite the maps from diagrams (\ref{zvyazok2}) in terms of these identifications.

\subsubsection{The  map $q_{C\elr_{\alpha}}$ revisited.}  
\label{se21e19} Let $E^{C\elr_{\alpha}}_w:\bZ_{A_0}\to
 \bZ_{{\zA}(\elr_{\alpha},w)} $ be the map induced by the closed embedding of the corresponding sets.
According to Sec \ref{Celra}, 
\begin{equation}\label{eqelra}
q_{C\elr_{\alpha}}=\sum_{w\in W_P} \be_{Pw} E^{C\elr_{\alpha}}_w.
\end{equation}

\subsubsection{The map $q_{K\elr_{-\alpha}}$}\label{aalf}
It follows that the map
$$
q_{K\elr_{-\alpha}}:\bigoplus\limits_{w\in \wa} \bZ_{{\zA}(K,w)}\to \bigoplus\limits_{w\in \wa} \bZ_{{\zA}(\elr_{-\alpha},w)}
$$ 
is a direct sum, over all $w\in \wa$, of the maps
$$
\bZ_{{\zA}(K,w)}\to \bZ_{{\zA}(\elr_{-\alpha},w)},
$$
over all $w\in \wa$.

\subsubsection{The map $q_{K\elr_\alpha}$} 
Let $w,w'\in \wa$ be such that ${\zA}(K,w)\supset {\zA}(\elr_{\alpha};w')$.
Let 
$E^{K\elr_{\alpha}}_{ww'}:\bZ_{{\zA}(K,w)}\to\bZ_{{\zA}(\elr_{\alpha};w')}$ be the map induced by this embedding.

We then have
$$
q_{K\elr_{\alpha}}=\sum\limits_{ww'} n^{K\elr_{\alpha}}_{ww'} E^{K\elr_{\alpha}}_{ww'}.
$$

\begin{Proposition}\label{sumat1}
1) $n^{K\elr_{\alpha}}_{ww}=(-1)^{|w|}$;
\\
2) for every  compact subset $L\in (P\cap \Pi)\times \Co$ and every $w\in \wa$, there are only
finitely many $w'\in \wa$ such that $n_{ww'}\neq 0$ and $L\cap {\zA}(\elr_{-\alpha};w')\neq \emptyset$;
\\
3)    If $n^{K\elr_{\alpha}}_{ww'}\neq 0$, then  we have a strict embedding ${\zA}(w',K)\subset {\zA}(w,K)$.
\end{Proposition}

\textsc{Proof.} Parts 1) and 3) follow from  Sec.\ref{qstr} and Prop. \ref{ipip}, part 2) follows 
from Prop.\ref{PiPHom}.  $\Box$

\subsection{ $\Sigma$ and $\cS$ are Hausdorff}

{Recall that $\Sigma$ was defined in section \ref{pSigma} and ${\cal S}$ in the section \ref{Sec9o}. }

Let us start with some general observations.

\subsubsection{Generalities on \'etale spaces}\label{etal}

Let $F$ be a sheaf of abelian groups on a {Hausdorff} topological space {$X$}.
 Call $F$ {\em rigid} if its 
\'etale space is Hausdorff.  The following facts are easy to check.

1) Let $U\subset X$ be a Hausdorff open subset. Then $\bZ_U$ is rigid.
Indeed, the corresponding  \'etale space is {$(\bZ\backslash\{ 0\})\times U\cup \{ 0\}\times X$}.

2) Every sub-sheaf  $F_1$ of a rigid sheaf  $F$ is rigid. Indeed, the \'etale space of $F_1$ is
identified  with a closed subspace of  a Hausdorff \'etale space of $F$.

3) Let $0\to A\to B\to C\to 0$ be an exact sequence of sheaves, where
$A,C$ are rigid. Then so is $B$. Indeed, Let $A'\to B'\stackrel{\pi}\to C'$ be the \'etale spaces of $A,B$, and $C$.
Let $b_1,b_2\in B'$. Suppose $\pi(b_1)\neq \pi(b_2)$; we then have separating neighborhoods
 $\pi(b_1)\in U_1$; $\pi(b_2)\in U_2$ so that $\pi^{-1}U_1,\pi^{-1}U_2$ separate $b_1$ and $b_2$.
Let now $\pi(b_1)=\pi(b_2)=c$ but $b_1\neq b_2$.  Since  $\pi$ is a local homeomorhisms, there are
neigborhoods $W_i$ of $b_i$ in $B'$ such that $W_i$ are projected homeomorhically into $C'$.
By possible shrinking we may achieve that $W_i$ project to the same open subset $U\in C'$; $c\in U$.
so that we have homeomorphisms $\pi^{-1}_i:U\to W_i$.  We then have
a  continuous map $\delta:U\to A'$, where $\delta(u)=\pi^{-1}_2u-\pi^{-1}_1u\in A_u\subset A'$.
Since $b_1\neq b_2$,$\delta(c)\neq 0$, so that we have a neighborhood $U'\subset U$ of $c$ on which
$\delta$ does not vanish. It now follows that the neighborhoods  $\pi^{-1}_iU'$  do separate $b_1$
and $b_2$.

4) Let $i_n:F_n\to F_{n+1}$, $n\geq 0$  be a directed sequence of embeddings, where
$F_0$ and all $F_{n+1}/i_nF_n$ are rigid. Then
 $F:=\displaystyle\varinjlim\limits_n F_n$ is also rigid.
Indeed, 3) implies that all $F_n$ are rigid.  Let $F'_n,F'$ be the \'etale spaces of $F_n,F$.
We have induced maps $F'_n\to F'$; $F'_n\to F'_{n+1}$ which induce
a map $\varinjlim F'_n\to F'$ which can be easily proven to be a homeomorphism. 
Since all the maps $F'_n\to F'_{n+1}$ are closed embeddings, it follows that $F'$ is Hausdorff.

5)  Let $p:Y\to X$ be a local homeomorphism, where $Y$ is Hausdorff. 
Let $\emptyset \ne U\subset V\subset X$ be open sets,  where $V$ is connected.  Suppose we are given a  section
$s:U\to Y$.  There exist at most one way to extend $s$ to $V$. Indeed, let $s_1,s_2:V\to Y$ be
extensions of $s$. Let us prove that the set $W:=\{v\in V:s_1(v)\neq s_2(v)\}$ is open. 
Indeed, let $v\in W$. The points $s_1(v)$, $s_2(v)$ can be separated by neighborhoods $U_1,U_2\subset Y$.
Let $\cU:={s_1}^{-1}U_1\cap {s_2}^{-1}U_2$; $\cU$ is a neighborhood of $v$.  It now follows that
$s_i(\cU)\subset U_i$, therefore $s_i(\cU)$ do not intersect; we have thus found an open neighborhood $\cU\subset W$ of $v$, hence $W$ is open.

Let us now prove that $W':=\{v\in V:s_1(v)=s_2(v)\}$ {is open}.  It is clear that $s_i(U)$ are open subsets of $Y$,
so that $W'=s_1(U)\cap s_2(U)$ is open.

Finally, $V=W\sqcup W'$ and $W'\neq \emptyset$. This implies $W=\emptyset$.

\subsubsection{Reduction to rigidity on $\Pi\cap P$}
Since $\cS\subset \Sigma$ is a connected component, it suffices to prove that $\Sigma$ is Hausdorff.
The latter reduces to  showing  that $p_{\Sigma}^{-1}((P\cap \Pi)\times \Co)$ is Hausdorff for every pair
of intersecting $\alpha$-strip $P$ and $-\alpha$-strip $\Pi$, which is equivalent
to the  rigidity of the sheaf $\Phi_0|_{(\Pi\cap P)\times \Co}$, which is isomorphic to $\Ker\mQ$.

\subsubsection{Filtration on $\Phi_0|_{\Pi\cap P\times \Co}$} 

Let us  choose an arbitrary
identification $ \bZ_{> 0}\stackrel{\sim}{\to} {\wa}$; $n\mapsto w_n$.
Define  a filtration on $\cG:=\Phi^C\oplus\Psi^K|_{\Pi\cap P\times \Co}$
by setting 
$$
\cG^n
:=\Phi^C|_{\Pi\cap P\times\Co}\oplus \bZ_{{\zA}(K,w_1)}\oplus \cdots\oplus \bZ_{{\zA}(K,w_n)}.
$$
It is clear that 
$$\Phi^C|_{\Pi\cap P\times \Co}=:\cG^0\subset \cG^1\subset \cdots\cG^n\subset\cdots\subset \cG$$
is an exhaustive filtration.
It is also clear that $\cG^n\subset \cG$ is a direct summand. Denote by $P_n^\cG:\cG\to\cG^n$ the projection.

Set
$$
F_n\Phi_0:=\Ker \mQ|_{\cG^n}.
$$
It follows that $F$ is an exhaustive filtration of $\Phi_0|_{\Pi\cap P\times \Co}$. 
By Sec.  \ref{etal} 2), it suffices to show that each sheaf $F_n$ is rigid.

\subsubsection{ Sheaf $F'_n\supset F_n$} \label{FprimNsuper}


We have the following projection onto a direct summand
$$
P_n:\Phi^{\elr_\alpha}_{\Pi\cap P}\oplus \Phi^{\elr_{-\alpha}}_{\Pi\cap P}\to
\bigoplus\limits_{m=1}^n \bZ_{{\zA}(\elr_\alpha;w_m)}\oplus \bZ_{{\zA}(\elr_{-\alpha};w_m)}=:\cL_n.
$$

Let $F'_n:=\Ker P_n \mQ|_{\cG^n}$. We have: $F_n$ is a sub-sheaf of $F'_n$, so 
that it suffices to show that each $F'_n$ is rigid.

\subsubsection{Further filtrations on $\cG^n,\cL_n,F'_n$}

Fix $n\in \Z_{>0}$.
 Let us re-label the words $w_1,w_2,\ldots,w_n$
to, say $\bw_1,\bw_2,\ldots,\bw_n$, so that the following holds true: 

{\em if $i>j$, then it is impossible that ${\zA}(K,\bw_i)$ is a proper subset of ${\zA}(K,\bw_j)$.}

Since we are dealing with only finitely many words, this is always possible. Let $j\leq n$.
Set $\bF^j\cG^n:=\bZ(K,\bw_1)\oplus\cdots\oplus \bZ(K,\bw_j)\subset \cG^n$.
Set $\bF^j\cL_n:=\bZ_{{\elr}_{\pm \alpha},\bw_1)}\oplus\cdots\oplus 
\bZ_{{\zA}({\elr}_{\pm \alpha},\bw_j)}
\subset {\cL_n}$. 
We also set $\bF^{n+1}\cG^n=\cG^n$; $\bF^{n+1}\cL_n=\cL_n$.
Let $\Gr^j\cG^n$; $\Gr^j\cL_n$ be the associated graded quotients.
 
Proposition \ref{sumat1} and Sec. \ref{aalf} imply that the map $P_n\mQ$ preserves the filtration
$\bF$: $P_n\mQ:\bF^j\cG^n\to \bF^j\cL_n$.
Set
$
\bF^jF'_n:=\Ker P_n\mQ|_{\bF^j\cG^n}.
$
It is clear that this way we get a filtration  on $F'_n$. Let $\Gr^jF'_n$ be the associated graded quotients.
Our problem now reduces to proving rigidity of $\Gr^j F'_n$ by Sec. \ref{etal}, 3).
Since $P_n\mQ$ preserves $\bF$, we have 
$$
\Gr^j F'_n\subset\Ker \Gr^j P_n\mQ:\Gr^j \cG^n\to \Gr^j\cL_n.
$$
By Sec \ref{etal} 2), the problem reduces to showing rigidity of $ \Ker \Gr^j P_n\mQ:\Gr^j \cG^n\to \Gr^j\cL^n.$
\subsubsection{Finishing the proof}
Let $j\leq n$. We then have $\Gr^j \cG^n=\bZ_{{\zA}(K,\bw_j)}$; 
$\Gr^j \cL_n=\bZ_{{\zA}(\elr_\alpha;\bw_j)}\oplus \bZ_{{\zA}(\elr_{-\alpha};\bw_j)}$.
By Sec. \ref{aalf}   and Proposition \ref{sumat1},
we have: $$\Gr^j P_n\mQ=(-1)^{|\bw_j|}E_{\bw_j}^{\elr_\alpha}\oplus E_{\bw_j}^{\elr_{-\alpha}},
$$
where the morphisms $$E_{\bw_j}^{\elr_{\pm \alpha}}:\bZ_{{\zA}(K,\bw_j)}\to \bZ_{{\zA}(\elr_{\pm \alpha};\bw_j)}
$$ are induced by the closed embeddings of the corresponding  sets.
It now follows that $\Ker \Gr^j P_n\mQ=\bZ_{{\zA}(\Int K;\bw_j)}$, which is rigid by Sec. \ref{etal},1).
 
Let now $j=n+1$. We have $\Gr^{n+1}\cL_n=0$; $\Gr^{n+1}\cG^n=\bZ_{A_0}$, so that
$$
\Ker \Gr^j P_n\mQ=\bZ_{A_0},
$$
which is also rigid, as a sheaf on $(\Pi\cap P)\times \C=A_0$, by Sec. \ref{etal},1). This finishes the proof.

\subsection{Surjectivity of  the projection $p_\cS:\cS\to X$.}
In this subsection we will prove 
\begin{Thm} The projection $p_\cS:\cS\to X$ is surjective.
\end{Thm}

Proof of this theorem will occupy the rest of this subsection.
We will construct an open subset $\cU\subset \Sigma$ such that 

1) $\cU$ projects surjectively onto $X$;

2) $\cU$ is connected;

3) $\cU\cap h(S_\alpha)\neq \emptyset$, where $h:S_\alpha\to \Sigma$ is as in (\ref{opredh}).

Conditions 2),3) imply that $\cU\subset \cS$, and Theorem follows.

Let us now construct $\cU$ and verify 1)-3).
\subsubsection{Constructing $\cU$}
We construct  $\cU$  stripwise.  We will freely use the notation from Sec \ref{Celra}.
Let $P$ be an $\alpha$-strip.  Define a closed subset
$$
A(P):=\bigcup\limits_{w\in W_P} {\zA}(\elr_\alpha,w)\subset P\times \Co\subset X\times \Co.
$$

Let $\cU:=X\times\Co\backslash \bigcup\limits_P A(P)$, where the union is taken over the set of
all $\alpha$-strips $P$. Denote by $j^X_\cU:\cU\to X\times \Co$ the open embedding.

 Let us now embed $\cU$ into $\Sigma$.   We have a natural embedding
$J_{\cU}:\bZ_\cU\to \bZ_{X\times \Co}=\Phi^{{C}}$.  As follows from (\ref{eqelra}), we have
$
q_{C\elr_a}J_\cU=0,
$
which implies that the map $J_\cU$ factors through $\Ker q_{C\elr_\alpha}:$ 
$$
J_\cU:\bZ_\cU\stackrel{J^q_\cU}\into \Ker q_{C\elr_\alpha}\to \Phi^C.
$$
 As follows from the diagram
(\ref{zvyazok2}), we have a natural embedding 
\begin{equation} \iota_q:\Ker q_{C\elr_\alpha}\into \Ker \mQ, 
\label{iotaqdef} \end{equation}
and we set 
\begin{equation}\label{jmQ}
J_\mQ:=\iota_qJ^q_\cU,
\end{equation}
which is an injection
 $J_\mQ:\bZ_\cU\into \Ker \mQ=\Phi_0$.

{To summarize, we have the following commutative diagram of sheaves on $X\times \C$:}
$$ { \xymatrix{ && (\Ker\mQ = \Phi_0) \ar@{^{(}->}[r] & \Phi^C \oplus \Phi^K. \\
\Z_{\cU}  \ar@/^1pc/[rru]^-{J_\mQ} \ar[r]^{J^q_\cU} \ar@/_1pc/[rr]_{J_\cU} & \Ker q_{C\elr_\alpha} \ar@{^{(}->}[ru]_{\iota_q} \ar[r] &
\Phi^C \ar@{^{(}->}[ru]
}}$$

The map $J_\mQ$ induces an embedding of the \'etale spaces: $\cU\times \bZ\to \Sigma$.
Let $j_\cU:\cU\to \Sigma$ be the restriction of this map onto $\cU\times 1\subset \cU\times \bZ$.
This map is a local homeomorphism and an embedding, therefore, $j$ is an open embedding. 
Let us  identify $\cU$ with $j_\cU(\cU)$. 

\subsubsection{Verifying 1)} Let $$P_\Sigma:\Sigma\stackrel{p_\Sigma}\to X\times \Co\stackrel{\pi_X}\to X$$
be the through map, where where $p_\Sigma$ is the same as in section \ref{pSigma}, and $\pi_X$ is the projection
onto a Cartesian factor.
We see that the composition $P_\Sigma  j_\cU$
coincides with the composition $\cU\stackrel{J_\cU^X}\to X\times \Co\stackrel{\pi_X}\to X$.
Let us check that this map is surjective. Indeed, let $x\in X$.
There are at most two $\alpha$-strips which contain $x$.
We therefore have: $\cU\cap x\times \Co$  is obtained from $x\times \Co=\Co$ by removing a finite number
of $\alpha$-rays, which is non-empty.
\subsubsection{Verifying 2)}

As the sets $W_P$ are finite, it  easily follows that

--- the sets $\cU(P):=P\times \Co\backslash A(P)$ are connected;

--- if $P_1\cap P_2\neq \emptyset$, then $\cU(P_1)\cap \cU(P_2)\neq \emptyset$.
This  implies that $\cU$ is connected. 

The rest of the subsection is devoted by verifying 3).
\subsubsection{Reformulation of 3)}
Recall that the map $h:S_\alpha\to\Sigma$ is induced by the map
$\cI_0:{\Z_{S_\alpha}}\to g^{-1}\Phi_0$, see (\ref{cio}). The injection $j_\cU:\cU\to \Sigma$ is
induced by the map $J_\mQ:\bZ_\cU\to \Ker \mQ=\Phi_0$, see (\ref{jmQ}).
Let $i_{\bx_0}:\Co\to X\times \Co$ be the embedding $i_{\bx_0}(s)=(\bx_0,s)$.
We have $g=i_{\bx_0}\pi_{S_\alpha}$.
Let us denote $\cU_{\bx_0}:=i_{\bx_0}^{-1}\cU$. Observe that $\cU_{\bx_0}$ is obtained
from $\Co$ by removing a finite number of $\alpha$-rays.

\begin{Lemma}\label{peresek} There exists  a non-empty open subset
$V\subset \cU_{\bx_0}$ such that: \\
{i})  the map $\pi_{S_\alpha}$ induces a homeomorphism $\pi_{S_\alpha}^{-1}V\to V$, so that 
we have $\pi_{S_\alpha}^{-1}\bZ_V=\bZ_{\pi_{S_\alpha}^{-1}V}$;
\\
{ii}) the following  diagram {of sheaves on $S_\alpha$} commutes
$$\xymatrix{
 \bZ_{\pi_{S_\alpha}^{-1}V}\ar[rr]^{j_{VS}}\ar[d]^{j_{V\cU}}&& \bZ_{S_\alpha}\ar[d]^{\cI_0}\\
 g^{-1}\bZ_{\cU}\ar[rr]^{{g^{-1}(}J_\mQ{)}}&&  g^{-1}\Phi_0}
$$
where the arrow $j_{VS}$ is induced by the open embedding $\pi_{S_\alpha}^{-1}V\subset S_\alpha$,
and the arrow $j_{V\cU}$ is the composition $\bZ_{\pi^{-1}_{S_\alpha}V}=
\pi_{S_\alpha}^{-1}\bZ_V\stackrel{*}\to \pi_{S_\alpha}^{-1}\bZ_{\cU_{\bx_0}}=
 g^{-1}\bZ_\cU$,
where the arrow  $*$ is induced by the open embedding $V\subset \cU_{\bx_0}$.
\end{Lemma}

Let us first explain how Lemma implies 3). Indeed, it follows from Lemma that {we have a commutative diagram of topological spaces}
\begin{equation} { 
     \xymatrix{ \pi^{-1}_{S_\alpha} V 
            \ar[rr]^-{h|_{\pi^{-1}_{S_\alpha}V}} 
            \ar[d]_{\pi_{S_{\alpha}} }
            && \Sigma \\
V \ar@{}[r]|{\subset} & {\cal U}_{{\mathbf x}_0} \ar[r]^{i_{{\mathbf x}_0}} & {\cal U} \ar[u]_{j_{\cal U}}  }, \label{oc11} 
   } 
\end{equation}
{where the counterclockwise composition $\pi^{-1}_{S_\alpha}V\to {\cal U}$ coincides with a component of the map of \'etale spaces of sheaves induced by $j_{V{\cal U}}$. } 

{Then \eqref{oc11} implies } that $h(S_\alpha)\cap j_\cU(\cU)\supset j_\cU(i_{\bx_0}V)$.

We will now prove the Lemma.

\subsubsection{Subset $W\subset S_\alpha$}

 Let $W:=\pi_{S_\alpha}^{-1}(\Co\backslash K)\subset S_\alpha$.
Denote by $J_W:\bZ_{W}\to \bZ_{S_\alpha}$ the map induced by the  
open embedding $j_W:W\subset S_\alpha$. Let us consider
the composition $hj_W$, which is induced by the map $\cI_0 J_W:\bZ_{W}\to g^{-1}\Phi_0$.

Denote by $\pi:\Phi_0\to \Phi^C\oplus \Phi^K$ the natural embedding
(recall that $\Phi_0=\Ker \mQ$).  Set \\
$\pi_{0K}:=\Pi_K\pi:\Phi_0\to \Phi^K$, where
 $\Pi_K:\Phi^C\oplus \Phi^K\to \Phi^K$ is the projection.

 Let us show 
\begin{Lemma}\label{obnulenie}
We have $
(g^{-1}\pi_{0K})\cI_0 J_W=0.
$
\end{Lemma}

\textsc{Proof.}
Indeed,  the map $\pi$ factors as 
$$
\Phi_0 \ \stackrel{\iota}{\to} \  \Phi \ = \ {(}\Cone \mQ{)}[-1] \  \stackrel{P_\Phi}\to \ \Phi^C\oplus \Phi^K \, ,
$$
where the last arrow is the canonical map.  Set $\pi_K:=\Pi_KP_\Phi$.
We have
$$(g^{-1}\pi_{0K})\cI_0=(g^{-1}\Pi_K)(g^{-1}\pi)\cI_0=(g^{-1}\Pi_K)(g^{-1}P_\Phi)g^{-1}\iota \cI_0
=(g^{-1}\pi_K)\cI,
$$
{where $\cI$ is as in section \ref{OkSec341}.} {Recall that in section \ref{OkSec341} we defined $\cI$ in such a way that u}nder the isomorphism $g^{-1}\Phi=g^!\Phi[2]$, the map $\cI$ corresponds by the conjugacy to
the map 
 $i_\Phi:g_!\bZ_{S_\alpha}[{-}2]\to \Phi${, where $i_\Phi$ was constructed in \eqref{iPhiConstrd}}.  

{We claim that:} 
\begin{equation} \text{\it The map $(g^{-1}\pi_K)\cI$ corresponds by the conjugacy to $\pi_K i_{\Phi}$.} \label{Nov14} \end{equation} 
{Indeed, the conjugate to }
$${ (g^{-1}\pi_K)\cI: \Z_{S_\alpha} \stackrel{\cI}{\to} g^{-1}\Phi \stackrel{g^{-1}\pi_K}{\to} g^{-1}\Phi^K 
} $$
{is defined as $nat[2]\circ (Rg_!g^!\pi_K) Rg_!\cI$, where $nat:Rg_! g^! \Phi^K \to \Phi^K$, and the statement \eqref{Nov14} reduces to commutativity of the diagram}
$$  \xymatrix{ Rg_! \Z_{S_\alpha} \ar[r]^-{Rg_!\cI} \ar[d]_-{i_\Phi[2]} &
Rg_! g^! \Phi[2] \ar[ld] \ar[rr]^-{Rg_! g^! \pi_K[2]} && Rg_!g^! \Phi^K[2] \ar[d]^-{nat[2]} \\
\Phi[2] \ar[rrr]_-{\pi_K[2]} &&& \Phi^K[2] };
$$
{but the triangle is commutative by the properties of adjoint functors, and the square commutes by functoriality of $Rg_! g^!$.}

  Denote by 
 $$\lambda:g_!\bZ_{W}[-2]\to 
g_!\bZ_{S_\alpha}[-2]$$ the map induced by $j_W${, i.e. $\lambda=g_!(J_W)[-2]$}.
The problem now reduces to showing that 
$\pi_K i_\Phi \lambda=0$.

  By the construction of
the map $i_\Phi$, the map $\pi_K i_\Phi$ factors as 
$g_!\bZ_{S_\alpha}[-2]\stackrel{p_K}\to \bZ_{\bx_0\times K}[-2]\stackrel {i_{\Phi^K}}\to \Phi^K$,
where $p_K$ is as in (\ref{eq145}),
so that  $\pi_K i_\Phi \lambda=i_{\Phi^K}p_K \lambda$. It is easy to see that $p_K \lambda=0$,
which finishes the proof.
$\Box$

\bigskip

It now follows that the map  $\cI_0 J_W:\bZ_{W}\to g^{-1}\Phi_0$ factors
as
$$
\bZ_{W}\stackrel{\cJ_W}\to g^{-1}\Ker q_{C\elr_\alpha}\to g^{-1}\Phi_0,
$$
where the right arrow is induced by the obvious embedding  ${\iota_q:}\Ker q_{C\elr_\alpha}\hookrightarrow \Phi_0${, cf.\eqref{iotaqdef},} 
coming from the definition $\Phi_0=\Ker \mQ$.

\subsubsection{Finishing the proof} 

Recall, see \eqref{jmQ}, that the map $J_\mQ:\bZ_\cU\to \Phi_0$
factors as 
$
J_\mQ:=\iota_qJ^q_\cU$.

Suppose that the susbet $V\subset \cU$ from Lemma \ref{peresek} satisfies: $\pi_{S_\alpha}^{-1}V\subset W$.
The statement {ii}) of
Lemma \ref{peresek}  now follows from the commutativity {(which is shown below)} of the following diagram
\begin{equation} \xymatrix{
\bZ_{\pi_{S_\alpha}^{-1}V}\ar[r]^{j_{VW}}\ar[d]^{j_{V\cU}}& \bZ_W\ar[d]^{\cJ_W}\\ 
g^{-1}\bZ_\cU \ar[r]^{(J^q_\cU)'}&g^{-1}\Ker q_{C\elr_\alpha}}
\label{BGodunov}
\end{equation}
where $j_{V\cU}$ is {the same as in the statement of Lemma \ref{peresek},} 
the map $j_{VW}$ is induced by the open embedding $\pi_{S_\alpha}^{-1}V\subset W$. The map $(J^q_\cU)'$
is induced by $J^q_\cU${, i.e. $(J^q_\cU)'=g^{-1}(J^q_\cU)$}. {Indeed, once the commutativity of \eqref{BGodunov} is known, we obtain the statement ii) by combining commutative diagrams as follows:}
$$ {\xymatrix{
 \Z_{\pi^{-1}_{S_\alpha}V} \ar[r]^-{j_{VW}} \ar[d]_-{j_{V{\cal U} } }
& 
\Z_W \ar[d]_{{\cal J}_W} 
\ar[rd]^{J_W} \\
g^{-1}\Z_{\cal U} \ar[r]^-{(J^q_\cU)'}  \ar[rd]_-{g^{-1}J_{\cal Q}}
&
g^{-1}\Ker q_{C\elr_\alpha} \ar[d]_-{g^{-1}\iota_q} & 
\Z_{S_\alpha} \ar[ld]^-{{\mathbf I}_0} \\
& g^{-1}\Phi_0
}} $$

{Let us now prove the commutativity of the diagram \eqref{BGodunov}.} We have an  injection $\kappa:\Ker q_{C\elr_\alpha}\to \Phi^C=\bZ_{X\times \Co}$
which induces an injection $\kappa':g^{-1}\Ker q_{C\elr_\alpha}\to g^{-1}\bZ_{X\times \Co}$.
The commutativity of the above diagram is equivalent to the commutativity of
\begin{equation}\label{kvadr}
\xymatrix{
\bZ_{\pi_{S_\alpha}^{-1}V}\ar[r]^{j_{VW}}\ar[d]^{j_{V\cU}}& \bZ_W\ar[d]^{\kappa' \cJ_W}\\ 
g^{-1}\bZ_\cU \ar[r]^{\kappa' (J^q_\cU)'}&g^{-1}\bZ_{X\times \Co}}
\end{equation}

Let us now define 
$$
V:= (\Co\backslash K)\cap \cU_{\bx_0}.
$$

Let us check that $V$ satisfies all the conditions:

a) $V$ is non-empty.  The set $\cU_{\bx_0}$ is obtained by removing
from $ \Co$ a finite number of $\alpha$-rays, which  implies non-emptyness
of $(\Co\backslash K)\cap \cU_{\bx_0}$.

b) $\pi_{S_\alpha}^{-1}V\subset W$ ---this is clear.

c) $\pi_{S_\alpha}:\pi_{S_\alpha}^{-1}V\to V$ is a homeomorphism ---clear.

d) Commutativity of (\ref{kvadr}). We have $g^{-1}\bZ_{X\times \Co}=\bZ_{S_\alpha}$.
It follows that the composition $\kappa'\cJ_W$ equals the map $\bZ_W\to\bZ_{S_\alpha}$
induced by the inclusion $W\subset S_\alpha$. Next, the map $\kappa J_\cU:\bZ_\cU\to \bZ_{X\times \Co}$
is induced by the open embedding $j_{\cU}:\cU\to X\times \Co$.  The commutativity now follows.
This finishes the proof.

\subsection{Infinite continuation in the direction of $K$}\label{beskon}
We need some definitions
\subsubsection{Parallelogram $\bU$}
 Let $\bU\subset \Co$  be an  open  parallelogram with vertices $A,B,C,$ and $D$, such that
$\vec{AB}$ and $\vec{DC}$ are collinear  to $e^{-i\alpha}$ and $\vec{BC}$ and $\vec{AD}$ are 
collinear to $e^{i\alpha}$.
\subsubsection{Small sets}
Let $\Gamma\subset \Co$. Call  $\Gamma$ {\em small} if for every point $c\in \Co$,
the intersection $\Gamma\cap c-K$ is a finite set. 
\begin{Claim}\label{smallset}
 Let $L\subset \Co$ be a bounded subset.
The set $\Gamma\cap (L-K)$ is  then also finite.
\end{Claim}
\textsc{Proof.}
 Assuming the contrary,
let $
\{\gamma_1,\gamma_2,\ldots,\gamma_n,\ldots\} \in \Gamma\cap (L-K)$
so that $\gamma_i=c_i-z_i$, $z_i\in K$, $c_i\in L$.  Since $L$ is bounded, the sequence $c_i$
has a convergent sub-sequence $c_{i_n}\to c$ for some $c\in \Co$. Let $\ve\in \Int K$. It follows, that
$c_{i_n}\in c+\ve-K$ for all $n$ large enough, which contradicts to smallness of $\Gamma$ 
$\Box$

\subsubsection{Theorem}

{Using notation of Sec.\ref{etalespacecS}, let}
$${p_{\cS,X} \ : \ \cS \hookrightarrow \Sigma \stackrel{p_\Sigma}{\to} X\times\C \stackrel{proj} X, } $$
$$ { \cS_z=p_{\cS,X}^{-1}(z), } $$
{and}
$$ { P_z \ : \ {\cal S}_z \stackrel{p_{\cS,X}}{\to} z\times \C = \C. } $$

\begin{Thm}\label{prodolzhenie1}
Suppose we have a section $\sigma$ of $P_z$:
$$\xymatrix{ \cS_z \ar[r]^{P_z} & \Co\\ & \bU \ar[ul]_{\sigma} \ar[u] }$$
Then there exists a small subset $\Gamma\subset \bU+K$  such that $\sigma$ extends to 
$(\bU+K) \backslash (\Gamma + \elr_{-\alpha})$ and $(\Gamma + \elr_{-\alpha})\cap \bU = \emptyset$.
\end{Thm}

{\bf Remark} For every bounded set $L$ there are  only finitely many $\gamma\in \Gamma$
such that $(\gamma+\elr_{-\alpha})\cap L\neq \emptyset$, as follows from Claim \ref{smallset}.

Before proving this theorem, let us observe that it easily implies Theorem  \ref{prodolzhenie}.
Indeed, given $\xu\in \Co$, we see that
$\cS^{\xu}$ is a disjoint union of all $\cS_{z}$, where $p_X(z)=\xu$, which reduces Theorem \ref{prodolzhenie}
to the current Theorem.
The rest of this subsection is devoted to its proof.

\subsubsection{Reformulation in terms of sheaves}

 By basic properties of an \'etale space of a sheaf, liftings $\sigma$ as in Theorem,  are in 1-to-1 correspondence with 
maps of sheaves $f_\sigma:\Z_\bU\to \Phi_0|_{z \times \Co}$. 

For every $w\in \wa$ and a fixed $z\in X$, set $\zA_z(K,w)=\zA(K,w)\cap (z\times \Co)\subset \Co$, where $\zA(K,w)$ are 
the same is in Sec \ref{akaw}
 We define $\zA_z(\elr_\alpha,w)$, $\zA_z(\elr_{-\alpha},w)$ in a similar way.

We then have the following maps:
$$  \Z_\bU \stackrel{f_{{\sigma}}}{\to} \raisebox{22pt}{\xymatrix{ \Z_{\C} \ar[r]^{\zq^{C \elr_\alpha}}
 \ar@{}[d]|{\oplus}
 & \bigoplus_w \Z_{\zA_z(\elr_{\alpha}, w)}  \\
\bigoplus_w \Z_{\zA_z(K, w ) } \ar[r]^{\zq^{K\elr_{-\alpha}}}\ar[ur]^{-\zq^{K\elr_\alpha}}&
 \bigoplus_w \Z_{\zA_z(\elr_{-\alpha}, w ) }   } }$$
where $\zq^{C \elr_\alpha}$, $\zq^{K\elr_\alpha}$,
$\zq^{K \elr_{-\alpha}}$ are  the restrictions of the maps $q^{C \elr_\alpha}$, $q^{K\elr_\alpha}$,
$q^{K \elr_{-\alpha}}$ onto $\bx_0\times \Co$. Let $\mQ_{\bx_0}$ be the restriction of the map $\mQ$ onto
$\bx_0\times\Co$, so that $\mQ_{\bx_0}$ is the sum of   $\zq^{C \elr_\alpha}$, $-\zq^{K\elr_\alpha}$,
and $\zq^{K \elr_{-\alpha}}$.
We now have 
\begin{equation} \mQ f_\sigma=0. \label{se29e11} \end{equation}

\subsubsection{Writing $f_\sigma$ in terms of its components}
We have components:
$$ f_\sigma(w) \ : \ \Z_\bU \to \Z_{\zA_z(K,w)} $$
$$ f_\sigma(0) \ : \ \Z_\bU \to \Z_{\Co} $$
we have (if $\bU\cap\zA_z(K,w) \ne \emptyset$):
$$ \hom(\Z_\bU; \Z_{\zA_z(K,w)}) \ = \ \Z \cdot g_w $$
where 
\begin{equation} g_w \ : \ \Z_\bU \to \Z_{\bU\cap \zA_z(K,w)} \to \Z_{\zA_z(K,w) } \label{se3e1129}\end{equation}
(the first arrow is induced by the closed embedding $\bU\cap {\zA}_{{z}}(K,w)\subset \bU$; 
the second arrow is an open embedding) \\
if $\bU\cap \zA_z(K,w) = \emptyset$, then $\hom(\Z_\bU, \Z_{\zA_z(K,w)} )=0$. 

So, 
\begin{equation} f_\sigma(w) = n_w \cdot g_w , \ \ \ \text{where} \ n_w \in \Z , \label{Se29vy1} \end{equation} 
and $f_\sigma(w) =0$ if $\bU\cap \zA_z(K,w)=\emptyset$.

Analogously, $hom(\Z_\bU,\Z_\C) =\Z\cdot g_0 $, so 
\begin{equation} f{(0)} = n_0\cdot g_0. \label{Se29vy2} \end{equation}

It also follows that:
\begin{Claim}\label{finitude} for every point $s\in \bU$ there are only finitely many $w$ such that
$f_{{\sigma}}(w)\neq 0$ and $s\in \zA_z(K,w)$.
\end{Claim}
\textsc{Proof} This follows from  consideration of the induced map
on stalks at {$s$}:
$$
(f_{\sigma})_s:(\bZ_\bU)_s=\bZ\to \bigoplus\limits_{w:s\in \zA_z(K,w)}\bZ=
( \bigoplus\limits_{w\in \wa}\zA_z(K,w))_s.
$$
The image of this map must be contained in the direct sum of only finitely many copies of $\bZ$,
the statement now follows.
$\Box$

\subsubsection{Restriction to a sub-parallelogram $\bV$} \label{SubParal}

Let $\bV\subset \bU$ be a parallelogram,
$\bV=AB'C'D'$, such that $B'\in (AB)$, $D'\in (AD)$ (so that $C'\in \bU$).

The  restriction
$$f_{\sigma,\bV}:= f_\sigma|_\bV \ : \ \Z_\bV \to\bZ_\bU
\stackrel{f_\sigma}{\to} \Z_{\C} \oplus \bigoplus_w \Z_{\zA_z(K,w)} $$
can thus be expressed as
$$ f_{\sigma,\bV} \ = \ {n_0\cdot g_0|_V \ + \ } \sum_{w\in \wa} n_w \cdot g_w|_\bV.
$$
Here $g_w|_\bV$ is the following composition:
$$ \Z_\bV \to \Z_\bU  \stackrel{g_w}{\to} \Z_{\zA_z(K,w)} $$
and $g_w$ is the same as in \eqref{se3e1129}. 

Let $S\subset \wa$ consist of all $w$ such that $n_w\neq 0$ and $g_w|_{\bV}\neq 0$. We can now rewrite
 \begin{equation} f_{\sigma,\bV} \ = \ \sum_{w\in S} n_w \cdot g_w|_\bV \label{SetSdefd} 
\end{equation} 

 Observe that 
\begin{equation}\label{nepusto}\text{$g_w|_\bV\neq 0$ iff
$\bV\cap \zA_z(K,w)\neq \emptyset$.}
\end{equation}
  Next, there  are 
only finitely many $w$ such that  $f(w)\ne 0$ and $\zA_z(K,w)\cap \bV \ne \emptyset$.
 Indeed, $\zA_z(K,w)\cap \bV\ne \emptyset$ implies $C'\in \zA_z(K,w)$, and we can set $z=C'$ in 
Claim \ref{finitude}. This shows that $S$ is a finite set.

{We comment that restricting from $\bU$ to $\bV$ was done in order to obtain this finiteness of $S$.}

\subsubsection{Proof of a weaker version  of the  Theorem}
We are going to prove the following statement: {\em there exists a small set $\Gamma\subset \bV+K$,
such that $\sigma|_{\bV\cap \cV}$ extends to $\cV$, where $\cV:=(\bV+K)\backslash (\Gamma+K)$.}

Define the extensions $\Z_{\bV+K} \stackrel{G_w}{\to} \Z_{\zA_z(K,w)} $ as follows:
$$ G_w \ : \ \Z_{\bV+K} \stackrel{c}{\to} \Z_{(\bV+K)\cap \zA_z(K,w)} \to \Z_{\zA_z(K,w)}, $$
where  the map $c$ is the restriction onto a closed subset and the second map is
 induced by the embedding of an open subset).

Let  $G_0:\bZ_{\bV+K}\to \bZ_\Co$ be the map coming from the open embedding of the 
corresponding sets.

Let $$
F_{\sigma,\bV}:=n_0G_0+\sum_{w\in S} n_w G_w:\bZ_{\bV+K}\to \Z_\Co\oplus \bigoplus_{w\in \wa} \Z_{\zA_z(K,w)},
$$ 
where the coefficients $n_w, n_0$ are the same as in \eqref{Se29vy1}, \eqref{Se29vy2}.
Let $J_\bV:\bZ_{\bV}\to\bZ_{\bV+K}$ be the map coming from the open embedding of the 
corresponding sets.
We have:
\begin{equation}\label{prodsech}
 f_{\sigma,\bV}= F_{\sigma,\bV} J_\bV.
\end{equation}
 Let us now find a a subset $\cV\subset \bV+K$ such that ${\mQ}\circ F_{\sigma,\bV}|_{\cV}=0$.
This vanishing along with (\ref{prodsech}) imply that $F_{\sigma,\bV}$ determines an extension
of $\sigma|_{\bV}$ onto $\cV$.

1) Consider the through map for some $w\in S$:
$$ \beta_w \ : \ \Z_\bV \stackrel{f_{\sigma,\bV}}{\to} \raisebox{22pt}{\xymatrix{ \Z_\C \ar[r] 
\ar@{}[d]|{\oplus}  &
 \bigoplus_{w\in \wa} \Z_{\zA_z(\elr_{\alpha};w)} \ar@{}[d]|{\oplus} \\
 \bigoplus_{w\in \wa} \Z_{\zA_z(K,w)} \ar[ur] \ar[r] &
 \bigoplus_{w\in \wa} \Z_{\zA_z(\elr_{-\alpha};w)} } }
\stackrel{\rho_w}{\to} \Z_{\zA_z(\elr_{-\alpha},w)} $$
$\rho_w$ is the projection onto a direct summand, and the middle map is $\mQ_z$.

By \eqref{se29e11}, $\beta_w=0$; on the other hand,  $\beta_w=n_w\cdot h_w$, where 
$$ h_w\ : \ \Z_\bV \stackrel{G_w}{\to} \Z_{\zA_z(K,w)} \stackrel{restr}{\to} \Z_{\zA_z(\elr_{-\alpha},w)}. $$
But 
$h_w=0$ iff $\bV\cap \zA_z(\elr_{-\alpha};w) =\emptyset$. 
So if $n_w\ne 0$, then \begin{equation} \bV\cap \zA_z(\elr_{-\alpha};w)=\emptyset. \label{se30con1} \end{equation} Since $w\in S$ 
and because of (\ref{nepusto}),
we  have \begin{equation} \bV\cap\zA_z(K;w)\neq \emptyset. \label{se30con2} \end{equation}
From \eqref{se30con1} and \eqref{se30con2} it follows that 
$(\bV+K)\cap \zA_z(\elr_{-\alpha};w)=\emptyset$. Hence,
we have 
\begin{equation}\label{rhow}
\rho_w \circ {\mQ} \circ F_{\sigma,\bV}: \Z_{(\bV+K)}\to \Z_{\zA_z(\elr_{-\alpha},w)}=0.
\end{equation}

Let us now consider the maps $\kappa\circ {\mQ}\circ F_{\sigma, \bV}$, where $\kappa$ is the projection
onto $\oplus_w \Z_{\zA_z(\elr_{\alpha},w)} $ as shown in the following diagram:
$$ \kappa\circ {\mQ} \circ F_{\sigma,\bV} \ : \ \Z_{\bV+K} \stackrel{F_{\sigma,\bV}}{\to} 
\raisebox{22pt}{\xymatrix{ \Z_\C
 \ar[r]^-{{q^{C\elr_\alpha}_z}} \ar@{}[d]|{\oplus}  & \bigoplus_{w\in \wa} \Z_{\zA_z(\elr_{\alpha};w)} \ar@{}[d]|{\oplus} \\ 
\bigoplus_{w\in \wa} \Z_{\zA_z(K,w)} \ar[ur] \ar[r] & \bigoplus_{w\in \wa} \Z_{\zA_z(\elr_{-\alpha};w)} } }
\stackrel{\kappa}{\to} \bigoplus_{w\in \wa} \Z_{\zA_z(\elr_{\alpha},w)} $$
 Let  $M_w:\bZ_\Co\to \bZ_{\zA_z(\elr_{\alpha};w)}$ be the components of
the map $q_{{z}}^{C \elr_\alpha}$. 
Let
$$\Delta=\{ w' \ : \ \exists w\in S \ : \ N_{ww'}\ne 0 \ \text{or} \ M_{w'}\ne 0 \} {\subset \wa}. $$
Here $S$ is as in \eqref{SetSdefd}, $N_{ww'}:=n_{\bA^{-1}(w);w'}$, and  $n_{\tilde w;w'}$ are the same as  in Prop. \ref{ipip}. {(Remark, however, that the statement of the Prop.\ref{ipip} is not used here. )}

For each $w'\in \wa$ let us write
$$ \zA_z(K,w') = d_{w'} + K. $$  Set $\Gamma:=\{d_{w'}:w'\in \Delta\}\subset \Co$.
As $S$ is finite (see end of section \ref{SubParal}), for any $ s\in \C$ there are only finitely many $w'\in \Delta \ : \ {\zA}(K,w')\ni s$. 
Equivalently there are only finitely many $w'$ such that $d_{w'}\in s-K$ so that
  $\Gamma$ is small .

Let $$\pi_w:\bigoplus_{w'\in \wa} \bZ_{\zA_z(\elr_\alpha,w')}\to \bZ_{\zA_z(\elr_\alpha,w)}
$$
be the projection. It follows that $\pi_w\kappa \mQ F_{\sigma,\bV}\neq 0$ only if $w\in \Delta$.
Set $\cV:=\bV+K\backslash (\Gamma+K)$.
It follows that  $\pi_w\kappa\mQ F_\bV|_{\cV} = 0 $, which implies $\kappa \mQ F_{\sigma,\bV}|_{\cV}=0$.
Taking into account (\ref{rhow}), we conclude $\mQ F_{\sigma,\bV}|_{\cV}=0$, i.e. 
$\sigma|_{\cV\cap \bV}$ extends onto  $\cV$, as we wanted.

\subsubsection{Proof of the theorem for $\bU$} 
Denote by $\sigma'$ the extension of $\sigma|_{\cV\cap \bV}$ onto $\cV$.
Observe that  the set $\cV\cap \bU$ is connected and that $\cV\cap \bV\subset\cV\cap\bU$.
 Thus, $\sigma$ and $\sigma'$
are two extensions of $\sigma|_{\bV\cap \cV}$ onto $\cV\cap \bU$. By Sec \ref{etal}
we have $\sigma|_{\cV\cap \bU}=\sigma'|_{\cV\cap\bU}$. Thus, $\sigma$ extends to
$\cV\cup \bU$ which is of the required type.
$\Box$

\section{Orthogonality criterion -- a simplified version} \label{simplified}
\def\Re{{\mathbb R}} 
\def\hom{{Hom}}

The goal of this section is to prove Theorem \ref{plosk} below.  This theorem will  only be used in the next Section
\ref{genstrip}. 
\subsection{Formulation of the Theorem}

Let $X$ be a smooth manifold.  We will work on a manifold $Y =X\times \R \times\R$.
Let us refer to points of $Y $ as $(x,s_1,s_2)\in X\times \R\times \R$.
Let $P_1,P_2:Y \to X\times\R$ be projections
$$
P_i(x,s_1,s_2)=(x,s_i).
$$

Let us refer to points of $T^*Y$ as $(x,s_1,s_2,\omega,a_1ds_1,a_2ds_2)$,
where $\omega\in T^*_xX$; $a_1ds_1\in T^*_{s_1}\R$; $a_2ds_2\in T^*_{s_2}\R$.
Let $\Omega_{Y} \subset T^*Y$ be the closed subset consisting of all points
$(x,s_1,s_2,\omega,a_1ds_1,a_2ds_2)$ where $a_1=0$ or $a_2=0$ (or both).
Let $\cC_{Y}\subset \DerCat(Y)$ be the full subcategory consisting of all objects microsupported
within $\Omega_{Y}$.
Let ${}^\perp \cC_{Y}$ be the left orthogonal complement to $\cC_{Y}$ (consisting of
all $F\in \DerCat(Y)$ such that $R\hom(F,G)=0$ for all $G\in \DerCat(Y)$).

\begin{Thm}\label{plosk}
$F\in {}^\perp \cC_{Y}$ iff $RP_{1!}F=RP_{2!}F=0$.
\end{Thm}

Let us start with proving that $F\in {}^\perp \cC_{Y}$ implies $RP_{1!}F=RP_{2!}F=0$.
Indeed, given any  $G\in \DerCat(X\times \R)$, we have
$$
R\hom(RP_{1!}F;G)=R\hom(F,P_1^!G).
$$
It is well known that every element   $(x,s_1,s_2,\omega,a_1ds_1+a_2ds_2)\in S.S.(p_1^!G)$
satisfies $a_2=0$, i.e. $P_1^!G\in \cC_{Y}$ and
$$
R\hom(RP_{1!}F;G)=R\hom(F,P_1^!G)=0.
$$
As $G$ is arbitrary, we conclude $RP_{1!}F=0$. One can prove the equality $RP_{2!}F=0$ in a similar way.

The rest of this section will be devoted to proving the opposite implication:

\begin{Thm}\label{plosk1} Let $F\in \DerCat(Y)$ satisfy
$RP_{1!}F=RP_{2!}F=0$.  Let $G\in \cC_{Y}$. Then $R\hom(F,G)=0.$
\end{Thm}

We start with introducing the major tool, namely a version of Fourier-Sato transform.
\subsection{Fourier-Sato Kernel}
Let $E$ be the dual real vector space to $\R^2$ so that we have a pairing
$<,>: \R^2\times E\to \R$. Let us use the standard coordinates $s_1,s_2$ on $\R^2$
and $\sigma_1,\sigma_2$ on $E$ so that $$<(s_1,s_2),(\sigma_1,\sigma_2)>=s_1\sigma_1+s_2\sigma_2.$$
Let $Y_2:=X\times \R^2 \times \R^2$. 
Define projections $\pi_1,\pi_2:Y_2\to Y$:
$$
\pi_1(x,s,s')=(x,s);
$$
$$
\pi_2(x,s,s')=(x,s'),
$$
where $s=(s_1,s_2)\in \R^2$ and $s'=(s'_1, s'_2)\in \R^2$. 

Let
$K\subset Y_2\times E 
$
be the following closed subset
$$
K=\{(y,s,s',\sigma)|\langle s-s',\sigma\rangle\geq 0\}.
$$
Let us also define the projections
$$
p_1  \ : \ Y_2\times E \ \to \  Y_2\stackrel{\pi_1}\to Y;
$$
$$
p_2 \ : \ Y_2\times E\ \stackrel{\pi_2\times \id_E}{\longrightarrow} \ Y\times E.
$$

We then have the following  functor: $\Psi: \DerCat(Y)\to \DerCat(Y\times E)$:
$$
\Psi(F):=Rp_{2*}R\ihom(\bZ_K; p_1^!F)
$$
which are modified versions of Fourier-Sato transform.
Let us establish certain properties of these functors (similar to those of Fourier-Sato transform).

\subsubsection{Properties of the modified Fourier-Sato transform.}
\begin{Lemma}\label{isholem}
Let $\pi_E:Y\times E\to Y$ be the projection. We  then have a natural isomorphism
$$
F\to R\pi_{E*}\Psi(F)[2].
$$
\end{Lemma}
\textsc{Proof} 
Let $p_E:Y_2\times E\to Y_2$ be the projection. We then
have
\begin{equation}
R\pi_{E*}\Psi(F)\sim R\pi_{2*}R\ihom(Rp_{E!}\bZ_K;R\pi_1^!F). \label{se14e336} 
\end{equation}
(Indeed, one uses $p_1=\pi_1\circ p_E$,  the adjunction formula for $p_{E!}$,  and  $\pi_E\circ p_2=\pi_{E}\circ \pi_2$. )

A simple computation shows that  we have 
$$
Rp_{E!}\bZ_K\cong \bZ_\Delta[-2].
$$
where $\Delta\subset Y_2$ is the diagonal, i.e. the set of all points of the form
$(x,s,s)$. The statement now follows. $\Box$

\subsubsection{Singular support estimation} 
Let us define the following set \begin{equation}\label{krestC}
C:=\{(\sigma_1,\sigma_2)|\sigma_1=0 \text{ or } \sigma_2=0\}\subset E.
\end{equation}
Let $U:=E\backslash C$.

\begin{Lemma} \label{se14L6}  Suppose $G\in \cC_Y$. Then
we have:
$$
\SS(\Psi(G))\cap T^*(Y\times U)\subset \{(x,s,\sigma,\omega,0,bd\sigma)\}\subset T^*(Y\times U),
$$
where $(x,s)\in X\times\R^2=Y$; $\sigma\in U$; $\omega\in T^*_xX$; $bd\sigma\in T^*_\sigma U$.
\end{Lemma}

\textsc{Proof.} 
 First of all, by ~\cite[Prop.5.3.9]{KS}, 
\begin{equation} \SS(\bZ_K)=\{ ((s,s',\sigma), \lambda d\langle s-s',\sigma\rangle) \ : \
 \lambda\langle s-s',\sigma\rangle =0, \ \lambda\ge 0, \ \langle s-s',\sigma\rangle \geq 0\}.  \label{SSPsi} \end{equation}

By ~\cite[proof of Prop.5.4.2]{KS},  
$ S.S. p_1^! G$  is contained in the following subset of $T^*(Y_2\times E)$:
$$
(x,s,s',\sigma,\omega, ads, 0\cdot ds', 0\cdot d\sigma),
$$
where $(x,s,\omega,ads)\in \Omega_Y$.

 Let us now check that
 \begin{equation}\label{inters}
S.S. p_1^! G\cap S.S.\bZ_K \subset \{ \text{zero section}\}.
\end{equation}
Suppose we have an element  $\eta$ in this intersection which does not belong to the zero section.
It should be of the form
as in (\ref{SSPsi}). Since $\eta\neq 0$, $\lambda>0$ and $\langle s-s',\sigma\rangle=0$.
We have $$
\lambda d\langle s-s',\sigma\rangle=\lambda\langle s-s',d\sigma\rangle+\lambda\langle ds-ds',\sigma\rangle.
$$
The $ds'$ component of $\eta$ is thus $-\lambda\langle ds',\sigma\rangle$. In order for $\eta\in 
\SS\pi_1^!G$, this component must vanish, which implies $\sigma=0$.
Analogously, $d\sigma$-component of $\eta$ must vanish as well, i.e. $s-s'=0$. This implies
that $\eta$ is in the zero section, contradiction. This proves (\ref{inters}).

It now follows that 
$$
\SS R\ihom(\bZ_K;p_1^!G)\subset \SS( p_1^!G)-\SS(\bZ_K)
$$
(where $ ``-"$ means subtraction in each fiber of $T^*(Y_2\times E))$, ~\cite[Cor.6.4.5]{KS}), 
i.e.
\begin{equation} \SS R\ihom(\bZ_K;p_1^!G) \ \subset \
 \left\{(x,s,s',\sigma,\omega,ads -  \lambda d\langle s-s',\sigma\rangle) \right\}  \label{se14e657} \end{equation}
where
\begin{equation}  (x,s,\omega,ads)\in \Omega_Y 
\label{se14e620}
\end{equation}
 and  $s,s',\sigma,\lambda$
satisfy the same conditions as in (\ref{SSPsi}).

Now let us estimate $$\SS Rp_{2*} R\ihom(\bZ_K;p_1^!G)=\SS(\Psi(G)).$$
By ~\cite[Lemma 3.3]{T08}, 
we have: if $(a')^0 d(s')^0\ne 0$ , then 
$$ (x^0,(s')^0,\sigma^0,\omega^0, (a')^0d(s')^0+b_0d\sigma^0) \not \in S.S. Rp_{2*} R\ihom(\bZ_K;p_1^!G) $$
as long as: \\

{\em  there exists $\varepsilon$ such that $R\ihom(\bZ_K;p_1^!G)$
 is nonsingular at all points
$ (x_\star,s_\star,s'_\star,\sigma_\star,\omega_\star, a_\star ds+ a'_\star ds'+b_\star d\sigma), $ where}
\begin{equation}\label{epsl}
\left\{ \begin{array}{cccc}
|x_\star-x^0|<\ve, & \text{any} \ s_\star\in \R^2, & |s'_\star-(s')^0|<\ve, & |\sigma_\star-\sigma^0|<\ve, \\ 
|\omega_\star-\omega^0|<\ve, &|a_\star|<\ve, & |a'_\star-(a')^0|<\ve, & |b_\star-b^0|<\ve  . \end{array} \right.
\end{equation}

Thus, the proof of the lemma \ref{se14L6} reduces to the following statement: 

{\it Let $(x^0,(s')^0,\sigma^0,\omega^0, (a')^0d(s')^0+b_0d\sigma^0) \in T^*(Y\times E)$ satisfy: \\
a) $\sigma^0=(\sigma^0_1,\sigma^0_2)$ is such that 
\begin{equation}\label{sigo}
\sigma^0_1\neq 0 \text{ and } \sigma^0_2\neq 0;
\end{equation}
b)  $(a')^0\neq 0$. \\
Then for some $\ve>0$  there are no solution $(x_\star,s_\star,s'_\star,\sigma_\star,\omega_\star, a_\star, a'_\star, b_\star)$ of the inequalities (\ref{epsl}) satisfying the conditions (coming from \eqref{se14e657}  )
\begin{equation} \left\{ \begin{array}{cccc} x_\star=x, & s_\star=s, & s'_\star=s', & \sigma_\star=\sigma,
\\
 \omega_\star=\omega, &  a_\star=a-\lambda \sigma, &  a'_\star=\lambda \sigma, & b_\star=-\lambda(s-s'),
\end{array} \right. \label{se14e706} \end{equation}
 such that condition of \eqref{SSPsi} and \eqref{se14e620} hold.}

Eliminating the variables with $\star$ and conditions on $x,\omega, b$, we must, for fixed $0$-variables find $\ve$ making the following list of conditions inconsistent:
\begin{enumerate}
\item \label{se15sprim} $|s'-(s')^0|<\ve$ 
\item \label{se15vs} $ |\sigma-\sigma^0|<\ve $
\item \label{se15a} $|a-\lambda \sigma|<\ve $
\item \label{se15aprim} $|\lambda \sigma - (a')^0 | < \ve$ 
\item \label{se15cross}  $a_1=0$ {or} $a_2=0$ 
\item \label{se15lamge0} $\lambda \ge 0$ 
\item \label{se15ortho} $\lambda\langle s-s',\sigma\rangle = 0$ 
\item \label{se15acute} $\langle s-s',\sigma\rangle \ge 0$
\end{enumerate}

Indeed, suppose there is a solution to this system of inequalities such that $a_1=0$. Then by condition \ref{se15a}, $|\lambda \sigma_1|<\varepsilon$, i.e. 
\begin{equation} |\lambda|<\frac{\varepsilon}{|\sigma_1|} \label{se15e200} \end{equation} 
 By condition \ref{se15vs}, \begin{equation}|\sigma|<|\sigma^0|+\varepsilon. \label{se15e202} \end{equation}
Combining condition \ref{se15aprim} with \eqref{se15e200} and \eqref{se15e202}, obtain
\begin{equation}
\varepsilon > | (a')^0-\lambda \sigma | \ge |(a')^0|- \lambda\cdot (|\sigma^0|+\varepsilon) \ge |(a')^0|- \frac{\varepsilon}{|\sigma_1|} (|\sigma^0|+\varepsilon) 
\label{se15e204} \end{equation}
If we choose $\varepsilon>0$ to satisfy (cf. condition a) ) 
\begin{equation} \varepsilon < \frac{1}{2} \min\{ |\sigma^0_1|, |\sigma^0_2|\} \label{se15e211} \end{equation}
then   \eqref{se15e204} yields
\begin{equation}
\varepsilon >  |(a')^0|- \frac{2\varepsilon}{|\sigma^0_1|} (|\sigma^0|+\varepsilon) 
\label{se15e214} \end{equation}

We have assumed $a_1=0$ above; if we assume $a_2=0$ (cf. condition \ref{se15cross}), we get an analogous inequality. Choosing $\ve>0$ to satisfy \eqref{se15e211} and to  violate both \eqref{se15e214} and its analog for $a_2=0$,  finishes the proof. 


 $\Box$

\subsubsection{}
\begin{Lemma} \label{nollo} Let $G\in Ob(\cC_Y)$. Then $\Psi(G)|_{Y\times U}=0$.
\end{Lemma}
\textsc{Proof}
Let $q:Y\times U\to X\times U$ be the projection $q(x,s,\sigma)=(x,\sigma)$. 
We have a natural map
$$
\iota:q^{-1}Rq_*(\Psi(G)|_{Y\times U})\to \Psi(G)|_{Y\times U}
$$
By virtue of lemma \ref{se14L6} and the fact that the fibers of $q$ are diffeomorphic to $\R^2$,
we see that  $\iota$ is an isomorphism.

It now remains to show that $Rq_*(\Psi(G)|_{Y\times U})=0$.

Let $K_U:=K\cap (Y_2\times U)$.
Let 
$q_1:Y_2\times U\to Y\times U$, $q_2:Y\times U\to Y$, $q_3:Y\times U\to X\times U$
be the projections
$$
q_1(x,s,s',\sigma)=(x,s',\sigma);
$$
$$
q_2(x,s,\sigma)=(x,s);
$$
$$
q_3(x,s,\sigma)=(x,\sigma).
$$
In this notation,
$$
Rq_*(\Psi(G)|_{Y\times U})=Rq_{3*}R\ihom_{Y\times U}(Rq_{1!}\bZ_{K_U};q_2^!G).
$$

Finally, we observe that $Rq_{1!}\bZ_{K\times U}=0$ (pointwise computation). $\Box$

\subsubsection{Representation of $G$} 

Let $i_C:C\subset E$ be the closed embedding; here $C$ is as in (\ref{krestC}).
Let $K_C:=K\cap (Y_2\times C)$. Let 
$$
p_1^C:Y_2\times C\to Y_2\stackrel{\pi_1}\to Y
$$
and
$$
p_2^C \ : \ Y_2\times C \ \stackrel{\pi_2\times \id_C}{\longrightarrow } \  Y\times C.
$$
Let $q^C:Y\times C\to Y$ be the projection.
Let $G\in \cC_Y.$
It now follows from Lemma \ref{nollo} that $\Psi(G)=(\id_Y \times i_C)_* (\id_Y \times i_C)^{-1}\Psi(G)$, which together with Lemma   \ref{isholem} yields a natural isomorphism
$$
G\cong Rq^C_*Rp_{2*}^CR\ihom_{Y_2\times C}(\bZ_{K_C};(p_1^{C})^!G)[2].
$$

So that we have an induced isomorphism
$$
R\hom(F,G)\cong R\hom(F;Rq^C_*Rp_{2*}^CR\ihom_{Y_2\times C}(\bZ_{K_C};(p_1^{C})^!G))[2].
$$
Let us rewrite the RHS.

First of all, set 
 $$\pi_2^C:=q^Cp_2^C  \ : \  Y_2\times C \ \to \ Y \ : \ (x,s,s',\sigma) \mapsto (x,s') . $$  We then have
$$
R\hom(F;Rq^C_*Rp_{2*}^CR\ihom_{Y_2\times C}(\bZ_{K_C};(p_1^{C})^!G)) 
$$
$$
=R\hom((\pi_2^C)^{-1}F;\ihom(\bZ_{K_C};(p_1^{C})^!G))
$$
$$
=R\hom((\pi_2^C)^{-1}F\otimes \bZ_{K_C};(p_1^{C})^!G)).
$$
Next, we factor $p_1^C=q^C\pi_1^C$, where
$$
\pi_1^C:Y_2\times C \  \stackrel{\pi_1\times \id_C}\longrightarrow \ Y\times C
$$
so that we can continue 
$$
R\hom((\pi_2^C)^{-1}F\otimes \bZ_{K_C};(p_1^{C})^!G))=
R\hom_{Y\times C}(R(\pi_{1}^C)_!((\pi_2^C)^{-1}F\otimes \bZ_{K_C});(q^C)^!G).
$$

Let us show that $\bF:=R\pi^C_{1!}((\pi_2^C)^{-1}F\otimes \bZ_{K_C})=0$ under assumtions on $F$ from Theorem
\ref{plosk1}. Indeed, let $(a,0)\in C$, $a\neq 0$. Then, for any $F\in \DerCat(Y)$, we have
$$
RP_{1!}F\cong \bF|_{Y\times (a,0)}.
$$
Similarly,
$$
RP_{2!}F\cong \bF|_{Y\times (0,a)}.
$$
Finally,
$$
RP_{0!} F\cong \bF|_{Y\times (0,0)},
$$
where $P_0:Y\times C\to Y$ is the projection.  Since $P_0$ passes through $P_1$,
all the restriction listed vanish under assumptions from Theorem \ref{plosk1}.
This concludes the proof.

\section{Orthogonality criterion for a  generalized strip}\label{genstrip}
\def\Re{{\mathbb R}}
\subsection{Conventions and notations}

Let $\alpha\in (0,\pi/2)$ be an acute angle, same as in {Sec.\ref{InDataS}}.

Set ${\bf e}=e^{-i\alpha}$; ${\bf f}=e^{i\alpha}$ 
so that ${\bf e},{\bf f}$ is a basis of $\C$ over $\R$ and every
complex number $z$ can be uniquely written as
$z=x{\bf e}+y{\bf f}$, $x,y\in \R$ so that we identify 

\begin{equation}\label{reco}
\C\stackrel\sim\to \R^2
\end{equation}
using the coordinates $(x,y)$.

 Define  a generalized strip which is  a  set of one of the following  types: \\
{\em First type:}
\begin{equation}\label{strip>}
{\bS} = \{ x{\bf e} + y{\bf f} \ : \ x>\gamma;\  y\in (A,B) \} \subset \R^2 = \C,
\end{equation} 
where $-\infty\leq \gamma<\infty$ and $-\infty\leq A<B\leq \infty$.

{\em Second type:}
\begin{equation}\label{strip<}
{\bS} = \{ x{\bf e} + y{\bf f} \ : \ x<\gamma;\  y\in (A,B) \} \subset \R^2 = \C,
\end{equation} 
where $-\infty< \gamma\leq \infty$ and $-\infty\leq A<B\leq \infty$.

\subsubsection{Convolution} 
Let $M,N$ be smooth manifolds
Define a convolution bi-functor
$$
* \ : \ \DerCat(M\times \R^2)\times \DerCat(N\times \R^2)\to \DerCat(M\times N\times \R^2)$$ as follows.
Denote 
\begin{equation}  A \ : \ M\times \R^2\times N\times  \R^2\to M\times N\times \R^2 \ : \ \ \ A(m,u,n,v)=(m,n,u+v) \label{MapAdef} \end{equation}
We now define
$$F*S:=RA_!(F\boxtimes^{\bL} S).$$

\subsubsection{The category $\cC_\bS$.} \label{CatCSdefd} 

Let $\Omega_\bS\subset T^*(\bS\times \Re^2)$ be a closed conic subset consisting of all points
$$
(x_1,y_1,x_2,y_2,a_1dx_1+b_1dy_1;a_2dx_2+b_2dy_2)
$$
where $(x_1,y_1)\in \bS$ and  $(a_1,b_1)=\pm(a_2,b_2)$ .

In terms of the complex coordinate $z=x\be+y\bff$ and the identification (\ref{reco}) we have:
$$
\Omega_\bS=\{(z,s,adz+bds|z\in  \bS,s\in \Co,a=\pm b\}.
$$

Let $\cC_\bS\subset \DerCat(\bS\times \Re^2)$ be the full subcategory consisting of
all objects microsupported within $\Omega_{\bS}$.
\subsubsection{Rays $l_+$ and $l_-$}
Let
$$l_+:=\{(x,0)|x\geq 0\} \subset \R^2 \ ; \ \ \ l_-:=\{(x,0);x\leq 0\}\subset \R^2, $$
\subsubsection{Projectors $P_\pm$} Let us define the following projectors
$ P_\pm \ : \bS\times \R^2\to \R^2$, where
  \begin{equation}\label{PpmDefined}
 P_\pm(x_1,y_1,x_2,y_2)=(x_1\pm x_2;y_1\pm y_2).
 \end{equation}

\subsection{Formutation of the criterion}

Our criterion is then as follows.

\begin{Proposition} \label{main}
Consider constant sheaves $\bZ_{l_\pm}\in \DerCat(\R^2)$. Let $F\in \DerCat({\bS}\times \R^2)$ and suppose that one of 
the natural maps
\begin{equation}\label{lplus}
\bZ_{l_+}*F\to \bZ_0*F=F
\end{equation}
\begin{equation}\label{lminus}
\bZ_{l_-}*F\to \bZ_0*F=F;
\end{equation}
is a quasi-isomorphism.

Suppose that both $RP_{+!}F=0$ and $RP_{-!}F=0$. Then $F\in \perpC_{{\bS}}$.
\end{Proposition}

The rest of this section is devoted to proving this criterion {\em under   the assumption \eqref{lplus}}.
The case \eqref{lminus} is treated in a fairly similar way and is omitted.
\subsection{Fourier-Sato decomposition}
 Denote by $E$ the dual  vector space to $\R^2$. We have the standard identification $E=\R^2$.
Let $\lang,\rang$  be the standard pairing  $E\times\R^2\to \R$. 
Let $Z\subset E\times\R^2$; $Z=\{(\zeta,u)|\lang \zeta, u\rang \geq 0\}$.
  
 As was explained above, we have the convolution 
$$
* \ : \ \DerCat(E\times \R^2)\times \DerCat(\bS\times \R^2)\to \DerCat(E\times {\bS}\times \R^2).
$$

For $F\in \DerCat({\bS}\times \R^2)$ set
\begin{equation}
\fs(F):=\bZ_Z* F\in \DerCat(E\times {\bS}\times \R^2), \label{FFdefd}
\end{equation} 
where $\bZ_Z\in \DerCat(E\times\R^2)$ is the constant sheaf on $Z$.
Notice that  $\fs(F)$ is an analog of (but is not directly equal to) the  Fourier-Sato transform of ~\cite[Ch.3.7]{KS}.

\begin{Lemma} (Fourier-Sato decomposition of  $F$) \label{LF1} Consider
 the projection $q: E\times 
\bS\times  \R^2 \to {\bS}\times \R^2$. Then for any  $F\in \DerCat(\bS\times \Re^2)$,  
we have a natural isomorphism
$$Rq_! {\mathbb F}({F})[2]\cong {F}.$$
 \end{Lemma}

\textsc{Proof.}
 Let us introduce the following projections (where, e.g., $p_{24}$ means the projection
onto the 2-nd and the 4-th factor):
$$ \xymatrix{ & E\times {\bS}\times \R^2 \times \R^2 \ar[ld]_{p_{123}} \ar[d]_{p_{23}} \ar[rd]_{p_{234}} 
 \ar[rrd]^{p_{24}} \\
E\times {\bS} \times \R^2 \ar[r]_-{q} & {\bS}\times \R^2 & {\bS}\times \R^2\times \R^2 
\ar[l]^-{\tilde p_{13}} \ar[r]^r & {\bS}\times \R^2 
} $$
 Introduce the following closed subset
$$Z'=\{ (\xi,z,x,y) \ : \ \langle \xi, x-y\rangle \ge 0 \}\subset E\times {\bS}\times \R^2\times \R^2.
$$

We can now rewrite:
$$ {\mathbb F}({ F}) = Rp_{123!} (\Z_{Z'} \otimes p^{-1}_{24}{F}), $$ 
hence
$$ Rq_! {\mathbb F}({F}) = R\tilde p_{13!} Rp_{234!} (\Z_{Z'}\otimes p_{24}^{-1} F) \ =  $$
(projection formula  ~\cite[Prop.2.5.13(ii)]{KS} is used)
$$ = \ R\tilde p_{13!} ( Rp_{234!}\Z_{Z'} \otimes r^{-1} F) $$
We have a  natural isomorphism
 $Rp_{234!}\Z_{{Z'}}\cong \bZ_{{\bS}\times \Delta}[-2]$, where $\Delta\subset \Re^2\times \Re^2$
is the diagonal.  The result now follows. $\Box$

\subsection{Transfer of the conditions ${R}P_{\pm!}F=0$ to ${\mathbb F}F$}






\begin{Claim} \label{clF2} Let  $F\in \DerCat({\bS}\times \R^2)$ satisfy $RP_{\pm!} F=0$. We then have
$R(\id_E \times P_{\pm})_! \fs(F)=0$.
\end{Claim}

\textsc{Proof.} 
Let us pick  a point $(\eta,s_0)\in E\times \R^2$ and  show that,
say, $R(\id_E \times P_+)_!\fs(F)|_{(\eta,s_0)}=0$.
We have:
$$
R(\id_E \times P_+)_!\fs(F)|_{(\eta,s_0)}=R\Gamma_c(E\times {\bS}\times\R^2;
(\id_E \times P_+)^{-1}\bZ_{(\eta,s_0)}\otimes^L \fs(F))
$$
$$
=R\Gamma_c(E\times {\bS}\times\R^2;\bZ_{(\id_E \times P_+)^{-1}(\eta,s_0)}\otimes RA_!(\bZ_Z\boxtimes  F))
$$
\begin{equation}\label{posl}
\stackrel{\text{\cite[Prop.2.5.13(ii)]{KS}}}{=}R\Gamma_c(E\times \R^2\times {\bS}\times \R^2;\bZ_{ A^{-1}P_+^{-1}(\eta,s_0)}\otimes
p_{12}^{-1}\bZ_Z\otimes p_{34}^{-1}F),
\end{equation}
where:
$$
p_{12}:E\times\R^2\times {\bS}\times \R^2\to E\times \R^2
$$
is the projection onto the first two factors;
$$
p_{34}:E\times\R^2\times {\bS}\times \R^2\to {\bS}\times \R^2
$$
is the projection onto the last two factors; and
finally,
$$ A \ : \ E\times \R^2 \times {\bS} \times \R^2 \
 \to \ E\times {\bS} \times R^2 \ :
 \ \ \ (\eta,s_1,z,s_2)\mapsto (\eta,z,s_1+s_2)$$
(as in  \eqref{MapAdef}).

We have:
$$
A^{-1}(\id_E \times P_+)^{-1}(\eta,s_0)=\{(\eta,s_1,z,s_2)|s_1+s_2+z=s_0\}.
$$

Note that 
$$
\bZ_{A^{-1}(\id_E \times P_+)^{-1}(\eta,s_0)}\otimes p_{12}^{-1}\bZ_Z 
=\bZ_{A^{-1}(\id_E \times P_+)^{-1}(\eta,s_0)}\otimes \bZ_{p_{12}^{-1}Z}=
\bZ_{(A^{-1}(\id_E \times P_+)^{-1}(\eta,s_0))\cap p_{12}^{-1}Z} 
$$
and put
$$
T:=(A^{-1}(\id_E \times P_+)^{-1}(\eta,s_0))\cap p_{12}^{-1}Z
=\{(\eta,s_1,z,s_2)|s_1+z+s_2=s_0; \lang \eta,s_1\rang\geq 0\}.
$$

Denote  by $i$ the restriction of $p_{34}$ to $T$:
$$
i \ : \ T\to {\bS}\times \R^2 \ : \ T\ni (\eta,s_1,z,s_2)\mapsto (z,s_2).
$$
We see that $i$ is a closed embedding and that
$$
i(T)=\{(z,s)|\lang \eta,s_0-s-z \rang\geq 0\}=P_+^{-1}K, \ \ K=\{w|\lang\eta,s_0-w\rang\geq 0\}\subset \R^2,
$$
where $P_+:{\bS}\times \R^2\to \R^2$ is as in \eqref{PpmDefined}.

We thus can continue our computation from (\ref{posl})
$$
=R\Gamma_c(E\times \R^2\times {\bS}\times \R^2;\bZ_T\otimes p_{34}^{-1}F)
$$
{(using that  $p_{34}^{-1} F \simeq p_{34}^! F [-4]$ since the fibers of $p_{34}$ are homeomorphic to $\R^4$ and that $Rp_{34!} p_{34}^{!} F \simeq F$)}
$$
=R\Gamma_c({\bS}\times \R^2; (Rp_{34!}\bZ_T)\otimes F{[-4]})=
R\Gamma_c({\bS}\times \R^2;\bZ_{i(T)}\otimes F{[-4]})=
$$
$$
= R\Gamma_c({\bS}\times \R^2;P_+^{-1}\bZ_K\otimes F{[-4]})=
$$
$$
\stackrel{\text{\cite[Prop.2.5.13(ii)]{KS}}}{=}R\Gamma_c(\R^2;\bZ_K\otimes RP_{+!}F{[-4]})=0.
$$
The equality $RP_{-!}{\mathbb F}F=0$ can be proven in the same way.
$\Box$

\subsection{Fourier-Sato decomposition for sheaves satisfying (\ref{lplus}) }

Define: 
\begin{equation} \Pi_+ = \ \{(\xi,\eta)\in E |\xi>0\} \ \subset E. \end{equation}

Suppose (\ref{lplus}) is the case. Then we have
\begin{equation}
\fs(F)\stackrel\sim\rightarrow \fs(\bZ_{l_+}*F)\stackrel\sim\to(\bZ_Z*\bZ_{l_+})*F. \label{eApr20}
\end{equation}

\subsubsection{Computing $\bZ_Z*\bZ_{l_+}$}

Introduce the following subset
$$
Z_+:=Z\cap {(}\Pi_+\times\Re^2{)}\subset \Pi_+\times \Re^2.
$$

\begin{Lemma} \label{Leq2412} We have an isomorphism
\begin{equation} 
\bZ_Z*\bZ_{l_+}=\bZ_{Z_+}. \label{eq2412}
\end{equation} 
\end{Lemma}

\textsc{Proof.} The inclusion $\{0\}\hookrightarrow l_+$ induces a map
\begin{equation}\label{lplusnol}
\bZ_Z*\bZ_{l_+}\to \bZ_Z*\bZ_0=\bZ_Z.
\end{equation}

It suffices to prove the following  two statements:

1) Let $x\in Z_+\subset E\times \Re^2$.  The map 
 \begin{equation}\label{rostok}
(\bZ_Z*\bZ_{l_+})_x\to (\bZ_{Z_+})_x=\bZ,
\end{equation}
induced by  (\ref{lplusnol}), is an isomorphism.

2) Let $x\in (E\times \Re^2)\backslash Z_+$. Then $(\bZ_Z*\bZ_{l_+})_x=0$.

\vskip2.5cm

{In preparation for the proof of 1) and 2),} for a point $x:=(\zeta,v)\in E\times \Re^2$, let us introduce 
a set
$$
K_x=\{(\zeta,u_1,u_2)|(\zeta,u_1)\in Z;u_2\in L_+;u_1+u_2=v\}\subset E\times \Re^2\times \Re^2,
$$
so that  we have
\begin{equation}\label{potoch1}
(\bZ_Z*\bZ_{L_+})_x=R^\bullet\Gamma_c(K_x,\bZ_{K_x}).
\end{equation}

Let
$$
L_x\{(\zeta,u_1,u_2)|(\zeta,u_1)\in Z;u_2=0;u_1+u_2=v\}\subset E\times \Re^2\times \Re^2
$$
so that 
\begin{equation}\label{potoch2}
(\bZ_Z*\bZ_{0})_x=R^\bullet\Gamma_c(L_x,\bZ_{L_x}).
\end{equation}

We have $L_x\subset K_x$ is a closed subset.  Under the identifications (\ref{potoch1}), (\ref{potoch2}),
the map \eqref{rostok} corresponds
to the restriction map
$$
R^\bullet\Gamma_c(K_x,\bZ_{K_x})\to R^\bullet\Gamma_c(L_x,\bZ_{L_x}).
$$

Let $v=(v_1,v_2)${, $\zeta=(\xi,\eta)$}. We then have
$$
K_x=\{((\xi,\eta),(x_1,v_2),(x_2,0)|\xi x_1+\eta y_1\geq  0;x_2\geq 0; x_1+x_2=v_1\}.
$$

The subset $L_x\subset K_x$ consists of all points with $x_2=0$.

The set $K_x$ is identified with the set 
$$
K'_x:=\{(x_1,y_1)\in \Re^2| \xi x_1+\eta y_1\geq  0;x_1\leq v_1\}.
$$
The set $L_x$ gets identified with the subset $L'_x$ of $K'_x$ consisting
of all points with $x_1=v_1$.

Let us check 1). Let $\pi:\Re^2\to \Re$ be the projection onto the second coordinate. 
It suffices to check that the natural map
$$
R\pi_! \bZ_{K'_x}\to R\pi_!\bZ_{L'_x}
$$
(induced by the embedding $L'_x\subset K'_x$) is an isomorphism. We further reduce the statement
so that it reads: the  following induced map on stalks at every point $y\in \Re$ is an isomorphism:
 \begin{equation}\label{promezh}
(R\pi_! \bZ_{K'_x})_y\to (R\pi_!\bZ_{L'_x})_y.
\end{equation}
We have
\begin{equation}\label{kxx}
(R\pi_! \bZ_{K'_x})_y\cong R\Gamma_c(K'_{xy};\bZ_{K'_{xy}});
\end{equation}
$$
(R\pi_! \bZ_{L'_x})_y\cong R\Gamma_c(L'_{xy};\bZ_{L'_{xy}});
$$
where
\begin{equation}\label{luchek}
K'_{xy}=\{(x_1,y)\in \Re^2| \xi x_1+\eta y\geq  0;x_1\leq  v_1\};
\end{equation}
$$
L'_{xy}=\{(x_1,y)\in \Re^2| \xi x_1+\eta y\geq  0;x_1= v_1\}.
$$  
The map (\ref{promezh}) corresponds to the natural map
\begin{equation}\label{promezh1}
R\Gamma_c(K'_{xy};\bZ_{K'_{xy}})\to R\Gamma_c(L'_{xy};\bZ_{L'_{xy}})
\end{equation}
induced by the closed embedding $L'_{xy}\subset K'_{xy}$.

 We have $\xi>0$ (because $x\in \Pi_+\times \Re^2$), in which case either  both $L'_{xy}$ 
and $K'_{xy}$ are empty sets, or $K'_{xy}$ is a closed segment and $L'_{xy}$ is its boundary
point, which implies  that  (\ref{promezh1}) and hence (\ref{promezh}) are isomorphisms.

Let us now check 2).  We have $\xi\leq 0$.
  It suffices to check that $(R\pi_!\bZ_{K_x})_y$=0 for all $y\in \Re$. 
Using (\ref{kxx}), we can equivalently rewrite this condition as follows:
$$
R\Gamma(K'_{xy};\bZ_{K'_{xy}})=0.
$$
 As follows from (\ref{luchek}), the condition $\xi\leq 0$ implies that $K'_{xy}$ is homeomorphic
to a closed ray, which implies the statement.
$\Box$.

Combining \eqref{eApr20} and \eqref{eq2412}, we immediately obtain:

\begin{Cor} \label{se15C7} Suppose $F\in \DerCat({\bS}\times \R^2)$ satisfies \eqref{lplus}. Then 
\begin{equation} \supp \fs(F) \ \subset \Pi_+\times \bS\times \Re^2. 
\label{J31b} 
\end{equation}
\end{Cor}

\vskip2.5pc
Motivated by the corollary \ref{se15C7}, set $$\fS'(F):=\fs(F)|_{\Pi_+\times \bS\times \Re^2}\in \DerCat(\Pi_+\times \bS\times \Re^2),$$
so that
\begin{equation}
\fS'(F)=\bZ_{Z_+}*F. \label{eApr20a}
\end{equation}

Let $\pi_+:\Pi_+\times \bS\times \Re^2\to \bS\times \Re^2$ be the projection.

Lemma \ref{LF1} and \eqref{J31b} imply the following isomorphism:
\begin{equation} \label{eqJan31}
F[-2]\sim R\pi_{+!}\fS'(F)=R\pi_{+!}(\bZ_{Z_+}*F).
\end{equation}

\subsubsection{Further reformulation}
Let us introduce a map  
$$ Q \ : \ \Pi_+\to \R \ ,  \ \ \ Q(\xi,\eta)=\eta/\xi.$$ 
Let also
$$q:\R\times {\bS}\times \R^2\to {\bS}\times \R^2$$
be the projection.
Finally, let us
 set 
$$ W:=\{(a,(x,y))| x+ay\geq 0\} \subset \R\times \R^2. $$
There is a commutative diagram with a Cartesian square:
\begin{equation} \xymatrix{ 
Z_+ \times {\bS}\times \R^2 \ar@{}[r]|-{\subset} & \Pi_+ \times \R^2 \times {\bS}\times \R^2  \ar[rr]^-{Q\times \id_{\R^2\times {\bS}\times \R^2}} \ar@{}[rrd]|{\Box} \ar[d]_{A} &&  \R\times \R^2 \times {\bS}\times \R^2 \ar[d]^{A} & W \times {\bS}\times \R^2 \ar@{}[l]|-{\supset}\\
& \Pi_+ \times{\bS}\times \R^2  \ar[rr]^-{Q\times \id_{{\bS}\times \R^2}} \ar[rd]_{\pi_+} && \R\times {\bS}\times \R^2 \ar[ld]^q,  \\
&& {\bS}\times \R^2 } \label{eqApr19} \end{equation}

\vskip3pc

The map $A$ in this diagram is induced by the addition $\Re^2\times \Re^2\to \Re^2$.


\begin{Lemma} \label{LEF2} 
i)  ``$\bZ_{Z_+}*F$ is constant along fibers of $Q\times \id_{{\bS}\times \R^2}$" in the sense that 
\begin{equation}
\bZ_{Z_+}*F\cong (Q\times \id_{{\bS}\times \R^2})^{-1} (\bZ_W*F); \label{eF2c}
\end{equation}
ii) If $F$ satisfies $(\ref{lplus})$, then there is a quasi-isomorphism
\begin{equation} F\cong  Rq_!(\bZ_W*F)[1]. \label{eF2} \end{equation}
\end{Lemma} 

\textsc{Proof}
From the definition of a constant sheaf as a pull-back of $\Z_{pt}$, we have $(Q\times \id_{\R^2})^{-1} \Z_{W\times{\bS}\times \R^2}=\Z_{Z_+\times {\bS}\times \R^2}$; and then, by the base change ~\cite[(2.5.6)]{KS} in the Cartesian square of \eqref{eqApr19}, we obtain \eqref{eF2c}. 

To prove \eqref{eF2}, write
$$ F \stackrel{\eqref{eqJan31}}{=} R\pi_{+!} (\Z_{Z_+}*F)[2] \stackrel{\eqref{eF2c}}{=} 
R\pi_{+!} (Q\times \id_{{\bS}\times \R^2})^{-1}(\Z_W * F)[2] = $$
$$ = R\pi_{+!} (Q\times \id_{{\bS}\times \R^2})^{-1} RA_! (\Z_W \boxtimes F )[2]
%
%
{=}  Rq_{!} R(Q\times \id_{{\bS}\times \R^2})_! 
(Q\times \id_{{\bS}\times \R^2})^{-1} RA_! (\Z_W \boxtimes F )[2] \ \stackrel{\text{\bf $Q^{-1}=Q^![- 1]$}}{=}  $$
$$ =   Rq_{!} R(Q\times \id_{{\bS}\times \R^2})_!
 (Q\times \id_{{\bS}\times \R^2})^{!}  (\Z_W * F )[1] {=} Rq_! (\Z_W *F )[1].$$
 $\Box$

\vskip3pc

\subsubsection{Rewriting the map (\ref{eF2})}

Define a map
$l:\R\times \R^2\to R$, where $R$ is another copy of $\R$, as follows: $l(a,x,y):=x+ay$.

Let
$$
L:\R\times {\bS}\times  \R^2\to \R\times {\bS}\times R;
$$
be given by $L(a,z,u)=(a,z,l(a,u))$.

Let
$W'\subset\R\times \R^2\times R$ be given by
$$
W'=\{(a,(x_1,y_1),t)|t-x-ay\geq 0\}.
$$

Let $$p_{\bS}: \R\times {\bS}\times \R^2\times R\to \R\times \R^2\times R;$$
$$p_{\R\times R}:\R\times {\bS}\times \R^2\times R\to {\bS} \times\R^2;$$
and $$p_{\R^2}:\R\times {\bS}\times \R^2\times R\to \R\times {\bS}\times R$$ be projections.

We have the following cartesian diagram: 
\begin{equation} \xymatrix{ & (a,u_1, z, u_2) \ar@{|->}[r] \ar@{}[d]|{\text{\rotatebox{-90}{$\in$}}} & (a,z,u_2,\ell(a,u_1+u_2)) \ar@{}[d]|{\text{\rotatebox{-90}{$\in$}}} \\
 (a,u_1, z, u_2)  \ar@{|->}[d] \ar@{}[r]|{\in} & \R \times \R^2 \times {\bS} \times \R^2 \ar[d]_{A} \ar[r]^{\tilde L} \ar@{}[rd]|-{\Box} & \R \times {\bS}\times \R^2 \times R \ar[d]^{p_{\R^2}} \ar@{}[r]|{\ni}& (a,z,u,t) \ar@{|->}[d] \\
(a,z,u_1+u_2) \ar@{}[r]|{\in} & \R \times {\bS} \times \R^2 \ar[r]^L & \R \times {\bS} \times R \ar@{}[r]|{\ni}& (a,z,t) \\
& (a,z,u) \ar@{|->}[r] \ar@{}[u]|{\text{\rotatebox{90}{$\in$}}}& (a,z,\ell(a,u)) \ar@{}[u]|{\text{\rotatebox{90}{$\in$}}} 
} \label{F2diag} \end{equation}
and $W\times \R^2_{u_2}\times {\bS}=\tilde L^{-1}(W'\times {\bS})$.


By the base change ~\cite[(2.5.6)]{KS} applied to the diagram \eqref{F2diag}, we have for all $F$ satisfying $(\ref{lplus})$:
\begin{equation}
\bZ_{W}*F=L^{-1}Rp_{\R^2!}(p_{\R\times R}^{-1}F\otimes p_{\bS}^{-1}\bZ_{W'}). \label{eF2b}
\end{equation}

Denote
$$
\Phi_F:=\bZ_{W}*F:=Rp_{\R^2!}(p_{\R\times R}^{-1}F\otimes p_{\bS}^{-1}\bZ_{W'}) \ \in \ \DerCat(\R\times {\bS}\times R).
$$
\subsubsection{Transferring Claim \ref{clF2} to $\Phi_F$}
Let
$P'_\pm:\R\times {\bS}\times R\to  \R\times R$ be given by
\begin{equation}
P'_\pm(a,(x,y),t)=(a,x+ay\pm t). \label{eqM10c}
\end{equation}

\begin{Lemma} \label{LeqF3} If $F\in \DerCat({\bS}\times \R^2)$ satisfies both \eqref{lplus} and $RP_{+!}F=0$ then 
\begin{equation} RP'_{+!} (\Phi_F)=0. \label{eqF3} \end{equation}
\end{Lemma}

Analogously, if $F$ satisfies both \eqref{lminus} and $RP_{-!}F=0$, then $RP'_{-!}(\Phi_F)=0$.

\textsc{Proof of Lemma \ref{LeqF3}.} Extend the diagram \eqref{F2diag} as follows: 
\begin{equation} \xymatrix{ &&\R\times \R^2 \times {\bS}\times \R^2 \ar[r]^{\tilde L} \ar[d]^-{A} & \R \times {\bS}\times \R^2 \times \R \ar[d]^-{p_{\R^2}} \\
E\times {\bS} \times \R^2 \ar[d]^{\id_E\times P_+} &
\Pi_+\times {\bS}\times \R^2 \ar[r]^-{Q\times \id_{{\bS}\times \R^2}} \ar[d]^{\id_{\Pi_+}\times P_+} \ar@{_{(}->}[l]_-{\iota\times \id}& \R\times {\bS}\times \R^2  \ar[r]^{L} \ar[d]^-{\id_{\R}\times P_+} & \R\times {\bS}\times R \ar[d]^{P'_+} 
\\
E\times  \R^2 &
\Pi_+\times \R^2 \ar[r]^-{Q\times \id_{ \R^2}} \ar@{_{(}->}[l]_-{\iota\times \id} & \R\times \R^2 \ar[r]^{L'} & \R\times R  \\
&& (a,w) \ar@{|->}[r] \ar@{}[u]|{\text{\rotatebox{90}{$\in$}}}& (a,\ell(a,w)) \ar@{}[u]|{\text{\rotatebox{90}{$\in$}}}
} \label{eqA20z} \end{equation}
where $\iota: \Pi_+\hookrightarrow E$ is the open inclusion.


We have $Z_+=Z\cap (\iota\times \id_{\R^2})\Pi_+$ and $\Z_{Z_+}=(i\times \id_{\R^2})^{-1}\Z_Z$. Thus by the base change ~\cite[(2.5.6)]{KS}, $\Z_{Z_+}*F\in \DerCat(\Pi_+\times {\bS}\times\R^2)$ is quasi-isomorphic to $(\iota\times \id_{{\bS}\times \R^2} )^{-1} (\Z_Z * F)$. 
Thus, 
$$ R(\id_{\Pi_+}\times P_+)_! (\Z_{Z_+} * F ) \ \stackrel{\text{\cite[(2.5.6)]{KS}}}{=} \ (\iota\times \id_{\R^2})^{-1} R(\id_E\times P_+)_! (\Z_{Z} * F ) \stackrel{\text{Claim \ref{clF2}}}{=} 0 .$$

But on the other hand,  
$$ {\mathbb F}(F) \ \stackrel{\eqref{eApr20a}}{=} \ \Z_{Z_+} * F \ \stackrel{\eqref{eF2c}}{=} \ (Q\times \id_{{\bS}\times \R^2})^{-1}(\Z_W * F ) \ \stackrel{\eqref{eF2b}}{=} \ (Q\times \id_{{\bS}\times \R^2})^{-1} L^{-1}\Phi_F   $$
hence 
$$ R(\id_{\Pi_+}\times P_+)_! (Q\times \id_{{\bS}\times \R^2})^{-1} L^{-1}\Phi_F = 0, $$
or applying the base change ~\cite[(2.5.6)]{KS} to the middle and right bottom squares of \eqref{eqA20z}, we have
$$(Q\times \id_{\R^2})^{-1} (L')^{-1} RP'_{+!} (\Phi_F)=0.$$
Since both maps $(Q\times \id_{\R^2})$ and $L'$ are locally trivial fibrations with a vector space as a fiber, we conclude that $RP'_{+!}\Phi_F=0$.  $\Box$
\subsection{ Rewriting the condition of orthogonality to $\cC$ }

Let $F$ satisfy the conditions of Proposition \ref{main} (assuming (\ref{lplus}).
Let  $H\in \cC_\bS$, where $\cC_\bS$ is  defined in section \ref{CatCSdefd}. 
 Proposition \ref{main} now reduces to  proving that $R\hom(F,H)=0$.

Let us investigate $R\hom(F,H)$ using the representation \eqref{eF2} of $F$. We have:
$$
R\Hom(F,H) \stackrel{\eqref{eF2}}{=} R\Hom(Rq_!(\Z_W * F), H)[-1] \stackrel{\eqref{eF2b}}{=}
R\Hom(Rq_!L^{-1}(\Phi_F);H) [-1]
$$
\begin{equation}\label{predvar}
=R\Hom_{\R\times {\bS}\times R}(\Phi_F;RL_*q^!H)[-1].
\end{equation}
Singular support estimate shows
that  
\begin{Proposition}
We have:
$$S.S. RL_*q^!H \ \subset \Omega_\cH, 
$$
where 
\begin{equation}
\Omega_\cH:=\bigcup_{\text{``+" and ``-"}}\{ (a,x_1,y_1,t, \R.(d(x_1+ay_1)\pm dt)+\R. da) \}  \label{eF8}
\end{equation} 
and where $a\in \R$, $(x_1,y_1)\in {\bS}$, $t\in R$.
\end{Proposition}
\textsc{Proof} 
Because $q$ is a projection on a direct factor, by ~\cite[Prop.3.3.2(ii)]{KS} we have $S.S. q^!H=S.S. q^{-1}H$ which in turn can be,  using ~\cite[Prop.5.4.13]{KS}, estimated by  (in the notation of that proposition) $^{t}q'(q_\pi^{-1}(S.S.(H)))$; thus 
$$ S.S. q^! H \ \subset \  \{ a,z,u, \alpha da + \upsilon du \ : \ \zeta = \pm \upsilon \}. $$
By ~\cite[Prop.5.4.4]{KS}, 
$$S.S. RL_*q^!H  \subset L_\pi({}^t {L'}^{-1} \{ a,z,u, \alpha da + \zeta dz + \upsilon du \ : \ \zeta= \pm \upsilon \} ).  $$
We have 
$$ \begin{array}{rcl}
T^*(\R_a\times {\bS}_z \times \R^2_{u=(x,y)} ) & \stackrel{{}^t L'}{\longleftarrow} & 
\R_a \times {\bS}_z \times \R^2_{u=(x,y)} \times_{(\R_a\times {\bS}_z \times \R_t) } T^*(\R_a \times {\bS}_z\times R_t ) 
\\ (a,z,u, \alpha da + \zeta dz + \xi dx + \eta dy) && (a,z,u,\alpha da + \zeta dz + \tau dt) \\
\upsilon=(\xi,\eta) && t=\ell(a,u) \\
dx+ady + yda & \leftrightarrow & dt .
\end{array} $$
Thus
$$S.S. RL_*q^!H  \ \subset \ L_\pi( \{ a,z,u, \alpha da + \zeta dz + \tau dt \ : \ \zeta= \pm \tau(1,a) \} ) \ = \   $$
$$ \ \ \ \ \ \ = \ \{ a,z,t, \alpha da + \zeta dz + \tau dt \ : \ \zeta= \pm \tau(1,a) \}  $$
which is equivalent to \eqref{eF8}.  $\Box$

Thus, Proposition \ref{main} follows from the following one:
\begin{Claim}\label{promtreb} Let $\Phi_F,\cH\in \DerCat(\Re\times \bS\times R)$ satisfy: $RP'_{\pm!}\Phi_F=0$ (where $P'_\pm$
are as in \eqref{eqM10c}); $S.S.\cH\subset \Omega_\cH$, where  $\Omega_\cH$ is as in \eqref{eF8}.
Then we have:
$$
R\hom(\Phi_F;\cH)=0.
$$
\end{Claim}
 
\subsection{Subdivision into 3 cases} We are going to subdivide the space $\Re\times \bS\times R$ with coordinates $(a,z,u)$
into 3 parts according to the sign of $a$. 

\subsubsection{Subdivision of $\Re\times \bS\times R$}
$$
U_+:=(0,\infty)\times\bS \times R\subset \Re\times \bS\times R
$$
$$
U_-:=(-\infty,0)\times\bS \times R\subset \Re\times \bS\times R;
$$
$$
U_0:=0\times\bS \times R\subset \Re\times \bS\times R.
$$
Denote 
$$
j_\pm:U_\pm\to \Re\times \bS\times R
$$
the corresponding open embeddings
and
by
$$
i_0:U_0\to \Re\times \bS\times R
$$
the corresponding closed embedding.

\subsubsection{Subdivision of $\Phi_F$}

Set 
$$
\Phi_\pm:=j_{\pm}^{-1}\Phi_F \ \in \ \DerCat(U_\pm) ;
$$
$$
\Phi_0:=i_0^{-1}\Phi_F  \ \in \ \DerCat(U_0).
$$
We have a distinguished triangle
\begin{equation}\label{promtrebtreugolnik}
\to j_{+!}\Phi_{+}\oplus j_{-!}\Phi_{-}\to \Phi_F\to i_{0!}\Phi_0\stackrel{+1}\to .
\end{equation}

Let 
$$P^{U_+}_\pm:=P'_{\pm}j_+ ; \ \ \ P^{U_-}_\pm:=P'_{\pm}j_- ; \ \ \  P^{U_0}=P'_{\pm}i_0 $$
 be the restrictions of $P'_\pm$ from \eqref{eqM10c} onto $U_+,U_-$, and $U_0$.
Base change theorem implies that
$$
P^{U_+}_{\pm!}\Phi_{+}=0;
$$
$$
P^{U_-}_{\pm!}\Phi_{-}=0;
$$
$$
P^{U_0}_{\pm!}\Phi_{0}=0.
$$

\subsubsection{Subdivision of $\cH$}
Let $\cH_\pm\in \DerCat(U_\pm)$; 
$$\cH_\pm:=j_{\pm}^{-1}\cH.$$
 Let $\cH_0\in \DerCat(U_0)$; 
$$\cH_0:=i_0^!\cH.$$
Let us estimate the microsupports of these objects.
Let 
$$
\Omega_{U_\pm}:=\Omega_\cH\cap T^*U_\pm\subset T^*U_\pm,
$$
where we assume the  embeddings $T^*U_\pm\subset T^*(\Re\times \bS\times R)$
induced  by $j_\pm$.

It is immediate that $\SS(\cH_\pm)\subset \Omega_{U_\pm}$.

Let 
$$
\Omega_0:=\bigcup_{\text{``+" and ``-"}}\{ (x_1,y_1,t, \R.(dx_1\pm dt) \} \subset T^*(\bS\times R),
$$
where, same as in \eqref{eF8},  $(x_1,y_1)$ are coordinates on $\bS$, and $t$ on $R$.

Corollary [KS] 6.4.4(ii) implies that
$$
\SS(\cH_0)\subset \Omega_0.
$$

\subsubsection{Subdivision of Claim (\ref{promtreb})}

By virtue of the distinguished triangle in (\ref{promtrebtreugolnik}),
Claim (\ref{promtreb})  gets split into showing the following vanishings:

$$
R\hom_{\Re\times \bS\times R}(j_{+!}\Phi_+;\cH)=R\Hom_{U_+}(\Phi_+;\cH_+)=0;
$$
$$
R\hom_{\Re\times \bS\times R}(j_{-!}\Phi_-;\cH)=R\Hom_{U_-}(\Phi_-;\cH_-)=0;
$$
$$
R\hom_{\Re\times \bS\times R}(i_{0}\Phi_+;\cH)=R\Hom_{U_0}(\Phi_0;\cH_0)=0.
$$

Our task now reduces to showing the following 3 statements:

\begin{Claim}\label{promtrebplus} Let $\Phi_+,\cH_+\in \DerCat(U_+)$.   Suppose $RP^{U_+}_{\pm!}\Phi_+=0$
and $\SS(\cH_+)\subset \Omega_{U_+}$. Then 
$$R\hom(\Phi_+,\cH_+)=0.
$$
\end{Claim}
\begin{Claim}\label{promtrebminus} Let $\Phi_-,\cH_-\in \DerCat(U_-)$.   Suppose $RP^{U_-}_{\pm!}\Phi_+=0$
and $\SS(\cH_-)\subset \Omega_{U_-}$. Then 
$$R\hom(\Phi_-,\cH_-)=0.
$$
\end{Claim}

\begin{Claim}\label{promtrebnol} Let $\Phi_0,\cH_0\in \DerCat(U_0)$.   Suppose $RP^{U_0}_{\pm!}\Phi_0=0$
and $\SS(\cH_0)\subset \Omega_{U_0}$. Then 
$$R\hom(\Phi_0,\cH_0)=0.
$$
\end{Claim}

\subsubsection{Furhter reduction} 

Let $\diamondsuit$ be one of the symbols: $+,-$, or $0$.
Let $I_+:=(0,\infty)$; $I_-:=(-\infty,0)$; $I_0:=\{0\}$.
Let
$$
Q'_\diamondsuit:U_\diamondsuit\times \bS\times R\to I_\diamondsuit \times \Re\times R
$$
be given by
$$
Q'_\diamondsuit(a,(x,y),t):=(a,x+ay,t)
$$
(in the case $\diamondsuit=0$ we assume $a=0$).
Denote by $\sV_\diamondsuit\subset  \Re\times\Re\times R$  the image of $Q'_\diamondsuit$.
Depending on $\bS$,  $\sV_\diamondsuit$  can be of one of the following types:

1)  For some linear function $f_\diamondsuit:I_\diamondsuit\to\R$,   
$$
\sV_\diamondsuit=\{(a,v,t)| a\in I_\diamondsuit; v>f(a);\}.
$$
In this case, set  $\sU_\diamondsuit:=I_\diamondsuit\times (0,\infty)\times R$; set
$$
Q_1:U_\diamondsuit\to \sU_\diamondsuit,
$$
$$
Q_1(a,(x,y),t):=(a,x+ay-f(a),t).
$$

2) For some linear function $f_\diamondsuit:I_\diamondsuit\to \Re$,
$$
\sV_\diamondsuit=\{(a,v,t)| a\in I_\diamondsuit; v<f(a)\}.
$$
In this case, set  $\sU_\diamondsuit:=I_\diamondsuit\times (-\infty,0)\times R$; set
$$
Q_1:U_\diamondsuit\to \sU_\diamondsuit;
$$
$$
Q_1(a,(x,y),t):=(a,x+ay-f(a),t).
$$

3) 
$$
\sV_\diamondsuit=I_\diamondsuit\times \Re\times R.
$$
In this case, set  $\sU_\diamondsuit:=I_\diamondsuit\times (-\infty,\infty)\times \Re$; set
$Q_1:U_\diamondsuit\to \sU_\diamondsuit$,
$$
Q_1(a,(x,y),t):=(a,x+ay,t).
$$

It is easy to see that  in each of the cases the map $Q_1$ is surjective; furthermore it is
a smooth fibration with its typical fiber
diffeomorphic to $\Re$.  We also see that the 1-forms from $\Omega_{U_\diamondsuit}$ vanish on
fibers of $Q_1$, which implies that the natural
map
$$
\cH_\diamondsuit\to Q_1^!RQ_{1!}\cH_\diamondsuit
$$
is an isomorphism.

Set 
$$
\cL_\diamondsuit:=RQ_{1!}\cH_\diamondsuit\in \DerCat(\sU_\diamondsuit).
$$

Define conic closed subsets  $\Omega_{\sU_\pm}\subset T^*\sU_\pm$
as follows:
$$
\Omega_{\sU_\pm}:=\bigcup_{\text{``+" and ``-"}}\{ (a,v,t, \R.(dv\pm dt)+\R.da \},
$$
where $(a,v,t)\in \sU_\pm\subset I_\pm\times\Re\times R$.
Define a conic closed subset $\Omega_{\sU_0}\subset  T^*\sU_0$:
$$
\Omega_{\sU_\pm}:=\bigcup_{\text{``+" and ``-"}}\{ (0,v,t, \R.(dv\pm dt) \}.
$$

It is easy to see that
$$
\SS(\cL_\diamondsuit)\subset \Omega_{\sU_\diamondsuit}.
$$

\subsubsection{} We have
$$
R\hom(\Phi_\diamondsuit;\cH_\diamondsuit)=R\hom(\Phi_\diamondsuit;Q_1^!\cL_\diamondsuit)=R\hom_{\sU_\diamondsuit}(RQ_{1!}\Phi_\diamondsuit;\cL_\diamondsuit).
$$

Set $G_\diamondsuit:=RQ_{1!}\Phi_\diamondsuit$.
Let $P^{\sU_\diamondsuit}_\pm:\sU_\diamondsuit\to \Re\times \Re$
be the restrictions of the following maps $\Re\times \Re\times R\to \Re\times \Re$:
\begin{equation}\label{otob}
(a,v,t)\mapsto (a,v\pm t).
\end{equation}
It now follows that
$$
RP^{\sU_\diamondsuit}_{\pm!}G_\diamondsuit=0.
$$

So, we can rewrite Claims \ref{promtrebplus}---\ref{promtrebnol}
as follows. 

\begin{Claim}\label{promtrebobyed}
Let $G_\diamondsuit,\cL_\diamondsuit\in \DerCat(\sU_\diamondsuit)$ satisfy:  
\begin{equation}\label{usl1}
RP^{\sU_\diamondsuit}_{\pm!}G_\diamondsuit=0;
\end{equation}
 $\SS(\cL_\diamondsuit)\in \Omega_{\sU_\diamondsuit}$.
Then $R\hom(G_\diamondsuit;\cL_\diamondsuit)=0$.
\end{Claim}
\subsection{The case $\sU_\diamondsuit=I_\diamondsuit\times (-\infty,\infty)\times R$}  
This case follows from Theorem \ref{plosk} below.
Below, we are going to consider the case $\sU_\diamondsuit=I_\diamondsuit\times (0,\infty)\times\Re$. The case
$\sU_\diamondsuit= I_\diamondsuit\times (-\infty,0)\times \Re $ is fairly similar.
\subsection{Proof of Claim \ref{promtrebobyed} for $\sU_\diamondsuit=I_\diamondsuit\times (0,\infty)\times \Re$}
As above, our major tool is development of a certain representation of $G$.
\subsubsection{Representation of $G$} 

Let $V_1\subset I_\diamondsuit\times\R \times  (0,\infty)\times \R$
be given by
\begin{equation} 
V_1=\{(a,u,v,t)|\ |t|<v\}. \label{eqMay6b}
\end{equation}

Let $V:=I_\diamondsuit\times \R\times (0,\infty)\times (0,\infty)$.
We have an identification 
$J:V\to V_1$, 
\begin{equation}\label{ji}
J(a,u,\xi_1,\xi_2)=(a,u,\frac{\xi_1+\xi_2}{2},\frac{\xi_1-\xi_2}{2}).
\end{equation}

Let $\cI_1:V_1\to I_\diamondsuit\times (0,\infty)\times\R$  be given by
\begin{equation}
\cI_1(a,u,v,t)=(a,v,u+t). \label{oct3e1118}
\end{equation}
Let $\cI=\cI_1J$:
$$
\cI(a,u,\xi_1,\xi_2)=(a,\frac{\xi_1+\xi_2}{2},u+\frac{\xi_1-\xi_2}{2}),
$$
so that 
$\xi_1=v+t$; $\xi_2=v-t$.

Let
$q_1,q_2:V\to I_\diamondsuit\times \R_{>0}\times\R_{>0}$,
\begin{equation}
q_i (a,u,\xi_1,\xi_2)=(a,u,\xi_i),\quad i=1,2. \label{oct3e1026}
\end{equation}

Let us summarize our notation in the following diagram (a wavy line indicates that a sheaf is defined over the given space):
$$\xymatrix{ 
&& (a,u,v,t) \ar@{|->}[rr] \ar@{}[d]|{\inthesetbelow} && (a,v,u+t) \ar@{}[d]|{\inthesetbelow} \\
X\times \R \times (\R_{>0}\times \R) \ar@{}[rr]|-{\supset} && V_1=\{(a,u,v,t)\, : \, |t|<v\} \ar[rr]^-{\cI_1} && I_\diamondsuit \times \R_{>0} \times \R \ar@{~}[r] & G \\
& H \ar@{~}[r] &  V= I_\diamondsuit \times \R \times \R_{>0}\times \R_{>0} \ar[rr]_-{q_i} \ar[rru]_{\cI} \ar[u]_{J} && I_\diamondsuit \times \R\times \R_{>0}.  \\
&& (a,u,\xi_1,\xi_2) \ar@{}[u]|{\inthesetabove} \ar@{|->}[rr] && (a,u,\xi_i) \ar@{}[u]|{\inthesetabove}
}$$

\begin{Claim}\label{pred}  Suppose that an object $G\in \DerCat(I_\diamondsuit\times (0,\infty)\times \R)$ satisfies
(\ref{usl1}) both with the sign ``+'' and with the sign ``-".  There exists  an object $H\in \DerCat(V)$ such that \\
1) both $Rq_{1!}H\sim 0$ and $Rq_{2!}H\sim 0$; \\
(2)$R\cI_!H\sim G.$
\end{Claim}

{\bf Remark.} Observe that (\ref{usl1}) reads as follows: $RP^1_{\pm!}G=0$,
where
\begin{equation}
P^1_\pm:I_\diamondsuit\times (0,\infty)\times \R\to \Re\times \Re\ \ : \ \ P^1_{\pm}(a,v,t)=(a,v\pm t), 
\label{oct3e1251}
\end{equation}
same as in (\ref{otob}).

Proof of this Claim will occupy the next subsection
\subsection{Proof of Claim \ref{pred}}
\subsubsection{Functors $r_1$ and $r_2$ and their properties}
For $F\in \DerCat(I_\diamondsuit\times  \R\times (0,\infty)\times (0,\infty))$ we have
natural maps (coming from the adjunction)
\begin{equation} F\to q_{1}^!Rq_{1!}F; \ \  F\to q_{2}^!Rq_{2!}F. \label{eqMay2a} \end{equation}
Let 
$r_1(F), r_2(F)$ be the cones of these maps so that we have
natural maps (in the conventions of ~\cite[Ch.1.4]{KS})
\begin{equation}\label{er1}
r_1(F)\to F[1]
\end{equation}
\begin{equation}\label{er2}
 r_2(F)\to F[1]. 
\end{equation}
We therefore have a composition map 
\begin{equation}\label{er1er2}
r_1r_2F\to F[2].
\end{equation}
\begin{Lemma}\label{ischez}
We have $Rq_{1!}r_1r_2=Rq_{2!}r_1r_2=0.$
\end{Lemma}
\textsc{Proof}
 First of all we observe that
\begin{equation} Rq_{1!}r_1\sim 0 , \ \ \  Rq_{2!}r_2\sim 0. \label{e9Fa} \end{equation}  
Indeed, the question boils down to 
showing that $Rq_{1!}$ applied to \eqref{eqMay2a} yields a
 quasi-isomorphism $Rq_{1!} F \stackrel{\sim}{\to} Rq_{1!} q^!_1 Rq_{1!} F$.

There is a natural transformation of endofunctors  on $\DerCat(I_\diamondsuit\times \Re\times (0,\infty))$:
$\ve:Rq_{1!} q_1^! \to \Id$  (since $Rq_{1!}$ is left
adjoint to $q_1^!$). Since $q_1$ is a projection along $(0,\infty)$, it is well known
that $\ve$ is an isomorphism of functors.
 By ~\cite[Ch.IV.1, Th.1(ii)]{MacLane}, there is a diagram 
$$ \xymatrix{Rq_{1!}F \ar[rd]_{\id} \ar[r] & Rq_{1!}q_1^! Rq_{1!} F \ar[d] \\ & Rq_{1!} F} $$
in which the vertical arrow is induced by $\ve$, which implies that the vertical arrow is an isomorphism,
hence, so is the horizontal arrow.  This finishes proof of (\ref{e9Fa}).

Secondly,  we have a natural quasi-isomorphism
\begin{equation} r_1r_2\sim r_2r_1. \label{e9Fb} \end{equation} 
Indeed, 
let us represent $q_1,q_2$ as convolution with kernels. Let  $A,B,C$ be smooth manifolds. 
We have the convolution bifunctor
 $\circ: \DerCat(A\times B)\times \DerCat(B\times C)\to \DerCat(A\times C)$ defined by
\begin{equation} F\circ G = R\pi_{AC!}(\pi_{AB}^{!}F\otimes \pi_{BC}^!G). \label{Fog} \end{equation}


Let $A=\R$, $B_1=B_2= (0,\infty)$,  $C=\pt$ so that $F$ is a sheaf on $A\times B_1\times B_2$, 
$q_1:A_1\times B_1\times B_2 \to A\times B_1 \times C$ is the projection
along $B_2$. 

We have   $Rq_{1!}F=F\circ \bZ_{B_2 \times C}$. 

Set  $q_1^{\diamondsuit}G\cong q_1^{-1}G[1]=G\circ \bZ_{C\times B_2}[1]$.

Let us construct an isomorphism (natural in $F$ and $G$)
$$
R\hom(Rq_{1!}F;G)\stackrel \sim\to R\hom(F;q_1^\diamondsuit G).
$$

Fix one of the two  maps  
$I:\Delta_!\bZ_{B_2}\to \bZ_{B_2\times B_2}[1]$  such that the induced
map $RP_!\Delta_!\bZ_{B_2}\to RP_!\bZ_{B_2\times B_2}[1]$ is an isomorphism,
 where $P:B_2\times B_2\to B_2$
is the projection along the second factor. We have an induced   map
$$
\alpha:F\stackrel{\cong}{\to} F\circ \Delta_!\bZ_{B_2}\stackrel I\to
 F\circ \bZ_{B_2\times B_2}[1] \stackrel{\cong}{\to} q_1^\diamondsuit Rq_{1!}F
$$
It follows that this map induces an isomorphism
\begin{equation}\label{odin}
Rq_{1!}F\to Rq_{1!}q_1^\diamondsuit Rq_{1!}F. 
\end{equation}

The induced map
\begin{equation}
R\hom(Rq_{1!}F;G)\to R\hom(q_1^\diamondsuit Rq_{1!}F;q_1^\diamondsuit G)
\stackrel{-\circ\alpha}\longrightarrow R\hom(F;q_1^\diamondsuit G) \label{eqMay6a}
\end{equation} is an isomorphism
for all $F,G$.   Indeed,
  the  right arrow is an isomorphism because of (\ref{odin}).
The left arrow is an isomorphism because we have an isomorphism  of functors $q_1^\diamondsuit G=G\boxtimes \bZ[1]$ and
the statement now follows from the Kuenneth formula.


Thus we have constructed an adjunction between the functors $q_1^\diamondsuit$ and $Rq_{1!}$ in the sense of ~\cite[Ch.IV.1]{MacLane}. In case $G=Rq_{1!}F$, the map \eqref{eqMay6a} sends $\id_{Rq_{1!}F}$ to $q_1^!(\id_{Rq_{1!}F})\circ \alpha =\alpha$, therefore $\alpha$ is the universal arrow associated to the adjunction ~\eqref{eqMay6a} in the sense of ~\cite[Ch.IV.1, p.81]{MacLane}; by the uniqueness of an adjoint functor, see ~\cite[Cor.1, Ch.IV.1, p.85]{MacLane} and its proof, this means that $\alpha$ coincides with the ``standard" adjunction map (coming from ~\cite[Ch.3.1]{KS}) up to some natural autoequivalence of the functor $q_1^! Rq_{1!}$. 
This means that we have a canonical isomorphism of functors $q_1^\diamondsuit\cong q_1^!$ so that
we won't make difference between $q_1^\diamondsuit$ and $q_1^!$
We have
 \begin{equation}\label{simil}
q_1^!Rq_{1!}F=F\circ (\bZ_{B_2\times C}\circ \bZ_{C\times B_2})[1]=F\circ \bZ_{B_2\times B_2}[1].
\end{equation}
The above consideration shows that 
 $r_1F=\Cone \alpha\simeq F\circ \cL_1$, where $\cL_1:=\Cone (I:\Delta_! \Z_{B_2} \to \Z_{B_2\times B_2}[1])$.

Analogously, $r_2 F \simeq F\circ \cL_2$, where $\cL_2:=\Cone (I:\Delta_! \Z_{B_1} \to \Z_{B_1\times B_1}[1])$.

Therefore,
$$ r_1 r_2 F \simeq F\circ [\cL_1\boxtimes \cL_2] \simeq r_2 r_1 F, $$ as we wanted.


We now have: $Rq_{1!}r_1r_2=0$ because of (\ref{e9Fa}) and

\begin{equation} Rq_{2!}r_1r_2 \ \stackrel{\eqref{e9Fb}}{=} 
\ Rq_{2!}r_2r_1 \ \stackrel{\eqref{e9Fa}}{=} \ 0. \label{eF10} \end{equation}
This accomplishes proof of Lemma. $\Box$

\subsubsection{Construction of  the object  $H$ and proof of the  Claim \ref{pred} 1)}

We set $\Phi=\cI^!G$ and $H:=r_1r_2(\Phi)$.
Lemma \ref{ischez} says that $Rq_{1!}H\sim 0$ and $Rq_{2!}H\sim 0$, which proves
 part 1) of the Claim \ref{pred}.
\subsubsection{Reduction of part 2) of the Claim \ref{pred}} 
Let us deduce part 2) of the Claim \ref{pred} from the following statement.

We have a map
$$
\iota_H:H=r_1r_2\Phi\to \Phi[2],
$$
where the right arrow is defined in (\ref{er1er2}).  
Let us apply the functor $R\cI_!$ to $\iota_H$ so as to get a map
\begin{equation}\label{sp}
R\cI_!H\to R\cI_!\Phi[2]
\end{equation}
\begin{Claim}\label{isopred}
The map (\ref{sp})
is an isomorhpism. 
\end{Claim}
This Claim  implies part 2) of the  Claim \ref{pred}. Indeed,
 we can rewrite (\ref{sp}) as follows.
$$
R\cI_!H\to R \cI_!\Phi[2]=R\cI_!\cI^!G[2]\stackrel\sim\to  G[2],
$$
 where the rightmost arrow is an isomorphism because $\cI$ is a smooth fibration 
with  fibers diffeomorphic to $\R^1$. 

We now pass to proving Claim \ref{isopred}.
\subsubsection{Subdivision into 3 cases}

The map (\ref{sp}) factors as
  $$R\cI_!r_1r_2(\Phi)\stackrel{\eqref{er1}}\to R\cI_!r_2(\Phi)[1]\stackrel{\eqref{er2}}\to R\cI_{I!}\Phi[2].$$

As $\cI^! G=\Phi$ and by ~\cite[Prop.1.4.4.(TR3)]{KS}, the cone 
of the right arrow is isomorphic to $R\cI_!q_2^!Rq_{2!}\cI^!G[2]$. Analogously, the cone 
of the left arrow
is
$ R\cI_!q_{1}^!Rq_{1!}r_2\Phi[1] $
which, by definition of $r_2$, is the cone of the  natural arrow
$$
R\cI_!q_1^!Rq_{1!}\cI^!G\to R \cI_!q_1^!Rq_{1!}Rq_{2}^!Rq_{2!}\cI^!G.
$$
Thus,  isomorphicity of \eqref{sp}  is implied by  the  following three vanishing statements:

1) $R\cI_!q_2^!Rq_{2!}\cI^!G\sim 0$

2)$R\cI_!q_{1}^!Rq_{1!}\cI^!G\sim 0;$

3)$R\cI_!q_1^!Rq_{1!}q_{2}^!Rq_{2!}\cI^!G\sim 0$.

\subsubsection{Proof of the 1-st and the 2-nd vanishing}

Let $V_2:=I_\diamondsuit\times \Re\times (0,\infty)^4$. Let $\pi_1,\pi_2:V_2$
be given by
$$
\pi_1(a,v,\xi_1,\xi_2,\xi_1',\xi_2')=(a,v,\xi_1,\xi_2)
$$
and 
$$
\pi_2(a,v,\xi_1,\xi_2,\xi_1',\xi_2')=(a,v,\xi_1',\xi_2')
$$

Let $L_2,\subset V_2$ be  a closed subset of the form:

$$
L_2:=\{(a,v,\xi_1,\xi_2,\xi_1',\xi_2')|\xi_2=\xi_2'\};
$$

\begin{Lemma} \label{LM19} For any $F\in \DerCat(V)$  we have
$$ q_2^! Rq_{2!} F = R\pi_{2!}(\Z_{L_i} \otimes {\pi_2}^{-1} F) .$$
\end{Lemma}
\textsc{Proof} Similar to proof of (\ref{simil}).
$\Box$

\vskip2pc

Let $X_2:=I_\diamondsuit \times ((0,\infty)\times \R) \times ((0,\infty)\times \R)$.
Let 
$\pi^X_1,\pi^X_2:X_2\to I_\diamondsuit\times (0,\infty)\times \R$ be the projections
along the 3rd and the 2nd factors respectively. Define closed subsets $L_\pm\subset X_2$:
$$ L_{\pm} = \{ (a,(s_1,t_1),(s_2,t_2))\in I_\diamondsuit \times ((0,\infty)\times \R) \times ((0,\infty)\times \R) \ : \ 
s_1 \pm t_1=s_2 \pm t_2 \} $$

\begin{Lemma} \label{LM20} For any $F\in \DerCat(I_\pm \times (0,\infty) \times \R)$, 
$$ (P^1_-)^{-1} R P^1_{-!} F = R\pi^X_{1!}(\Z_{L_-} \otimes {\pi^X_2}^{-1} F) ,$$
where the map $P^1_-$ was defined in \eqref{oct3e1251}.
\end{Lemma}

\textsc{Proof.} The proof is analogous to the proof of lemma \ref{LM19}. $\Box$

\vskip2.5pc

We  now have
$$
R\cI_!q_2^!Rq_{2!}\cI^!G[-2]\sim R\cI_!q^{-1}_{2} Rq_{2!}\cI^{-1}G
$$
\begin{equation}
\sim R\pi'_{1!}(\bZ_{L_2}\otimes (\pi'_2)^{-1}G), \label{eqM10a}
\end{equation}
where $\pi'_i=\cI\pi_i:V_2\to I_\diamondsuit\times (0,\infty)\times \Re$,
as easily follows from Lemma \ref{LM19}.

Let us define the following map
$$
J_2: I_\diamondsuit\times \Re\times ((0,\infty)\times \Re)\times ((0,\infty)\times \Re)
\to I_\diamondsuit\times ((0,\infty)\times \Re)\times ((0,\infty)\times \Re)=X_2
$$
as follows:
$$
J_2(a,v,(s_1,t_1),(s_2,t_2))=(a,s_1,v+t_1,s_2,v+t_2).
$$
Let us also define a map (which is a closed embedding)
$$
K_2:V_2\to I_\diamondsuit\times \Re\times ((0,\infty)\times \Re)\times ((0,\infty)\times \Re)
$$
as follows:
$$
K_2(a,v,\xi_1,\xi_2,\xi_1',\xi_2'):=
(a,v,\frac{\xi_1+\xi_2}{2};\frac{\xi_1-\xi_2}2,\frac{\xi'_1+\xi'_2}{2};\frac{\xi'_1-\xi'_2}2).
$$

It follows that $\pi'_1=\pi^X_1J_2K_2$; $\pi'_2=\pi^X_2J_2K_2$

We can now rewrite \eqref{eqM10a} as follows:

$$
R\cI_!q_2^!Rq_{2!}\cI^!G[-2]\sim R\cI_!q^{-1}_{2} Rq_{2!}\cI^{-1}G
$$
\begin{equation}\label{LM21}
\sim R\pi^X_{1!}((RJ_{2!}RK_{2!}\bZ_{L_2})\otimes (\pi^X_2)^{-1}G), 
\end{equation}

Let
$$
L'_2\subset I_\diamondsuit\times \Re\times ((0,\infty)\times \Re)\times ((0,\infty)\times \Re)
$$
be a closed subset consisting of all points $(a,v,s_1,t_1,s_2,t_2)$ with $s_1-t_1=s_2-t_2$.

It is easy to see that $K_2(L_2)\subset L'_2$ is an open embedding. Indeed, $K_2(L_2)$ consists of all points
$(a,v,s_1,t_1,s_2,t_2)$ with $s_1-t_1=s_2-t_2$ , $s_1>|t_1|$, $s_2>|t_2|$.

 Therefore, we have
a map $RK_{2!}\bZ_{L_2}\to \bZ_{L_2'}$ which 
induces a map
\begin{equation}\label{Meq20b}
R\pi^X_{1!}((RJ_{2!}RK_{2!}\bZ_{L_2})\otimes (\pi^X_2)^{-1}G)\to
 R\pi^X_{1!}((RJ_{2!}\bZ_{L_2'})\otimes (\pi^X_2)^{-1}G).
\end{equation}
The cone of this arrow equals
$$
 R\pi^X_{1!}(M\Lotimes (\pi^X_2)^{-1}G),
$$
where
$$
M\sim RJ_{2!}\bZ_{N},
$$
and $N=L'_2\backslash K(L_2)$. Let us now  show by a pointwise computation that $M\sim 0$.
Indeed, let $\alpha:=(a,\sigma_1,\tau_1,\sigma_2,\tau_2)\in X_2)$ be a point.
Let us consider
$ H^\bullet(M_\alpha)=H^\bullet_c(J_2^{-1}\alpha;\bZ)$.

If $\sigma_1-\tau_1\neq \sigma_2-\tau_2$, then $J_2^{-1}\alpha=\emptyset$. If
$\sigma_1-\tau_1=\sigma_2-\tau_2=h$,
then $J_2^{-1}\alpha$
gets identified with the  set  of all $v\in\Re$ satisfying: {\em either $\sigma_1\leq |\tau_1-v|$
or $\sigma_2\leq |\tau_2-v|$.} Let us denote this set by $Y_\alpha\subset\Re$. 
It follows that $Y_\alpha$ consists of all points $v$ satisfying: $h+v\leq 0$ or
$h+v\geq 2\sigma$, where $\sigma$ is the maximum of $\sigma_1$ and $\sigma_2$.
In other words, $Y_\alpha$ is a disjoint union  of two closed rays so that $H^\bullet_c(Y_\alpha,\bZ)=0$.
This shows that $M\sim 0$.

The map \eqref{Meq20b} is therefore a quasiisomorphism. In view of \eqref{eqM10a}, the first 
vanishing will be shown once we prove that 
\begin{equation} R\pi^X_{1!}((RJ_{2!}\bZ_{L'_2})\otimes (\pi_2^X){-1}G)\sim 0. \label{eqM203p}
\end{equation}
But $RJ_{2!} \Z_{L'_2}=\Z_{L^1_-}[-1]$, and hence the l.h.s. equals
  $(P^1_-)^{-1} RP^1_{-!} G [-1]$ which is zero by \eqref{usl1}. 


%

The second vanishing is shown analogously.

{\em Proof of the third vanishing}
Define the following subset $$L\subset I_\diamondsuit\times \R\times ((0,\infty)\times \R)\times ((0,\infty)\times \R)):$$
$$
L=\{(a,v,s_1,t_1,s_2,t_2)| (a,v,s_1,t_1),(a,v,s_2,t_2)\in V\};
$$

Similar to the proof of the 1-st vanishing, one shows that
$$
R\cI_!q_1^!Rq_{1!}q_{2}^!Rq_{2!}\cI^!G[-3]\sim R \pi^X_{1!}((RJ_{2!}\bZ_{L})\otimes (\pi_2^X)^{-1}G),
$$
where $$
J_2:I_\diamondsuit\times \R\times ((0,\infty)\times \R)\times ((0,\infty)\times \R))\to  X_2
$$
and 
$$
\pi_1^X,\pi_2^X:I_\diamondsuit\times \R\times ((0,\infty)\times \R)\times ((0,\infty)\times \R))\to 
I_\diamondsuit\times (0,\infty)\times\Re
$$
are the same as in the proof of the 1-st vanishing.

Observe that $$J_2(L)=\{(a,(s_1,t_1),(s_2,t_2))|\ |t_1-t_2|<s_1+s_2\}.$$
the projecion $L\to J_2(L)$ is a smooth fibration whose fibers are diffeomorphic to $\Re^1$; we now see that
$$
RJ_{2!}\bZ_{L}\sim \bZ_{J_2(L)}[-1]\in \DerCat(X_2).
$$

We therefore need to show that
$$
R  \pi_{1!}^X(\bZ_{J_2(L)}\otimes (\pi_2^X)^{-1}G)\sim 0
$$

The complement to $J_2(L)$ in $X_2$ consists of two components
$$X_2\backslash J_2(L)=M_+\sqcup M_-,
$$
where
$$
M_+=\{\{(x,(s_1,t_1),(s_2,t_2))|\ t_1-t_2\geq s_1+s_2\}
$$
and
$$
M_-=\{\{(x,(s_1,t_1),(s_2,t_2))|\ t_1-t_2\leq -s_1-s_2\}
$$

We thus have a distinguished triangle 
$$
\to R\pi_{1!}(\bZ_{J_2(L)}\otimes \pi_2^{-1}F)\to
 R\pi_{1!}(\bZ_{X_2}\otimes \pi_2^{-1}G)
\to  R\pi_{1!}(\bZ_{M_+}\otimes \pi_2^{-1}G)\oplus R\pi_{1!}(\bZ_{M_-}\otimes \pi_2^{-1}G)\to 
$$
which comes from a short exact sequence
$$
0 \to \bZ_{J_2(L)}\to \bZ_{X_2}
\to \bZ_{M_+}\oplus \bZ_{M_-}\to 0.
$$ 

The second term of this triangle
is quasi-isomorphic to
$$
\pi^{-1}R\pi_!G,
$$
where  $\pi:I_\diamondsuit\times (0,\infty)\times \R\to{I_\diamondsuit}$
 is the projection. 
It follows that $R\pi_!G\sim 0$  because $\pi$ passes through  $P^1_+$ (as well as $P^2_-$)
from \eqref{usl1}.

We thus need to show that  $R\pi_{1!}^X(\bZ_{M_\pm}\otimes (\pi_2^X)^{-1}G)\sim 0$.

Introduce the following subsets $N_\pm\subset  I_\diamondsuit\times  ((0,\infty)\times \R)\times \R$:

$$
N_+=\{(a,(s_1,t_1),y)|\ t_1\geq s_1+y\}
$$
and
$$
N_-=\{(a,(s_1,t_1),y)|\ t_1\leq -s_1-y\}.
$$

Let $q_1: I_\diamondsuit\times ((0,\infty)\times \R)\times \R\to (0,\infty)\times \R$ and
$q_2:I_\diamondsuit\times ((0,\infty)\times \R)\times \R\to \R$ be projections.
We then have
$$
R\pi_{1!}^X(\bZ_{M_\pm}\otimes (\pi_2^X)^{-1}G)\sim  Rq_{1!}(\bZ_{N_\pm}\otimes q_2^{-1}RP^1_{\pm!} G)\sim 0
$$
because $RP^1_{\pm!} G=0$  by (\ref{usl1}).

This completes the proof of the 3rd vanishing as well as the proof of Claim \ref{pred}

\subsection{Finishing proof of Claim \ref{promtrebobyed}} 

Let $I_\diamondsuit\times \R_{>0}\times \R$, the target of the map $\cI_1$ from \eqref{oct3e1118}, have coordinates $(a,v,\eta)$.

Let $G,H,\cI$ be as in Claim \ref{pred} and let $H'$ be a sheaf on $I_\diamondsuit \times \R_{>0} \times R$ 
microsupported on the set 
\begin{equation}\label{supnew}
\bigcup_{\text{``+" and ``-"}} (a,v,\eta,\R. d(v\pm \eta)+\R.da).
\end{equation}
We then have
$$
R\hom(G,H')\sim R\hom(R\cI_!H,H')\sim R\hom(H,\cI^!H').
$$

By ~\cite[Prop.5.4.5(i)]{KS}, it follows from \eqref{supnew} that 

\begin{equation}
S.S. (\cI^! H') \ \subset \ \{ (a,u,\xi_1,\xi_2, bda+wdu+\tau_1 d\xi_1 + \tau_2 d\xi_2 \ : \ \tau_1=0 \ \text{or} \ \tau_2=0 \} . \label{newM18}
\end{equation}

Set $A'=H$, $B'=\cI^! H'$.

Let also $q_1,q_2:I_\diamondsuit\times \R\times (0,\infty)\times (0,\infty)\to  I_\diamondsuit\times \R\times (0,\infty)$
be projections as in \eqref{oct3e1026}: $q_i(a,u,\xi_1,\xi_2)=(a,u,\xi_i)$. 

We then have $Rq_{i!}A'=0,$  $i=1,2$, by Claim \ref{pred},1), and we have the estimate \eqref{newM18} for $B'$. 

Let us identify diffeomorphically  $\R\to (0,\infty)$. Under this identification, 
we have
two sheaves
$A,B$ on $Y\times \R\times\R$, where $Y=I_\diamondsuit\times \R$,
such that

1) $Rp_{1!}A=Rp_{2!}A\sim 0$, where $p_1,p_2:Y\times \R\times \R\to \R$ are projections;

2) $B$ is microsupported on the set of points
$(y,u_1,u_2,\omega+v_1du_1+v_2du_2)$,
where $\omega\in T^*_yY$ $u_1,u_2\in \R$; $v_1=0$ or $v_2=0$ (or both).

By Theorem \ref{plosk}, $R\hom(A,B)=0$, which finishes the proof of Claim \ref{promtrebobyed}, 
as well as Proposition \ref{main}.







\section{Proof of Theorem \ref{CKl}} \label{PfSemior} 
In section \ref{StrucObPhi} -\ref{phikonec}, we have  constructed
objects $\Phi^K,\Phi^{\elr_\alpha},\Phi^{\elr_{-\alpha}}$, as well as maps
$i_{\Phi^K}:\bZ_{\bx_0\times K}[-2]\to \Phi^K$,
$i_{\Phi^{\elr_{\alpha}}}:\bZ_{\bx_0\times \elr_\alpha}[-2]\to \Phi^{\elr_\alpha}$, and
$i_{\Phi^{\elr_{-\alpha}}}:\bZ_{\bx_0\times \elr_{-\alpha}}[-2]\to \Phi^{\elr_{-\alpha}}$.
In order to finish the proof of Theorem \ref{CKl}, it now remains to prove:

1)  Each of the objects  $\Phi^K,\Phi^{\elr_\alpha},\Phi^{\elr_{-\alpha}}$  belongs
to $\cC$, to be done in  Sec \ref{ince}.

2) Cones of the maps $i_{\Phi^K},i_{\Phi^{\elr_\alpha}},i_{\Phi^{\elr_{-\alpha}}}$
are in ${}^\perp\cC$, to be done in Sec \ref{prooforthogonal}

We only consider the case of $\Phi^K$ (and the map $i_{\Phi^K}$), because  the arguments for the remaning cases
are very similar.

Proof of 2) is based on the orthogonality criterion
  of the previous section (Proposition \ref{main}).

\subsection{Proof of $\Phi^K\in {\cal C}$.} \label{ince}

 Consider  open subsets
 $\Sigma_\ell\subset X$, where $\Sigma_\ell$ is the union 
 of two neighboring open strips $\Int P_1$, $\Int P_2$ and their common boundary ray $\ell$.
It is clear that $\Sigma_\ell$ form an open covering of $X$.

Let us consider the restriction
 estimate $\Phi^K|_{\Sigma_\ell\times \C}$.  It suffices to show that
$$
\SS(\Phi^K|_{\Sigma_\ell\times \C})\subset \Omega_X\cap T^*(\Sigma_\ell\times \C)
$$
for each  element $\Sigma_\ell$ of the open covering. 
Let us fix the notation: let $\Sigma_\ell=\Int P_1\sqcup \Int P_2\sqcup \ell$;
let $P_i':=\Int P_i\sqcup \ell$, $i=1,2$, be the closure of $P_i$ in $\Sigma_\ell$.
Set for brevity
$$
F:=\Phi^K|_{\Sigma_\ell\times \C}.
$$
 Finally, we introduce the following sheaf on $\Sigma_\ell\times \Co$:
 $$\Lambda^{K\pm}_{\Sigma_\ell}:=\bZ_{ \{ z\in\Sigma_\ell \ :  \ s\pm z \in K\}}.$$

Let us now suppose for definiteness that $\ell$ goes to the left. 
 As follows from the construction of $\Phi^K$  in Sec \ref{definitionphikp},\ref{oboznvt}, we have 
identifications ($i=1,2$):
$$
F|_{P'_i\times \Co}=
(\Lambda^{K+}_{\Sigma_\ell}*S_+\oplus\Lambda^{K-}_{\Sigma_\ell}*S_-)|_{P'_i\times \Co}.
$$
as well as a gluing map (\ref{oct3e3a}):
$$
\Gamma^{P_1P_2}_{\Phi^K}:(\Lambda^{K+}_{\Sigma_\ell}*S_+
\oplus\Lambda^{K-}_{\Sigma_\ell}*S_-)|_{\ell\times \Co}\to(\Lambda^{K+}_{\Sigma_\ell}*S_+
\oplus\Lambda^{K-}_{\Sigma_\ell}*S_-)|_{\ell\times \Co}
$$

When restricted onto $\Lambda^{K+}_{\Sigma_\ell}*S_+|_{\ell\times \Co}$, this map
becomes the identity. This readily implies that we have an embedding
$$
\Lambda^{K+}_{\Sigma_\ell}*S_+\into F,
$$
whose restriction onto each $P'_i$ is just the identical embedding onto the direct summand.
We can construct a surjection $F\to \Lambda^{K-}_{\Sigma_\ell}*S_-$ in a similar way.
All together, we get a short exact sequence
$$
0\to \Lambda^{K+}_{\Sigma_\ell} *S_+\to F\to  \Lambda^{K-}_{\Sigma_\ell}* S_-\to 0,
$$

The marginal terms of this sequence do clearly 
have their singular support inside $\Omega_X \cap T^*(\Sigma_\ell\times \C)$, cf.\eqref{DefOmega}, hence so 
does the middle term $F$. This finishes the proof.

\subsection{Proof of orthogonality} \label{prooforthogonal} 

In this subsection, we prove that the cone of the map $i_{\Phi^K}$ is in ${}^\perp\cC$. 
We will exhibit an increasing   exhaustive filtration $F$ of $\Phi^K$ such that   the map $i_\Phi$ 
factors through $F^0\Phi^K$.  Our statement then reduces to showing that 
  $Cone({\cal F}_0\to F^0\Phi^K)$, as well as all successive quotients of $F^{i+1}\Phi^K/F^i\Phi^K,$ $i\geq 0$,
 belong to ${}^\perp {\cal C}$.

\subsubsection{Regular sequences}
\begin{figure} \includegraphics{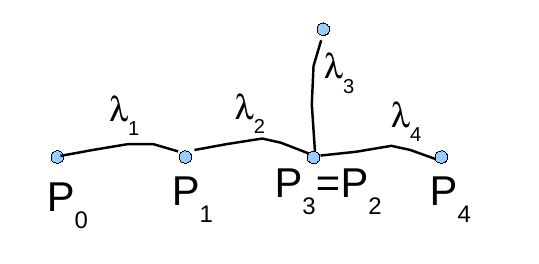} 
\caption{A regular sequence -- Notation \ref{i19no1}.} \label{CoShWKBp65}  \end{figure}

\begin{Notation} \label{i19no1} {\rm Let $\lambda_n\lambda_{n-1}\cdots \lambda_1$ be a nonempty 
sequence of bounday $\alpha$-rays.

Call this sequence {\em regular} if for each $k\geq 1$ the rays $\lambda_k$ and $\lambda_{k+1}$
are different and belong to the closure of a (unique) $\alpha$-strip $P_k$, fig.\ref{CoShWKBp65}.  
 We also assume that $P_0$ is the initial strip (i.e. $\bx_0\in P_0$.} \end{Notation}

Note that,  in general, a ray can occur in a regular sequence several times.

\subsubsection{Admissible rays} We will freely use the notation from Sec. \ref{fik}, such as $\cL^\alpha$,
$W$, $\Lambda^{K\pm}$.
  
 Let $w\in \wa$ be of the form $\ell_m\ell_{m-1}\cdots \ell_1 \{ L \ \text{or} \ R\}$ and let
$\ell\in \cL^\alpha$ be a boundary $\alpha$-ray. We call $\ell$ {\em $\lambda,w$-admissible},
if there exists a $k$ such that $\ell=\lambda_k$ and
 and $\ell_m\ell_{m-1}\cdots \ell_1$ is a subsequence
of $\lambda_{k}\lambda_{k-1}\cdots \lambda_1$ 
(i.e. there is an increasing sequence $\kappa_1<...<\kappa_m$ such
 that $\ell_1=\lambda_{\kappa_1}$, ..., $\ell_m=\lambda_{\kappa_m}$).

\begin{Remark} \label{Rk19no3}{ \rm Let $w=l_ml_{m-1}\cdots (L\text{ or }R)$.
 If   $\ell_m=\ell$, then this condition is equivalent
to $\ell_m\ell_{m-1}\cdots \ell_1$ being a subsequence of $\lambda$;
it $\ell_m\neq \ell$, then the condition  is equivalent to 
$\ell \ell_m\ell_{m-1}\cdots \ell_1$ being  a subsequence of $\lambda$.}
\end{Remark}

\subsubsection{Subset $P_{\lambda,w}$}
Let $P$ be an $\alpha$-strip.
We  define an open subset $P_{\lambda,w}\subset P$ as  follows.

1) if every boundary ray of $P$ is not $\lambda,w$-admissible, then we set $P_{\lambda,w}:=\emptyset$.
  
2) otherwise (there are $\lambda,w$-admissible boundary rays of $P$) we define $P_{\lambda,w}$ as the union
of $\Int P$ with all  $\lambda,w$-admissible boundary rays of $P$.
\subsubsection{Subsheaves $\Lambda^{K\pm}_{P,\lambda,w}$}
 Let 
$j:=j_{\lambda,w}^P:P_{\lambda,w}\times \C \to P \times \C$ be the open embedding.

 As in {Sec.\ref{LambdaU}}, let  $\Lambda^{K\pm}_{P}=\Z_{\{(z,s): \ z\in P, \ s\pm z\in K\}}$.  

Accordingly, we can define  subsheaves 
$$
\Lambda^{K\pm}_{P,\lambda,w}:=j_!j^!\Lambda_{P}^{K\pm}\subset 
\Lambda_{P}^{K\pm}\in \DerCat(P\times \Co).$$

Observe that $\Lambda^{K\pm}_{P,\lambda,w}=0$ if $P$ has no $\lambda,w$-admissible boundary rays.
\subsubsection{Subsheaves $\Phi^{K,\lambda}_{P}\subset \Phi^K_P$}
We have an identification
$$
\Phi^K|_P=\bigoplus\limits_{w\in \wa_\pravo} S_w*\Lambda_P^{K-}\oplus 
\bigoplus\limits _{w\in \wa_\levo} S_w*\Lambda_P^{K+}.
$$

For each regular sequence $\lambda$ (where $\lambda$ stands for $\lambda_n\lambda_{n-1}\ldots \lambda_1$),
let us construct a sub-sheaf $\Phi^{K,\lambda}\subset \Phi^K$ as follows.
Set 
\begin{equation}
\Phi^{K,\lambda}_P:=\bigoplus_{w\in \wa_\pravo} S_w*\Lambda^{K-}_{P,\lambda,w}\oplus
 \bigoplus_{w\in \wa_\levo} 
S_w*\Lambda^{K+}_{P,\lambda,w} \label{jn22sum}
\end{equation} 
We have an obvious embedding
$$
\Phi^{K,\lambda}_P\to \Phi^K_P.
$$

\subsubsection{Sheaves $\Phi^{K,\lambda}_P$ match on the intersections}
Let $P$ and $P'$ be two intersecting $\alpha$-strips; let $\ell=P\cap P'$. We then have two sub-sheaves of
$\Phi^K_\ell$, namely $\Phi^{K,\lambda}_P|_{\ell\times \Co}$ and $\Phi^{K,\lambda}_{P'}|_{\ell\times \Co}$.
Let us check that these  two subsheaves  do in fact coincide:

\begin{Claim} $$
\Phi^{K,\lambda}_P|_{\ell\times \Co}=\Phi^{K,\lambda}_{P'}|_{\ell\times \Co}$$
\end{Claim}

\textsc{Proof} Let $w\in \wa$. Consider the following sheaf:
$\Lambda^{\pm}_{P,w}:=\Lambda^{K\pm}_{P,\lambda,w}|_{\ell\times \Co}$. By definition, 
$\Lambda^{\pm}_{P,w}=0$ unless $\ell$ is $\lambda,w$-admissible, in which case
$\Lambda^{\pm}_{P,w}=\Lambda^{K\pm}|_\ell$. 

Let $\wc(\ell,\lambda)\subset \wa$ be the subset consisting of all $w$, where $\ell$ is $\lambda,w$-admissible.
Let $\wc(\ell,\lambda)=\wc(\ell,\lambda)_\levo\sqcup S(\ell,\lambda)_\pravo$, where
$\wc(\ell,\lambda)_\levo=\wc(\ell,\lambda)\cap \wa_\levo$;
 $\wc(\ell,\lambda)_\pravo=\wc(\ell,\lambda)\cap \wa_\pravo$.

It now follows that  $\Phi^{K,\lambda}_P|_{\ell\times\Co}$, as a subsheaf of 
$\Phi^K_P|_{\ell\times \Co}=\bigoplus_{w\in \wa_\levo} S_w*\Lambda^{K+}_\ell\oplus
\bigoplus_{w\in \wa_\pravo} S_w*\Lambda^{K-}_\ell$, coincides with the following its direct summand:
$$
\Phi^{K,\lambda}_P|_{\ell\times\Co}=\Phi(\ell,\lambda):=\bigoplus_{w\in \wc(\ell,\lambda)_\levo} S_w*\Lambda^{K+}_\ell\oplus
\bigoplus_{w\in \wc(\ell,\lambda)_\pravo} S_w*\Lambda^{K-}_\ell.
$$

Analogously, we have an equality
  $$
\Phi^{K,\lambda}_{P'}|_{\ell\times\Co}=\Phi(\ell,\lambda)$$
of subsheaves of $$\bigoplus_{w\in \wa_\levo} S_w*\Lambda^{K+}_\ell\oplus
\bigoplus_{w\in \wa_\pravo} S_w*\Lambda^{K-}_\ell=\Phi^K_{P'}|_{\ell\times \Co}.
$$

It now suffices to check that the sub-sheaf $\Phi(\ell,\lambda)$ is preserved by the gluing
map $\Gamma^{P P'}_{\Phi^K}$ from Sec \ref{oboznvt}. By definition of $\Gamma^{P P'}_{\Phi^K}$,
 it suffices to check:
{\em let $w\in \wc(\ell,\lambda)$  and suppose $\ell w\in \wa$ (meaning that the  leftmost ray of the word $w$
goes in the opposite direction to $\ell$); then $\ell w\in \wc(\ell,\lambda)$}. Indeed, 
$w\in \wc(\ell,\lambda)$, $\ell w\in \wa$ is equivalent to $\ell w$ being a sub-sequence of $\lambda$, which is the
same as $\ell w\in \wc(\ell,\lambda)$.
$\Box$

\bigskip

This Claim implies that {\em there is a unique sub-sheaf $\Phi^{K,\lambda}\subset \Phi^K$ such that
$\Phi^{K,\lambda}_P=\Phi^{K,\lambda}|_{P\times \Co}$ for all $\alpha$-strips $P$.}

\subsubsection{Definition of a filtration on $\Phi^K$}

\begin{Notation} \label{i19no225} {\rm Choose and fix an infinite  regular sequence 
\begin{equation}
\ldots \lambda_n\lambda_{n-1}\ldots \lambda_2\lambda_1 \label{jn16eq400}
\end{equation}
such that  

---{\em every ray occurs in this sequence 
infinitely many times;}

---{\em the ray $\lambda_1$ is adjacent to the  $\alpha$-strip $\bP_0$  containing $\bx_0$.}

Denote by $\lambda^{(n)}$ 
 the subsequence $\lambda_n\lambda_{n-1}\ldots \lambda_2\lambda_1$.  } \end{Notation}

Set $F^n\Phi^K:=\Phi^{K,\lambda^{(n)}}$.
Let us check
\begin{Claim} We have $F^n\Phi^K\subset F^{n+1}\Phi^K$.
\end{Claim}
\textsc{Proof.}
It suffices to check that $F^n\Phi^K|_{P\times \Co}\subset F^{n+1}\Phi^K|_{P\times \Co}$
for every strip $P$ (as sub-sheaves of $\Phi^K_P$).  It suffices to check that
$P(\lambda^{(n)},w)\subset P(\Lambda^{(n+1)},w)$ for all $w$, which follows from:
if a ray $\ell$ is $\lambda^{(n)},w$-admissible, then $\ell$ is $\lambda^{(n+1)},w$-admissible.
This follows from the definition of $\lambda,w$-admissibility.
$\Box$

\begin{Claim}  Subsheaves $F^n\Phi^K$ form an exhaustive filtration of $\Phi^K$.
\end{Claim} 
\textsc{Proof.}  It suffices to check that $\bigcup F^n\Phi^K|_{P\times \Co}=\Phi^K_P$.
This is implied by:  for every $w\in \wa$ and every boundary ray $\ell$ of $P$, there exists an 
$n>0$ such that $\ell\in P_{\lambda^{(n)},w}$, equivalently: $\ell$ is  $\lambda^{(n)},w$-admissible.
Let us prove this statement. By the construction of $\lambda$, every finite sequence of rays, is a subsequence
of $\lambda^{(n)}$ for $n$ large enough (because every ray occurs in the sequence $\{\lambda_i\}_{i=1}^\infty$
infinitely many times). Let  $w=\ell_m\cdots\ell_1(L\text{ or } R)$,  then  the sequence $\ell\ell_m\cdots\ell_1$
(if $\ell\neq\ell_m$) or $\ell_m\cdots\ell_1$ is a subsequence of $\lambda^{(n)}$ for some $n$, 
meaning that $\ell$ is $\lambda,w$-admisssible.
$\Box$

\subsubsection{Computing $F^1\Phi^K$}\label{F1} In this subsection,  $P_*$ denotes the strip adjacent to $\lambda_1$ 
and different from $P_0$. 
We assume that $\lambda_1$ goes to the right  and that $P_0$ is above $P_*$ (all other cases are treated
in a similar way).

Let us  give an explicit description of $F^1\Phi^K$. 
First of all, a ray $\ell$ is $\lambda^{(1)},w$-admissible iff  $\ell=\lambda_1$ and $w$ is one of the following
$L,R,\lambda_1L$. 
Therefore, $P_{\lambda^{(1)},w}\neq \emptyset$  iff: $P$ contains $\lambda_1$, that is
$P=P_0$ or $P=P_*$, and  $w$ is one of $L,R,\lambda_1L$.  In each of this cases
$P_{\lambda^{(1)},w}=\Int P\cup \lambda_1$.

Thus, $F^1\Phi^K$ is supported on $\Sigma:=\Int P_0\cap \lambda_1\cap\Int P_*$.  Let $P'_0=\Int P_0\cup
\lambda_1$; $P'_*=\Int P_*\cup \lambda_1$.
We have
$$F^1\Phi^K|_{P'_*\times \Co}=A_*\oplus B_*;$$
$$F^1\Phi^K|_{P'_0\times \Co}=A_0\oplus B_0,$$
where 
$A_*=S_R*\Lambda_{P'_*}^{K-}$; $A_0=S_R*\Lambda_{P'_0}^{K-}$;
$B_*=S_L*\Lambda_{P'_*}^{K+}\oplus S_{\lambda_1L}*\Lambda_{P'_*}^{K-}$;
$B_0=S_L*\Lambda_{P'_0}^{K+}\oplus S_{\lambda_1L}*\Lambda_{P'_0}^{K-}$
The gluing map $\Gamma^{P_0P_*}_{\Phi_K}$ maps $A_0|_{\lambda_1\times \Co}$
into $A_*|_{\lambda_1\times \Co}$ and $B_0|_{\lambda_1\times \Co}$
into $B_*|_{\lambda_1\times \Co}$, therefore, the sheaves $A_*$ and $A_0$ get glued into
a sheaf $A$ on $\Sigma$, and $B_*$ and $B_0$ into a sheaf $B$ so that
$F^1\Phi^K=A\oplus B$. One also sees that $A=S_R*\Lambda^{K-}_\Sigma$.
Let $j:\Int P_0\to\Sigma$ be the open embedding.
\subsubsection{The map $i_\Psi$ factorizes through $F^1\Phi^K$}\label{factF1} Keeping  the assumptions  of the previous
subsection,
let us now construct the factorization of  the map $i_\Psi:\bZ_{\bx_0\times K}[-2]\to \Phi^K$
through $F^1\Phi^K$. 
The cases when $\lambda_1$ goes to the left of $P_*$ is above $P_0$ are  treated in a similar way.

 Let $j:\Int P_0\times \Co\to X\times \Co$ be the open embedding.
By definition, $i_\Psi$ factors as
\begin{equation}\label{facipsi}
\bZ_{\bx_0\times K}[-2]\to j_!(S_L*\Lambda_{\Int P_0}^{K+}\oplus S_R*\Lambda_{\Int P_0}^{K-})\to \Phi^K,
\end{equation}
where the first arrow is induced by  the  following maps  in $\bD(\Int P_0\times \Co)$:
$$\iota_L:\bZ_{\bx_0\times K}[-2]\to \bZ_{\{(z,s)|z\in \Int P_0,s+z\in \bx_0+K\}}=
S_L*\Lambda_{\Int_{P_0}}^{K+};
$$
$$\iota_R:\bZ_{\bx_0\times K}[-2]\to \bZ_{\{(z,s)|z\in \Int P_0,s-z\in -\bx_0+K\}}=
S_R*\Lambda_{\Int_{P_0}}^{K-},
$$
which are induced by the closed codimension 2 embeddings of the corresponding sets.
 
The right arrow in  (\ref{facipsi}) factors through $F^1\Phi^K$ as follows. Let as decompose
$j=j_1j_0$, where $j_0:\Int P_0\times \Co\to \Sigma\times\Co$ and $j_1:\Sigma\times\Co\to X\times \Co$
are the open embeddings. We have natural maps
$i_A:j_{0!}(S_L*\Lambda_{\Int P_0}^{K+})\to A$; $i_B:j_{0!}(S_R*\Lambda_{\Int P_0}^{K-}\to B$.
Whence a map
$$
i_A\oplus i_B:j_{0!}(S_L*\Lambda_{\Int P_0}^{K+}\oplus
 S_R*\Lambda_{\Int P_0}^{K-})\to A\oplus B=F^1\Phi^K|_{\Sigma\times \Co}.
$$
The right arrow in (\ref{facipsi}) is then obtained by applying $j_{1!}$ to $i_A\oplus i_B$.
For future references, let us consider
$\Cone(\bZ_{\bx_0\times K}[-2]\to F^1\Phi^K)$, which  is supported
on $\Sigma\times \Co$. We now see that
  $$\Cone(\bZ_{\bx_0\times K}[-2]\to F^1\Phi^K)|_{\Sigma\times \Co}$$
 is isomorphic to 
the Cone of the following
composition map in $\bD(\Sigma\times \Co)$:
\begin{equation}\label{compossec}
\bZ_{\bx_0\times K}[-2]\to j_{0!}(S_L*\Lambda_{\Int P_0}^{K+}\oplus
 S_R*\Lambda_{\Int P_0}^{K-})\to A\oplus B,
\end{equation}
where the right arrow is $i_A\oplus i_B$, and the left arrow is induced by $\iota_L\oplus \iota_R$.

\subsubsection{Computing successive quotients of the filtration}

Let us compute the quotients $\cG^n:=F^n\Phi^K/F^{n-1}\Phi^K$, $n\geq 2$. Our computation will result
in  decompositions (\ref{decolevo}), (\ref{decopravo})

For that purpose, we choose an $\alpha$  strip $P$ 
 and compute the restriction  $\cG^n_P:=F^n\Phi^K/F^{n-1}\Phi^K|_{P}.$

Set
$$
P(n,w):=P_{\Lambda^{n},w}\backslash P_{\Lambda^{n-1},w}\subset P.
$$
 $P(n,w)$ is a locally closed subset of $P$ so that we can define
 the following sheaves on $P\times \Co$:
$$
\Lambda^{K\pm}_{P(n,w)}=\bZ_{\{(z,s)|z\in P(n,w);s\pm z\in K\}}.
$$

We have an identification
$$
\cG^n_P=\bigoplus\limits_{w\in \wa_\levo} S_w*\Lambda^{K+}_{P(n,w)}\oplus 
\bigoplus\limits_{w\in \wa_\pravo} S_w*\Lambda^{K-}_{P(n,w)}.
$$

Let us now describe the sets $P(n,w)$. Below, for a $w\in \wa$, we set $\trim(w)$ to be the word
$w$ with its rigthmost letter (L or R) removed.

{\em Step 1} Consider all the situations when $\Int P\subset P(n,w)$

This occurs iff $\Int P$ is part of $P_{\lambda^{(n)},w}$
but not $P_{\lambda^{(n-1)},w}$.
This is equivalent to the following:

Condition I:
 $n$ is the minimal number satisfying:

(1) $\lambda_n$  is a boundary ray of $P$;

(2) $\trim(w)$ is a subsequence of $\lambda^{(n)}$.

Let us reformulate these conditions. Introduce the following notation.
For a word $w$ set $M(w)$ to be the minimal number such that $\trim(w)$ is a subsequence of
$\lambda^{(M(w))}$. For a word $w${, $w\ne \{ R\}, \{L\}$, } we also write $w=lw'$, 
where $l$ is the leftmost ray of $w$.

Let us split our consideration  into two cases: 

A) $l=\lambda_n$, (meaning that $\trim(w)$ is non-empty);

B)$\trim(w)$ is empty or $l\neq \lambda_n$.

Case A).
The combination Condition I+Case  A) is equivalent to the following combination:

A) (i.e. $l=\lambda_n$), and

A1) $M(w)=n$, and

A2) $\lambda_n$ is a boundary ray of $P$.

It follows that {\em given a boundary ray $r$ of $P$ different from $\lambda_n$,  such an $r$ is not
$\lambda^{(n)},w$-admissible}:  the admissibility would mean that  the word $rw$ is a subsequence
of  $\lambda^{(n)}$ (see remark \ref{Rk19no3})); 
since $r\neq \lambda_n$, $rw$ is also a subsequence of $\lambda^{(n-1)}$,
which implies $M(w)<n$, contradiction.

Thus, in this case we have $P(n,w)=\Int P\cup \lambda_{n}$. 

Case B)

Let us give an equivalent reformulation of the combination.
\begin{Lemma}
Condition I and case  B).
It is equivalent to the following combination:

B) and

B1) $\lambda_n$ is a boundary strip of $P$, and 

B2) $M(\lambda_{n}w)=n$,  and

B3) If $\trim(w)$ is non-empty, then $l$ is not a boundary ray of $P$, and, finally,

B4) $M(rw)\geq n$ for any boundary ray $r$ of $P$.
\end{Lemma}
\textsc{Proof.}
Let us first derive B1)-B4) from Condition I and B):

 B1) is just the condition  (1);  

B2): (2) and B) imply $M(\lambda_nw)\leq n$.  If $M(\lambda_nw)<n$,  then $n$ is not the
minimal number satisfying (1) and (2);

Violation of B3) implies that $n-1$ satisfies (1) and (2) --- contradiction. \\
Violaton of B4) implies that $M(rw)<n$; since  the number $M(rw)$ satisfies (1) and (2), we have a contradiction.

Let us now derive Condition I from B) and B1)-B4).

B1,B2 imply that $n$ satisfies (1) and (2). Suppose $n$ is not minimal, i.e there exists $p<n$
such that $\lambda_p$ is a boundary ray of $P$ and $M(w)\leq p$.
B3 implies that $\lambda_p$ is different from the leftmost ray of $w$. Therefore, $M(\lambda_pw)\leq p$,
which is prohibited by B4. 
$\Box$

Let us now introduce a one more condition $B5$.

 Let $P_{n-1}$ be (a unique) $\alpha$-strip
which is adjacent to both $\lambda_n$ and $\lambda_{n-1}$.  Let $P_*$ be the 
other $\alpha$-strip adjacent to $\lambda_n$. 
 
The condition B5 is as follows:

B5)$P=P_*$.

Let us prove that 
\begin{Lemma}
Combination Condition I+ B is equivalent to the combination $B,B2,B5.$
\end{Lemma}
 \textsc{Proof.} Let us first prove that
B,B1-B4 imply B5.
Since $\lambda_n$ is a boundary ray of $P$, the only alternative to B5 is
 $P=P_{n-1}$.  Then  $\lambda_{n-1}$ is a boundary
ray of $P$ and $M(\lambda_{n-1}w)\leq n-1$ which contradicts to B4.

Let us prove that {\em $B,B2,B5$ imply $B1,B3,B4$}.

B1: By B5 $P_*=P$, and $\lambda_n$ is a boundary ray of $P_*$;

B3,B4:
  B2 implies that for all $p\in [M(w);n-1]$, $\lambda_p\neq \lambda_n$. This implies that
 {\em $P_*$ is not adjacent to any of $\lambda_p$ with
$p\in [M(w);n-1]$} Indeed, suppose $P_*$ is adjacent to such a $\lambda_p$.
 Consider the graph $\Gamma$ whose vertices are strips and
and whose edges are rays. We have  two non-intersecting paths between $P_{n-1}$ and $P_*$:
one of them is $\lambda_n$, we also have a path between $P_{n-1}$ and $P_*$
 in the  connected graph composed of the edges
 $\lambda_{n-1}\lambda_{n-2},\cdots ,\lambda_p$, 
which contradicts to $\Gamma$ being a tree.  

The just proven statement implies B3 and

B4')
$M(rw)>n$ for every boundary
ray of $P=P_*$ which differs from $\lambda^{(n)}$. 

Finally, B2) and B4') imply B4), which finishes the proof.
$\Box$
\bigskip

Finally, we conclude from B4', that in the situation Condition 1+B we have:
$$
P(n,w)=\Int P\sqcup \lambda_n.
$$

\vskip2.5pc

{\em Step 2} Let us now examine the case (call it case C) when 
$
P(n,w)
$
is a non-empty union of boundary rays of $P$.
Since $P_{\lambda^{(n-1)},w} \subset P_{\lambda^{(n)},w}$, this is equivalent to
 $P_{\lambda^{(n-1)},w}$  being a proper (in particular, non-empty)
subset of $P_{\lambda^{(n)},w}$.  As follows from definitions, this is equivalent
to:

i') there is a $\lambda^{(n-1)},w$-admissible ray of $P$;

ii') There exists a boundary ray $r$ of $P$ such that $r$ is $\lambda^{(n)},w$-admissible,
but not $\lambda^{(n-1)},w$-admissible.

By Remark \ref{Rk19no3}, the condition i') is equivalent to: 

i'') there exists a boundary ray $r$ of $P$ 
such that either $r$ is the leftmost ray of $w$ and $M(w)\leq n-1$, or $r$ is not
the  leftmost ray of $w$ and $M(rw)\leq n-1$. 

In any case, i') implies that $M(w)\leq n-1$.

Also by Remark \ref{Rk19no3}, the condition ii')  is equivalent to the following one

ii'')  There exists a boundary ray $r$ of $P$ such that  
either 

a)  $r$ is not  the leftmost ray of $w$ and $M(rw)=n$;

or

b)  $r$ is the leftmost ray of $w$ and $M(w)=n$.

The case b) contradicts to i'), which implies $M(w)\leq n-1$.

The condition a) implies $r=\lambda_n$ and hence $\lambda_n$ is one and 
the only ray in $P_{\lambda^{(n)},w}$.

We thus can reformulate: 

The case C occurs iff

i') holds and

ii-$\alpha$) $\lambda_n$ is a boundary ray of $P$; 

ii-$\beta$)$\lambda_n$ is not the leftmost ray of $w$;

ii-$\gamma$) $M(\lambda_nw)=n$.

In the case C we have  $P(n,w)=\lambda^n$.

From ii-$\gamma$ we conclude that 
\begin{equation} \lambda_p\neq \lambda_n\ \ \ \text{for all} \ p\in [M(w);n-1].  \label{i27eq3} \end{equation}

The condition i' is equivalent to
\begin{equation} \exists p\in [M(w),n-1] \ \ \ : \ \lambda_p \ \text{ is adjacent to } \ {\bar\vs}. \label{i27eq4} \end{equation}

Let us show that  $P=P_{n-1}$: \\
Indeed, by ii-$\alpha$, the only alternative is $P=P_*$.  In this case, analogously to the proof of 
B5$\Rightarrow$B4, the property \eqref{i27eq3} implies that
 $P_*$ is not adjacent to any of $\lambda_p$ with
$p\in [M(w);n-1]$, and that contradicts \eqref{i27eq4}. 

Thus, we have the following condition which is equivalent to i' and ii' (the proof of the converse is trivial):
 
C1) $P=P_{n-1}$; $\lambda_n$ is not the leftmost ray of $w$ and $M(\lambda_nw)=n$.

In this case $P(n,w)=\lambda_n$.

\bigskip

Let us summarize our findings. Introduce the following notation. 
 Let $\wa_{n,\levo}$ be the set of all words $w$ in $\wa_\levo$ 
such that the leftmost ray of $w$ is not $\lambda_n$ and $M(\lambda_nw)=n$.
Let $\wa_{n,\pravo}$ be the similar thing. 

We then have the following three cases when
the set $P(n,w)$ is non-empty:

-- Conditions $A,A1,A2$ is satisfied. Equivalently, the following conditions are the case:

a1) $P=P_{n-1}$ or $P=P_*$;

a2) $w=\lambda_nu$, where $u\in \wa_{n,\levo}$ 
if $\lambda_n\in {\cal L}_\pravo$, and $u\in \wa_{n,\pravo}$ if  $\lambda_n\in {\cal L}_\levo$.

In this situation
$P(n,w)=\Int P\cup \lambda_n.$

--- B,B2,B5 are satisfied. Equivalently: $P=P_*$ ;  $w\in \wa_{n,\levo}$ 
if $\lambda_n\in {\cal L}_\pravo$, and $w\in \wa_{n,\pravo}$ if  $\lambda_n\in {\cal L}_\levo$. Then
$P(n,w)=\Int P_*\cup \lambda_n.$
\\
--- C1 is satisfied. Equivalently: 

b1) $P=P_{n-1}$; 

b2) $w\in \wa_{n,\levo}$ 
if $\lambda_n\in {\cal L}_\pravo$, and $w\in \wa_{n,\pravo}$ if  $\lambda_n\in {\cal L}_\levo$.

In this situation, we have
$P(n,w)=\lambda_n$.
\subsubsection{Description of $\cG_n$}\label{oct21nn}
In particular, we see that 
 the sheaf $\cG_n=F^n\Phi^K/F^{n-1}\Phi^K$ is supported on the union
  $\Int P_{n-1}\cap \lambda_n\cap \Int P_*$.

  Let $P'_*:=\Int P_*\cup \lambda_n$. We will now  describe the 
restriction of $\cG_n$ onto $P'_*$.  

 Suppose that $\lambda_{n}\in {\cal L}_\levo$. We then have 
$$
\cG_n|_{P'_*\times \Co}=\bigoplus\limits_{w\in \wa_{n,\pravo}}( S_{w}*\Lambda^{K-}_{P'_*}\oplus
S_{\lambda_nw}*\Lambda^{K+}_{P'_*})\oplus \bigoplus\limits_{w\in \wa_{n,\levo}} S_w*\Lambda^{K+}_{P'_*}.
$$

For $w\in \wa_{n,\pravo}$, we denote
$$
B_w^{P'_*}:=S_{w}*\Lambda^{K-}_{P'_*}\oplus
S_{\lambda_nw}*\Lambda^{K+}_{P'_*};
$$
for $w\in \wa_{n,\levo}$, we set 
$$
A_w^{P'_*}:=S_w*\Lambda^{K+}_{P'_*}.
$$
so that we can rewrite
$$
\cG_n=\bigoplus\limits_{w\in \wa_{n,\pravo}}B_w^{P'_*}\oplus\bigoplus\limits_{w\in \wa_{n,\levo}} A_w^{P'_*}.
$$

In the case $\lambda_{n}\in {\cal L}_\pravo$, change all signs and all orientations:
we have
$$
\cG_n=\bigoplus\limits_{w\in \wa_{n,\levo}}B_w^{P'_*}\oplus\bigoplus\limits_{w\in \wa_{n,\pravo}} A_w^{P'_*},
$$
where
for $w\in \wa_{n,\levo}$, we denote
$$
B_w^{P'_*}:=S_{w}*\Lambda^{K+}_{P'_*}\oplus
S_{\lambda_nw}*\Lambda^{K-}_{P'_*};
$$
for $w\in \wa_{n,\pravo}$, we set 
$$
A_w^{P'_*}:=S_w*\Lambda^{K-}_{P'_*}.
$$

\vskip2.5pc

(2) Let $P'_{n-1}$ be the union of the interior of $P_{n-1}$ and $\lambda_n$.

We then have   in the case $\lambda_n\in {\cal L}_\levo$:
$$
\cG_n|_{P'_{n-1}\times \Co}=\bigoplus\limits_{w\in \wa_{n,\pravo}}  B_w^{P'_{n-1}}\oplus 
\bigoplus\limits_{w\in \wa_{n,\levo}} A_w^{P'_{n-1}},
$$

where for $w\in \wa_{n,\pravo}$ we set
$$
B_w^{P'_{n-1}}:= S_{w}*\Lambda^{K-}_{\lambda_n}\oplus
S_{\lambda_nw}*\Lambda^{K+}_{P'_{n-1}};
$$
for $w\in \wa_{n,\levo}$ we set
$$
A_w^{P'_{n-1}}:= S_w*\Lambda^{K+}_{\lambda_n}.
$$

If $\lambda_n\in {\cal L}_\pravo$, then one has to change all the directions and all the signs:

$$
\cG_n|_{P'_{n-1}\times \Co}=\bigoplus\limits_{w\in \wa_{n,\levo}}  B_w^{P'_{n-1}}\oplus 
\bigoplus\limits_{w\in \wa_{n,\pravo}} A_w^{P'_{n-1}},
$$

where for $w\in \wa_{n,\levo}$ we set
$$
B_w^{P'_{n-1}}:= S_{w}*\Lambda^{K+}_{\lambda_n}\oplus
S_{\lambda_nw}*\Lambda^{K-}_{P'_{n-1}};
$$
for $w\in \wa_{n,\pravo}$ we set
$$
A_w^{P'_{n-1}}:= S_w*\Lambda^{K-}_{\lambda_n}.
$$

Analyzing the gluing maps, we see that
$$
A_w^{P'_*}|_{\lambda_n\times \Co}=A_w^{P'_{n-1}}|_{\lambda_n\times\Co}
$$
as sub-sheaves of $\cG_n|_{\lambda_n\times\Co}$ and similarly for $B_w$.
Therefore, we have well defined sub-sheaves $A_w,B_w$  of $\cG_n$:
$A_w$ is defined by the conditions:
$$
A_w|_{P'_*\times\Co}=A_w^{P'_*};
$$
$$
A_w|_{P'_{n-1}\times\Co}=A_w^{P'_{n-1}},
$$
and similarly for $B_w$.

Let us stress that  
$B_w|_{\Int P_{n-1}\cup \lambda_n \cup \Int P_n}$ is {\it not} isomorphic to 
the direct sum of  $S_w *\Lambda^{K+}_{\Int P_{n-1}\cup\lambda_n\cup \Int P_*}$ and $S_{\lambda_n w}*
\Lambda^{K-}_{\Int P_{n-1}
\cup \lambda_n \cup \Int P_*}$ 

We have in the case $\lambda_n\in {\cal L}_\levo$:
\begin{equation}\label{decolevo}
\cG_n=\bigoplus\limits_{w\in \wa_{n,\pravo}} B_w\oplus\bigoplus\limits_{w\in \wa_{n,\levo}} A_w;
\end{equation}
if $\lambda_n\in {\cal L}_\levo$, then we have:

\begin{equation}\label{decopravo}
\cG_n=\bigoplus\limits_{w\in \wa_{n,\levo}} B_w\oplus\bigoplus\limits_{w\in \wa_{n,\pravo}} A_w.
\end{equation}

\subsubsection{Reduction of the orthogonality property} As was explained in Sec \ref{factF1},
 the map
map $i_{\Phi^K}$ factors as
$\bZ_{\{z=\bx_0,s\in K\} }[-2]\to F^1\Phi^K\to \Phi^K.$

It therefore suffices to prove that $A_w,B_w$  belong to
 ${}^\perp \cC^{\Sigma}$, where $\Sigma = \Int P_{n-1}\cup \lambda_n \cup\Int P_*$ and that 
and 
$\Cone(\bZ_{\{z=\bx_0,s\in K\} }[-2]\to F^1\Phi^K)\in {}^\perp\cC^X$. As was explained in Sec \ref{F1},
the sheaf $F^1\Phi^K$ is supported on $\Sigma':=\Int P_0\cap\lambda_1\cap \int P_*$,
so that it suffices to show that
$$
\Cone(\bZ_{\{z=\bx_0,s\in K\} }[-2]\to F^1\Phi^K)|_{\Sigma'\times \Co}\in {}^\perp\cC^{\Sigma'}
$$
We do it in the rest of the section.

\subsubsection{Conventions} 

Suppose that the ray $\lambda_n$ is directed to the right so that 
$\lambda_n =\hat c(\lambda_n)+\R_{>0}.e^{i\alpha}$; the case of the opposite direction is similar. 


%

\begin{figure} \includegraphics{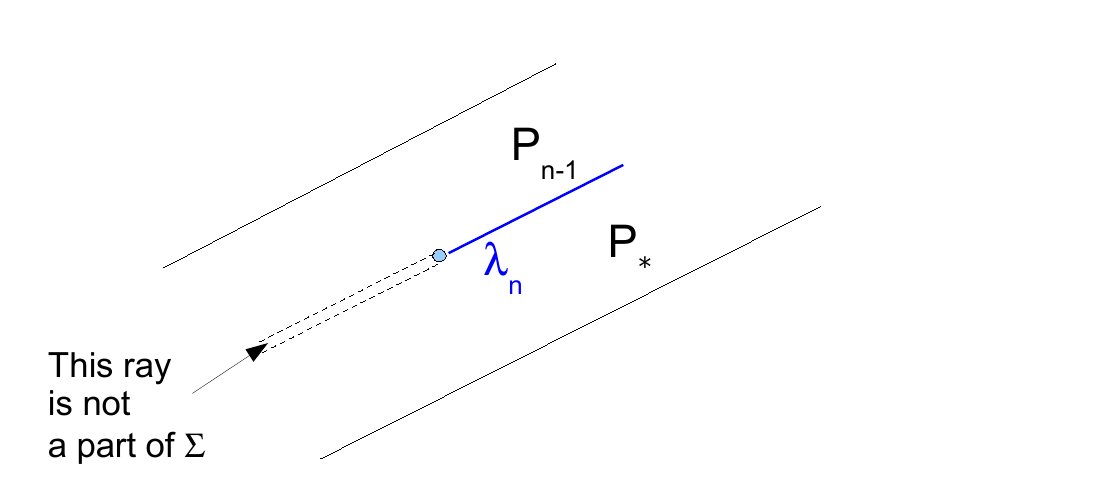} \captionempty \label{CoShWKBp64} \end{figure}

Assume the situation is as on figure \ref{CoShWKBp64}, namely, we assume that $P_{n-1}$ is above $\lambda_n$ and $P_*$ is below $\lambda_n$. The argument for the opposite situation is similar.

Define
$$ U := \{ \hat c(\lambda_n)+xe^{i\alpha}+ye^{-i\alpha} \in  \Sigma \ : \  x,y\in \R \ \text{and} \ x>0\}; $$ 
$$ V:= \{ \hat c(\lambda_n)+xe^{i\alpha}+ye^{-i\alpha} \in  \Sigma \ : \  x,y\in \R \ \text{and} \ x\leq 0 \}. $$

\subsubsection{Orthogonality of $A_w$} \label{OrthoAw}
Because of the assumptions above, we have $w\in \wa_\pravo$ and
$$
A_w=S_w*\Lambda_{P_*'}^{K-},
$$
where
$$
\Lambda_{P_*'}^{K-}=\bZ_{\{(z,s):z\in P_*';s-z\in K\}}.
$$ 
We have a short exact sequence:
\begin{equation}
0\to S_w*\Lambda^{K-}_{U\cap P'_*}\to A_w\to S_w*\Lambda^{K-}_{V\cap P'_*}\to 0, \label{jn3eq4}
\end{equation} 
where $\Lambda^{K\pm}_U:=\bZ_{(s,z)| z\in U; s\pm z\in K\}}$ and similarly for $\Lambda^{K\pm}_{V\cap P'_*}$.

(Note that in the case $\lambda_n\in {\cal L}_\levo$ we need to consider
 a sequence analogous to \eqref{jn3eq4} with $\Lambda^{K-}$ instead of $\Lambda^{K+}$.)

The problem is thus reduced to proving that
\begin{equation} S_w*\Lambda^{K-}_{U\cap P'_*}, \ \ S_w*\Lambda^{K-}_{V\cap P'_*} \ \ \in \
 \ {}^\perp \cC^{\Sigma} \label{jul6eq10}. \end{equation}

Now let us use the following consideration: if $j:U\times \C \to \Sigma\times \C$ is an open
 inclusion and if $F\in {}^\perp \cC^U$, then 
$j_!F \in {}^\perp \cC^\Sigma$ bacause $RHom(j_!F;G)\cong RHom(F;G|_{U\times \C})$. In application 
to the situation at hand, this allows us to reduce \eqref{jul6eq10} to proving
\begin{equation}
S_w*\Lambda^-_{U\cap P'_*}|_{U}\in {}^\perp\cC^U \label{jn30eq401}
\end{equation}
and
\begin{equation}
S_w*\Lambda^-_V|_{P_*}\in {}^\perp \cC^{P_*} \label{jn30eq402}
\end{equation}
which we are going to do using Proposition \ref{main}.

\textsc{Proof of \eqref{jn30eq401}.} 
Denote $F:=S_w*\Lambda^-_{U\cap P'_*}|_{U}$. We have
$F=\Z_S$, where $ S=\{ (z,s): z\in U\cap P_*', s-z\in\hatc(w)+ K\}.$

Next, $U=\{\hatc(\lambda_n)+xe^{i\alpha}+ye^{-i\alpha}|x>0; y\in I\}$, where
$I$ is a generalized open interval containing 0, so that $U$ is a generalized strip and
we can apply Proposition \ref{main}.
 
We have $U\cap P_*'=\{\hatc(\lambda_n)+xe^{i\alpha}+ye^{-i\alpha}|x>0;y\geq 0;y\in I\}$.

  Let us now check that $F$ satisfies all the assumptions of Prop. \ref{main}, which will show
that $F\in {}^\perp\cC^U$.

Namely, we need to show:
 a)  the map $\Z_{\elr_\alpha}*F\to \Z_{\{0\}}*F=F$, induced by the  embedding $0\in \elr_\alpha$,
 is an isomorphism, 

b)  $RP_{+!}F=0$;

c)  $RP_{-!}F=0$.

\underline{Proof of a)} is easy: the word $w$ contains at least one letter, hence $S_w$ is 
a convolution of $\ge 1$ sheaves of the type $\Z_{\{s\in a+K\}}$, $a\in \C$.
 But the map $\beta:\Z_{\elr_\alpha}*\Z_{\{s\in a+K\}} \stackrel{\simeq}{\to}
 \Z_0 * \Z_{\{s\in a+K\}}$, induced by the inclusion $0\in \elr_\alpha$, is an isomorphism. 

\underline{Proof of b)}  It suffices to check that $(RP_{+!}F)_t=0$ for every point $t\in \Co$.
We have $(R^\bullet P_{+!}F)_t=H^\bullet_c(P_+^{-1}t\cap S;\bZ)$. Denote $W_t:=P_+^{-1}t\cap S$.
The space $W_t$ consists of all points $(z,s)$, where $z\in U\cap P_*'$;
$s+z\in K$; $s-z=t$. Since $s=z+t$, we can exclude $s$: the space $W_t$ gets identified
with a  closed subset  $W'_t\subset U$  consisting of all points $z\in U\cap P_*'$ such that $2z+t\in \hatc(w)+K$.
Let us write $\hatc(w)-t-2\hatc(\lambda_n)=2(x_0e^{i\alpha}+y_0e^{-i\alpha})$.
We then see that $W'_t$ consists of all points $\hatc(\lambda_n)+xe^{i\alpha}+ye^{-i\alpha}$,
where $x>0;y\geq 0;y\in I; x\geq x_0;y\geq y_0$.
It is now easy to see that for all $x_0,y_0$, we have $H^\bullet_c(W_t,\bZ)=0$.

\begin{figure} 
\includegraphics{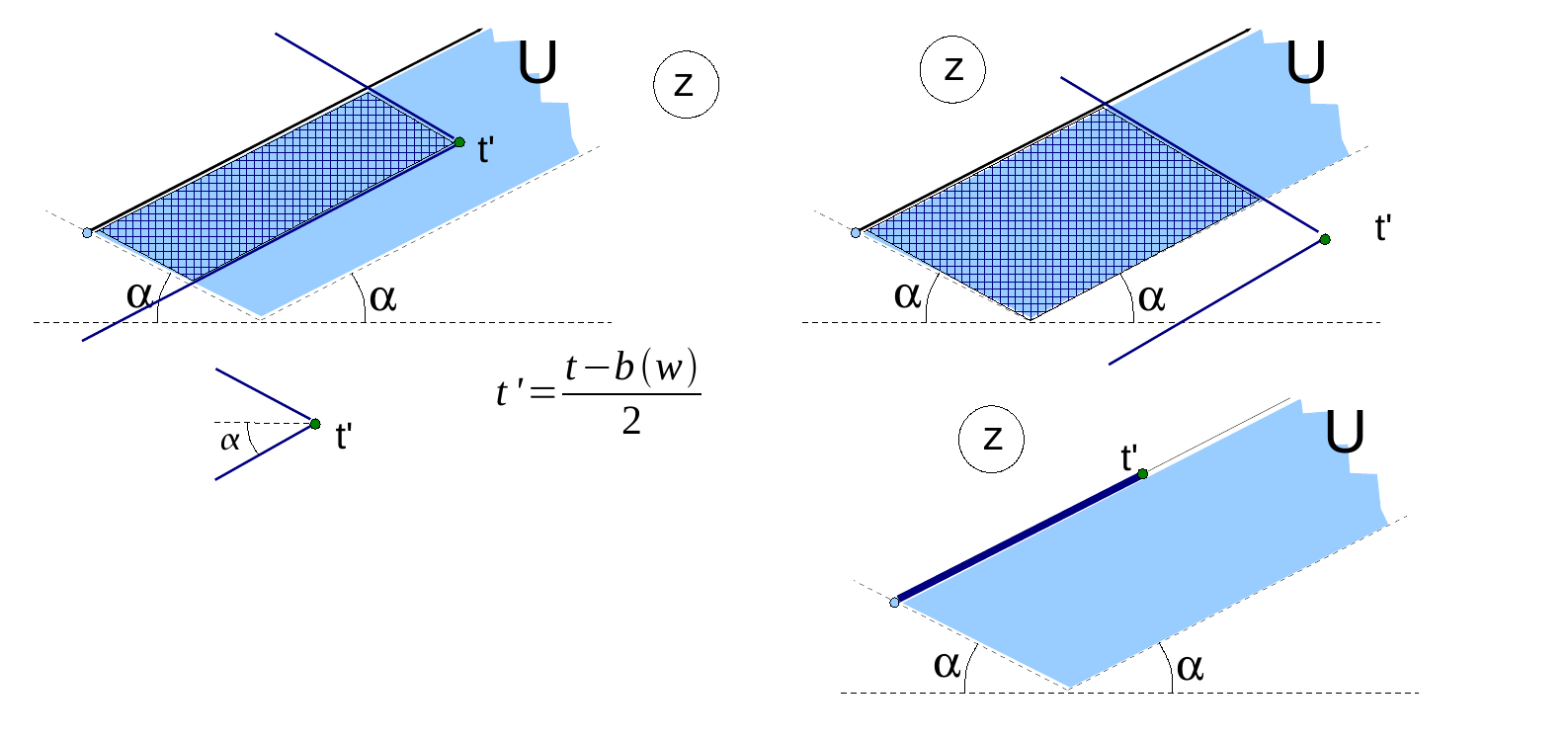}
\caption{Proof of \eqref{jn30eq401}, part b). } \label{CoShWKBp75}
\end{figure}

\underline{Proof of c)} Similar to above, we need to show that $H^\bullet_c(V_t;\bZ)=0$, where 
$V_t=P_-^{-1}t\cap S$, for
all $t\in \Co$. If $t\notin \hatc(w)+K$,  $V_t=\emptyset$. Otherwise,  $V_t$ gets identified with 
$U\cap P'_*$ i.e. the set of all points  $(x,y):x>0;y\geq 0;y\in I$. The statement now follows.

\textsc{Proof of  \eqref{jn30eq402}.}
  Set $G_1:= S_w * \Lambda^-_{V\cap P'_*}$.  We have
$$
V\cap P'_*=\{\hatc(\lambda_n)+xe^{i\alpha}+ye^{-i\alpha}|x\leq 0; y\in I; y> 0\}.
$$
In particular, $V\cap P'_*\subset \Int P_*$.  Similar to above, it suffices to show that
$G:=G_1|_{\Int P_*\times \Co}\in \cC^{\Int P_*}$.
Since $\Int P_*$ is a generalized strip, we can apply Proposition \ref{main}.
Let us check the assumptions of this Proposition.

We have $G=\bZ_T$, where $T\subset \Int P_*\times \Co$ consists of all points
$(z,s)$, where $z=\hat(\lambda_n)+xe^{i\alpha}+ye^{-i\alpha}$; $x\leq 0$; $y<0$; $y\in I$;
$s-z\in {\hat c}(w)+K$.   
 
a) We see that the natural map $\bZ_{\elr_\alpha}*G\to \bZ_0*G=G$ is  clearly an isomorphism.

 b)$RP_{+!}G|_t=0$ for all $t$.  This is equivalent to $H^\bullet_c(W'_t,\bZ)=0$, where
$W'_t=P_+^{-1}t\cap T$.  Similar to above,
the set $W'_t$ gets identified with the set of all $(x,y)$, where $x\leq 0$; $y<0$; $y\in I$;
$x\geq x_0;y\geq y_0$ for some numbers $x_0,y_0$, the statement follows.

c) We need to chech that $H^\bullet_c(V'_t,\bZ)=0$, where $V'_t=P_-^{-1}t\cap T$.
We see that $V'_t=\emptyset$ for all $t\notin \hatc(w)+K$. Otherwise,
$V'_t$ gets identified with $T$.

\subsubsection{Orthogonality of $B_w$}

Let $U,V$ be the same subsets of $\Sigma$ as above.
We see that $\Sigma\backslash U=V=V_1\sqcup V_2$, where $V_1\subset \Int P_*$, $V_2\subset \Int P_{n-1}$.
  
For any locally closed subset $C\subset \Sigma$ we set
$B_C:=B_w\otimes \bZ_{C\times \C_s}\in \DerCat(\Sigma \times \C_s)$. We then have
a distinguished triangle
$$
\stackrel{+1}\to B_{V_1}\oplus B_{V_2}\to B_w\to B_U\stackrel{+1}\to .
$$

Similarly to section \ref{OrthoAw}, it suffices to prove
that
\begin{equation}\label{ortbu}
B'_U:=B_U|_{U\times \C} \in {}^\perp \cC^U;
\end{equation}
\begin{equation}\label{ortbv1}
B_{V_1}|_{\iP_{*}\times \C}\in {}^\perp\cC^{\Int P_{*}},
\end{equation}
\begin{equation}\label{ortbv2}
B_{V_2}|_{\iP_{n-1}\times \C}\in {}^\perp\cC^{\Int P_{n-1}},
\end{equation}

It is clear that  $U$, $V_1$,and $V_2$ are generalized strips so that we can apply Prop. \ref{main}.

{\em Proof of (\ref{ortbu})}
Let $\bP_1:=U\cap P_{n-1}$; $\bP_2:=U\cap P_{*}$ so that $\bP_1,\bP_2\subset U$ are closed subsets
and $\bP_1\cap \bP_2=\lambda_n$. 

As above, we have 
$$
U=\{\hatc(\lambda_n)+xe^{i\alpha}+ye^{-i\alpha}| x>0;y\in I\},
$$
where $I\subset\Re$ is a generalized open interval containing 0.
The subset  $\bP_1$ is given by $y\geq 0$, and $\bP_2$ by $y\leq 0$.

We have identifications
$$
B_1:=B'_U|_{\bP_1\times \Co}=S_w*\Lambda_{\lambda_n}^{K+}\oplus S_{\lambda_nw}*\Lambda_{\bP_1}^{K-};
$$
$$
B_2:=B'_U|_{\bP_2\times \Co}=S_w*\Lambda_{\bP_2}^{K+}\oplus S_{\lambda_n w}*\Lambda_{\bP_2}^{K-}.
$$
Whence induced identifications
\begin{equation}\label{iden1}
B_1|_{\lambda_n\times \Co}=S_w*\Lambda_{\lambda_n}^{K+}\oplus S_{\lambda_nw}*\Lambda_{\lambda_n}^{K-}
\end{equation}
\begin{equation}\label{iden2}
B_2|_{\lambda_n\times \Co}=S_w*\Lambda_{\lambda_n}^{K+}\oplus S_{\lambda_nw}*\Lambda_{\lambda_n}^{K-}
\end{equation}
The gluing map
$$
B_1|_{\lambda_n\times \Co}\to B_2|_{\lambda_n\times \Co}
$$
is induced by $\Gamma^{P_{n-1}P_*}_{\Phi^K}$ and equals 
$$
\Gamma=\Id +n\in \End(S_w*\Lambda_{\lambda_n}^{K+}\oplus S_{\lambda_nw}*\Lambda_{\lambda_n}^{K-}),
$$
where the only non-zero component of $n$
is 
$$
n^{+-}:S_w*\Lambda_{\lambda_n}^{K+}\to S_{w}*S_{\lambda_n}*\Lambda_{\lambda_n}^{K-}=
S_{\lambda_nw}\Lambda_{\lambda_n}^{K-}
$$
is defined by means of the map $\nu^K_{\lambda_n}$ from (\ref{jn29eq142}).

Let $i_k:\bP_k\to U$, $k=1,2$ and $i_0:\lambda_n\to U$ be closed embeddings.
Denote by $\iota_1:i_{1!}B_1\to i_{0!}(S_w*\Lambda_{\lambda_n}^{K+}\oplus 
S_{\lambda_nw}*\Lambda_{\lambda_n}^{K-})$ the natural isomorphism coming from the identification
(\ref{iden1}). Similarly, we have a map  $\iota_2:i_{2!}B_1\to i_{0!}(S_w*\Lambda_{\lambda_n}^{K+}\oplus 
S_{\lambda_nw}*\Lambda_{\lambda_n}^{K-})$, coming from (\ref{iden2}).
We can rewrite the above consideration in terms of the following short exact sequence of sheaves of abelian groups
\begin{equation}\label{seqex}
0\to B'_U\to i_{1!}B_1\oplus i_{2!}B_2\to i_{0!}(S_w*\Lambda_{\lambda_n}^{K+}\oplus
 S_{\lambda_nw}*\Lambda_{\lambda_n}^{K-})\to 0.
\end{equation}

Where the left arrow is induced by the direct sum of the obvious restriction maps and the right arrow
is $-\Gamma\iota_1\oplus \iota_2$. 
Let us denote the components of this map
$$
-\Id:i_{0!}S_w*\Lambda_{\lambda_n}^{K+}\to i_{0!}S_w*\Lambda_{\lambda_n}^{K+};
$$
$$
-\nu:i_{0!}S_w*\Lambda_{\lambda_n}^{K+}\to i_{0!}S_{\lambda_nw}*\Lambda_{\lambda_n}^{K-};
$$
$$
-r_1:i_{1!}S_{\lambda_nw}*\Lambda_{\bP_1}^{K-}\to i_{0!} S_{\lambda_nw}*\Lambda_{\lambda_n}^{K-};
$$
$$
r_2^+:i_{2!}S_w*\Lambda_{\bP_2}^{K+}\to  i_{0!}S_w*\Lambda_{\lambda_n}^{K+};
$$
$$
r_2^-:i_{2!}S_{\lambda_nw}*\Lambda_{\bP_2}^{K-}\to i_{0!} S_{\lambda_nw}*\Lambda_{\lambda_n}^{K-}.
$$

Consider the complex $B''$ composed of the 2 last terms of the sequence (\ref{seqex}), which is quasi-isomorphic
to $B'_U$. This complex has a filtration by the following subcomplexes:

$F^1B''$ is as follows:
 $$i_{0!}S_w*\Lambda_{\lambda_n}^{K+}\stackrel {-\nu}\to i_{0!}S_{\lambda_n w}*\Lambda_{\lambda_n}^{K-}\to 0;
$$
$F^2B''$ is as follows:
$$
i_{0!}S_w*\Lambda_{\lambda_n}^{K+}\oplus i_{2!}S_w*\Lambda_{\bP_2}^{K+}\to
 i_{0!}(S_w*\Lambda_{\lambda_n}^{K+}\oplus
 S_{\lambda_nw}*\Lambda_{\lambda_n}^{K-})\to 0
$$

We finally set $F^3B''=B''$.
The associated graded quotients are as follows:
$F^2/F^1$ equals $\Cone r_2^+[-1]$, which is quasi-isomorphic to $S_w*\Lambda_{\Int P_2}^{K+}$.

$F^3/F^2$ equals 
$$
i_{1!}S_{\lambda_nw}*\Lambda_{\bP_1}^{K-}\oplus i_{2!}S_{\lambda_nw}*\Lambda_{\bP_2}^{K-}.
$$

We will need a one more exact sequence.  We have subsheaves (direct summands)
$$
S_{\lambda_nw}*\Lambda_{\bP_1}^{K-}\subset B_1;\quad S_{\lambda_nw}*\Lambda_{\bP_2}^{K-}\subset B_2.
$$
Since the map $\Gamma$ induces identity on $S_{\lambda_nw}*\Lambda_{\lambda_n}^{K-}$,
the two subsheaves glue into  a subsheaf  $S_{\lambda_nw}*\Lambda_U^{K-}\subset B'_U$.
It is clear that we have a short exact sequence:
\begin{equation}\label{seqdva}
0\to S_{\lambda_nw}*\Lambda_U^{K-}\to B'_U\to i_{2!}S_w*\Lambda_{P_2}^{K+}\to 0.
\end{equation}

Let us now check the conditions of Prop \ref{main}. The isomorphicity of the map
 $\bZ_{\elr_\alpha}*B'_U\to B'_U$ can be checked directly. 

Let us now show that $RP_{+!}B'_U=0$. Because of the exact sequence (\ref{seqdva}),
it suffices to prove that $RP_{+!}S_w*\Lambda_{P_2}^{K+}=0$ and 
$RP_{+!}S_{\lambda_nw}*\Lambda_U^{K-}=0$.  This can be checked pointwise in a way similar to the previous 
subsection.  

Let us now check that $RP_{-!}B'_U=0$. It suffices to show that $RP_{-!}$, when applied to all associated graded
quotients of the filtration $F$ on $B''$, produces zero. The latter can be done pointwise in a way similar to the previous
sections.

Proof of (\ref{ortbv1}), (\ref{ortbv2}) is very similar to the previous subsection.

\subsubsection{Orthogonality of $\Cone(\bZ_{z=\bx_0,s\in K}[-2]\to F^1\Phi^K)$}
{The aim of this subsection is to prove that}
\begin{equation}
{\Cone(\bZ_{z=\bx_0,s\in K}[-2]\to F^1\Phi^K) \in {}^\perp {\cal C}^{\Sigma'}.} \label{i13eq4n}
\end{equation}

We will freely use the notation and the results from  Sec \ref{F1},\ref{factF1}.
As was mentioned above, $\Cone(\bZ_{z=\bx_0,s\in K}[-2]\to F^1\Phi^K)$ is supported
on $\Sigma\times\Co$, where $\Sigma=\Int P_0\cup\lambda_1\cup\Int P_*$.
The restriction $\Cone(\bZ_{z=\bx_0,s\in K}[-2]\to F^1\Phi^K)|_{\Sigma\times \Co}$ is isomorphic to
the Cone of the composition arrow in (\ref{compossec}). 
Denote the cone of the left arrow  in (\ref{compossec})  by $\Gamma_1$
 and the cone of the right arrow by $\Delta$.
Observe that  $\Gamma_1=j_{0!}\Gamma$,
where $\Gamma=\Cone(\iota_L\oplus \iota_R)$; $\Gamma\in D(\Int P_0\times\Co)$.
The problem now reduces to showing 
 that $\Gamma\in {}^\perp\cC^{\Int P_0}$ and $\Delta\in {}^\perp\cC^{\Sigma}$.

Denote  $A_L:=\Coker i_A$; $B_R:=\Coker i_B$. Observe that $A_L$ is of the form $A_w$ with $w=L$,
and $B_R$ is of the form $B_w$ with $w=R$, where $A_w,B_w$ are as defined in Sec \ref{oct21nn}.
It is also clear that $\Delta\cong A_L\oplus B_R$. As follows from the previous two subsections,
$A_L,B_R\in {}^\perp\cC^{\Sigma}$, hence, same is true for $\Delta$. 
Let us now  show that   $\Gamma\in  {}^\perp\cC^{\Int P_0}$.

By Prop.\ref{main}, it suffices to check statements a),b),c) below:

\underline{a)}  $\Gamma*\Z_{\{s\in e^{i\alpha}\R_{\ge 0} \}}\to \Gamma$ is an isomorphism:
it suffices to check that a similar map applied to each of  $\bZ_{\bx_0\times K}[-2]$, $S_L*\Lambda^{K+}_{\Int P_0}$,
and $S_R*\Lambda^{K-}_{\Int P_0}$ is an isomorphism, which is straightforward.

\underline{b)} $RP_{+!}\Gamma=0$. 
 It is enough  to check $RP_{+!}{\cal G}_k=0$, $k=1,2$,  
 where $$ {\cal G}_1= S_R * \Lambda^-_{\iP_0}=\Z_{\{ (z,s) : z\in \iP_0, \, s-z\in -\bx_0+K \}},$$
 ${\cal G}_2 =Cone(\Z_{\bx_0\times K}[-2] \to S_L * \Lambda^+_{\iP_0}$ and where
 $$S_L * \Lambda^+_{\iP_0}=\Z_{\{ (z,s) : z\in \iP_0, \, s+z\in \bx_0+K \}}.
$$

\begin{figure} \includegraphics{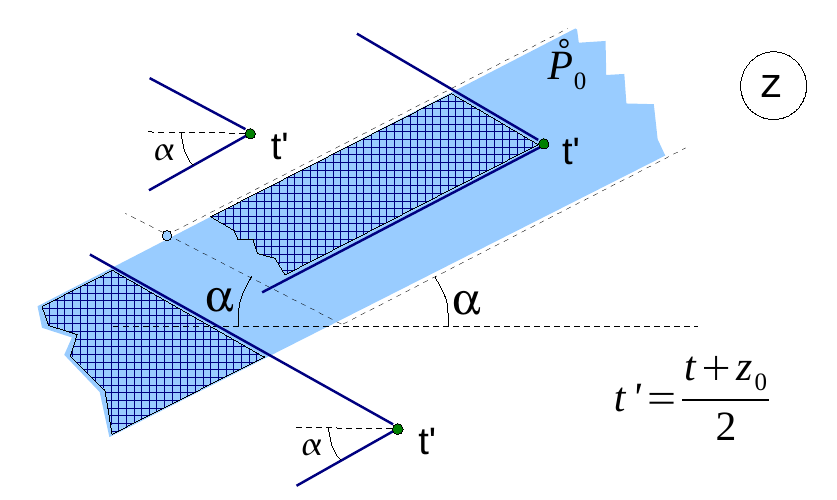} \caption{Proof of \eqref{i13eq4n}, Step b-i)} \label{CoShWKBp74} \end{figure}

\underline{b-i)} $RP_{+!}{\cal G}_1=0$. Indeed, by the base change, let us pass to the fiber of $P_+$ over $t\in \C$ and calculate $R\Gamma_c(\Z_{W_1})$ where $W_1=\{ (z,s)\in \C: \, z\in \iP_0, \, s-z\in -\bx_0+ K \ z+s=t \}$. 
Eliminating $s$ makes $W_1=\{ z\in \C : z\in \iP_0,\, z\in \frac{t+\bx_0}{2}-K \}$. For different values of $t$ this set is sketched on fig. \ref{CoShWKBp74}.
\\
Thus, $W_1$ is either empty or homeomorphic to a closed half-plane, so the
 result follows. 
 
\underline{b-ii)} $RP_{+!}{\cal G}_2=0$. Indeed, by the base change, let us pass to the fiber of $P_+$ over $t\in \C$ and calculate $R\Gamma_c(\Z_{W'_2})[-2]\to R\Gamma_c(\Z_{W_2})$, where $W'_2=\{ (z,s)\in \C: \, z=\bx_0, \, s\in K \ z+s=t \}$,  $W_2=\{ (z,s)\in \C: \, z\in \iP_0, \, s+z\in \bx_0+ K \ z+s=t \}$. 
Eliminating $s$ makes 
$$ \begin{array}{ccc} \text{if} \ t-\bx_0\in K:  & W'_2=\{\bx_0\} & W_2=\{ z\in \C : z\in \iP_0 \} \\  
\text{otherwise:}   & W'_2=\emptyset & W_2=\emptyset \end{array} $$
and the map $R\Gamma_c(\Z_{W'_2})[-2]\to R\Gamma_c(\Z_{W_2})$ is the obvious quasi-isomorphism. 

\underline{c)} $RP_{-!}\Gamma=0$. This can be shown similarly to $RP_{+!}\Gamma=0$.

\section{Identification of $\Phi^K$ and $\Psi^K$} \label{psiphi}

We are going to construct an identification as in (\ref{opredipsiphi}).
Namely, we will construct a map
$$
I_{\Psi\Phi}:\Psi^K\to \Phi^K
$$
such that \begin{equation}\label{soglas}
i_\Phi=I_{\Psi\Phi}i_\Psi,
\end{equation}
 where $i_\Phi:\cF_0^K\to \Phi^K$
is the map (\ref{intophik}) and $i_\Psi:\cF_0^K\to \Psi^K$ is the map (\ref{intopsik}).

The goal of this section is to give an explicit desciption of  $I_{\Psi\Phi}$.
This can be done as follows.
Let $P$ be a closed $\alpha$-strip. Let $\Pi$ be a closed  $(-\alpha)$-strip such that $P\cap \Pi\neq \emptyset$.
We then have identifications
$$
\iota_{\Phi P}|_{(\Pi\cap P)\times \Co}:\Lambda^+*S_+\oplus\Lambda^-*S_-|_{(\Pi\cap P)\times \Co}
=(\Phi^K|_{P\times \Co})|_{(\Pi\cap P)\times \Co}
=\Phi^K|_{(\Pi\cap P)\times \C}$$
$$
\iota_{\Psi \Pi}|_{(\Pi\cap P)\times \Co}:\Lambda^+*S_+\oplus\Lambda^-*S_-|_{(\Pi\cap P)\times \Co}=
(\Psi^K|_{\Pi\times \Co})|_{(\Pi\cap P)\times \Co}
=\Psi^K|_{(\Pi\cap P)\times \C}$$

meaning that the restriction $I_{\Psi\Phi}|_{(\Pi\cap P)\times \Co}$ can be
 rendered as an automorphism
$
J_{\Pi P}
$
of\\ 
$\Lambda^{K+}*S_+\oplus\Lambda^{K-}*S_-|_{(\Pi\cap P)\times \Co}$
 in the abelian category of sheaves on $(\Pi\cap P)\times \Co$, so that we have:
\begin{equation}\label{ippk1}
I_{\Psi\Phi}|_{(\Pi\cap P)\times \Co}=
\iota_{\Phi P}|_{(\Pi\cap P)\times\Co}J_{\Pi P}\iota^{-1}_{\Psi \Pi}|_{(\Pi\cap P)\times \Co}.
\end{equation}
 
We are now motivated for the next subsection.
\subsection{Endomorphisms of $\Lambda^{K+}*S_+\oplus\Lambda^{K-}*S_-|_{(P\cap \Pi)\times \Co}$}\label{pipa}
We will do the study in a slightly more general context. Let $Y$ be a  locally closed connected subset of $\Co$.
For a $c\in \Co$, set $$A_c^{\pm}:=\{(x,s)|s\pm x\in c+K\}\subset Y\times \Co.$$
Let $W^\pm$ be sets; set $W:=W^+\sqcup W^-$. Let $\bc_W:W\to\Co$ be   a function.  Let $w\in W_+$. Set
$A_w:=A_{\bc(w)}^+$. For  $w\in W_-$ we set $A_w:=A_{\bc(w)}^-.$
Define the following sheaves on $Y\times \Co$:  
$$
S_{W}:=\bigoplus\limits_{w\in W} \bZ_{A_{w}}.
$$

Let $\bc_i:W_i\to \Co$; $W_i=W_i^+\sqcup W_i^-$, $i=1,2$; $\bc_{W_i}:W_i\to \Co$; and let us study a group
$
\hom_{Y\times \Co}(S_{W_1};S_{W_2}).
$

We have
\begin{equation}\label{produxt}
\hom_{Y\times \Co}(S_{W_1};S_{W_2})
\stackrel\sim\to\prod_{w_1\in W_1}\hom_{Y\times \Co}(\bZ_{A_{w_1}};S_{W_2})
\end{equation}

Let us focus on $\hom_{Y\times \Co}(\bZ_{A_{w_1}};S_{W_2})$.
We have an embedding
$S_{W_2}\hookrightarrow\prod_{w_2\in W_2} \bZ_{A_{w_2}}$ which induces an embedding
$$
\iota:\hom_{Y\times \Co}(\bZ_{A_{w_1}};S_{W_2})\hookrightarrow
 \hom_{Y\times \Co}(\bZ_{A_{w_1}};\prod_{w_2\in W_2} \bZ_{A_{w_2}})
$$
\begin{equation}\label{sumprod}
=\prod_{w_2\in W_2}\hom_{Y\times \Co}(\bZ_{A_{w_1}};\bZ_{A_{w_2}}).
\end{equation}
Let us now compute
 $$\hom_{Y\times \Co}(\bZ_{A_{w_1}};\bZ_{A_{w_2}})=H^0(A_{w_2};A_{w_2}\backslash A_{w_1}).
$$
We have a homeomorphism $A_{w_2}\cong Y\times K$ so that $A_{w_2}$ is connected and
$
H^0(A_{w_2};A_{w_2}\backslash A_{w_1})
$
is zero unless $A_{w_2}\backslash A_{w_1}$ is empty, in which case it equals $\bZ$.
In other words, we have an isomorphism
$
\ve_{w_1w_2}:\bZ\stackrel\sim\to \hom_{Y\times \Co}(\bZ_{A_{w_1}};\bZ_{A_{w_2}})
$
if $A_{w_2}\subset A_{w_1}$; otherwise, $\hom_{Y\times \Co}(\bZ_{A_{w_1}};\bZ_{A_{w_2}})=0$.
Set $e_{w_1w_2}:=\ve_{w_1w_2}(1)$.  

Every element 
$$
\nu\in \prod_{w_2\in W_2}\hom_{Y\times \Co}(\bZ_{A_{w_1}};\bZ_{A_{w_2}})
$$
can be uniquely written as
$$
\sum\limits_{w_2} \nu_{w_1w_2}e_{w_1w_2},
$$
where the sum is taken over all $w_2$ such that $A_{w_2}\subset A_{w_1}$ and
$\nu_{w_1w_2}$ are arbitrary integers. 

\begin{Claim} 
The element  $\nu$  lies in the image of (\ref{sumprod}) iff
for every compact subset $L\subset A_{w_1}$:
\begin{equation}\text{there are only finitely many $w_2$ such 
that $\nu_{w_2w_1}=0$ and $A_{w_2}\cap L\neq 0$.}\label{condL}
\end{equation}
\end{Claim}
\textsc{proof}  We will use the following notation. For every $w\in W_1$ or $w\in W_2$, let us denote by
 $\bfun_{w}\in \Gamma(Y\times\Co;\bZ_{A_{w}})$  the canonical section, such that
for every $y\in Y\times\Co$, the stalk $(\bfun_w)_y$ generates the group $(\bZ_{A_{w}})_y$,
which is equal to $\bZ$ if $y\in A_{w}$ and to zero otherwise.

We have
 $$\nu(\bfun_{w_1})=\sum\limits_{w_2\in W_2} n_{w_2w_1} \bfun_{w_2}\in 
\Gamma(Y\times\Co;\prod\limits_{w_2\in W_2} \bZ_{A_{w_2}}).$$

Let us now suppose that $\nu$ lies in the image of (\ref{sumprod}). This implies that
the restriction $\nu(\bfun_{w_1})|_{L}\in \Gamma(L;\bigoplus\limits_{w_2\in W_2}\bZ_{A_{w_2}}).$
Since $L$ is compact, we have an isomorphism
$$
\bigoplus\limits_{w_2\in W_2}\Gamma(L; \bZ_{A_{w_2}})\to 
 \Gamma(L;\bigoplus\limits_{w_2\in W_2}\bZ_{A_{w_2}}).
$$
Given a section  $\sigma\in \Gamma(L;\bigoplus\limits_{w_2\in W_2}\bZ_{A_{w_2}})$,
denote by $\sigma_{w_2}\in \Gamma(L;\bZ_{A_{w_2}})$ the corresponding component of $\sigma$.
We have: $\sigma_{w_2}=0$ for almost all $w_2\in W_2$.
We have $\nu(\bfun_{w_1})_{w_2}=n_{w_2w_1}\bfun_{w_2}|_L.$ The element on the RHS does not vanish
iff $n_{w_2w_1}\neq 0$ and  $L\cap A_{w_2}\neq \emptyset$, which implies the statement.

Conversely, let us assume that for any $L$ there only are finitely many $w_2\in W_2$ such that $n_{w_2w_1}\neq 0$ and
 $L\cap A_{w_2}\neq \emptyset.$  It suffices to show that 
$$\nu(\bfun_{w_1})\in \Gamma(Y\times \Co;\bigoplus\limits_{w_2\in W_2}\bZ_{A_{w_2}})\subset
\Gamma(Y\times \Co;\prod\limits_{w_2\in W_2} \bZ_{A_{w_2}}).
$$
Let us choose an open covering of $Y\times \Co$ by precompact sets $U_a$ (i.e. the closure $L_\alpha$ of each
$U_a$ in $Y\times \Co$ must be compact).
It suffices to show that $\nu(\bfun_{w_1})\in \Gamma(U_a;\bigoplus\limits_{w_2\in W_2}\bZ_{A_{w_2}})$
for each $U_a$.  Then it suffices to show
that $\nu(\bfun_{w_1})\in \Gamma(L_a;\bigoplus\limits_{w_2\in W_2}\bZ_{A_{w_2}}).$
In fact, $\nu(\bfun_{w_1})\in \Gamma(L_a;\prod\limits_{w_2\in W'_2}\bZ_{A_{w_2}})$,
where $W'_2$ consists of all $w_2$  satisfying $n_{w_2w_1}\neq 0$, $A_{w_2}\cap L_a\neq 0$,
which is finite, whence the statement.
$\Box$.

As follows from the proof of  the Claim, $\nu$ belongs to the image of (\ref{sumprod}) iff the condition
(\ref{condL}) is satisfied for a family of compact sets $L_a$ whose interiors cover $X\times\Co$.

\begin{Prop} \label{PiPHom}  Elements from
$\hom_{X\times \Co}(S_{W_1};S_{W_2})$  are in 1-to-1 correspondence
with
 the  sums $$ \sum_{w_1\in W_1,w_2\in W_2, A_{w_2}\subset A_{w_1}} n_{w_1w_2} e_{w_1w_2} , $$
 satisfying:

there exists a family of compact subsets $L_a\subset X\times \Co$ such that the sets $\Int L_a$ cover
$X\times \Co$, and:
given a $ w_1\in W_1$ and any $L_a$, there are only 
finitely many $w_2\in W_2$ such that $n_{w_1w_2}\ne 0$  and
$L_a\cap A_{w_2}\neq \emptyset$.
\end{Prop}

\subsubsection{Filtration on $\hom_{Y\times \Co}(S_{W_1};S_{W_2})$ }\label{filr}

 Let $\ve\in K$. Let $T_\ve:Y\times \Co\to Y\times \Co$ be the shift  $(x,s)\mapsto (x,s+\ve)$.
We have $T_\ve(A_c)\subset A_c$, for every $\varepsilon\in K$, whence    an induced map
$$
\tau_\ve:\bZ_{A_c}\to T_{\ve!}\bZ_{A_c}=\bZ_{T_\ve(A_c)}.
$$
These maps give rise to a map
$$
\tau_\ve:S_{W_1}\to T_{\ve!}S_{W_1}.
$$

It is easy to see that $T_{\ve!}S_{W_1}=S_{W'_1}$, where $W'_1=W_1$ and $\bc_{W'_1}=\bc_{W_1}+\ve$,
so that Proposition \ref{PiPHom} applies to $T_{\ve!}S_{W_1}$.

We say 
 that $f\in F^\varepsilon \hom_{X\times \Co}(S_{W_1};S_{W_2})$
 if $f$ factors as 
$ f=g\tau_\ve$ for some $g:T_{\ve!}S_{W_1}\to S_{W_2}$.
Using Proposition \ref{PiPHom}, one can check that such a $g$ is unique, if exists.

We write $f\equiv f'\mod F^\ve$ if $f-f'\in F^\ve\hom(S_{W_1},S_{W_2}).$

We also write 
$
f\equiv f'
$
if $f\equiv f'\mod F^\ve$ for some $\ve\in \Int K$.

Let us prove that the filtration $F$ is complete in the following sense. 
Let $f_n\in \hom(S_{W_1};S_{W_2})$ be a sequence of  homomorphisms. 
Let us call $f_n$ {\em a Cauchy} sequence if:
$$
\forall \ve\in K \  \exists N(\ve): \forall n,m\geq N(\ve): f_n\equiv f_m\mod F^{\ve}.
$$
 
We say that {\it $f_n$ converges to $f$} if
$$
\forall \ve\in K  \ \exists N(\ve): \forall n\geq N(\ve): f\equiv f_n\mod F^{\ve}.
$$

\begin{Claim} Every Cauchy sequence $f_n$ converges  to a unique limit $f$.
\end{Claim}

\textsc{Proof}.
Let us first construct $f$.
Decompose
$ f_n = \sum_{w_1,w_2\in W} (f_n)_{w_1w_2}e_{w_1w_2}.$
Let $y\in X\times \Co$ and let $n,m\geq N(\ve)$.  Since $f_n-f_m$ passes through $\tau_\ve$, we deduce that
$(f_n)_{w_1w_2}-(f_m)_{w_1w_2}\neq 0$ only if 
 $A_{w_2} \subset T_{\ve}A_{w_1}$. For every $w_1,w_2$ there exists $\ve_{w_1w_2}$
such that  this condition is violated, meaning that for $n,m\geq N(\ve_{w_1w_2})$,
$(f_n)_{w_1w_2}=(f_m)_{w_1w_2}=:f_{w_1w_2}$.

The data $f_{w_1w_2}$ define a homomorphism $f$ by virtue of Proposition \ref{PiPHom}.
If $f'$ is another limit, it follows that $f-f'\equiv F^\ve$ for all $\ve$ which implies
$f_{w_1w_2}=f'_{w_1w_2}$ for all $w_1,w_2$, that is $f=f'$.
$\Box$

\vskip2.5pc

In particular, let $\gamma \in \End(A_{W})$, $\gamma=Id +n$ and assume that for some $k>0$,
$n^k\in F^\varepsilon$ for 
some $\varepsilon\in \Int K$ ,then $\gamma$ is invertible, and we can set
 $\gamma^{-1} = Id - n + n^2 - n^3 +... $ (the sequence of partial sums of this series is Cauchy).

We conclude with several Lemmas for the future use.
\subsubsection{Lemma on composition}  As above, let $P$ be an $\alpha$-strip and let
$\Pi$ be a $-\alpha$-strip.
Let $Y=\Pi\cap P$ and supose  $Y$ is   a bounded subset of $\Co$, so  that the  closure of $Y$
 is a parallelogram; let us denote
its vertices $ABCD$ so that $AC$ is one of the two diagonals and $\vec{AC}\in K$.
It then follows that   the closure of $P\cap \Pi$ equals $A+K\cap C-K$. Denote $\ve:=\vec{AC}$.
\begin{Lemma}\label{plusminus} Let $W_{1}^-=W_2^+=\emptyset.$
And 
let $f:S_{W_1}\to S_{W_2}
 $
and 
$g:S_{W_2}\to S_{W_1}.
 $
Then  $gf\equiv 0\mod  F^{2\ve}$ 
and $fg\equiv 0\mod F^{2\ve}.$
\end{Lemma}
\textsc{Proof.}

Let $f_{w_1w_2}e_{w_1w_2}$, $g_{w_2w_1}e_{w_2w_1}$ be components of $f,g$.

Let us consider the compositions $
f_{w_1w_2}e_{w_1w_2} g_{w'_2w_1}e_{w'_2w_1}
$
In order for this composition to be non-zero, there should be
$$
A_{w_2}\subset A_{w_1}\subset A_{w'_2}.
$$ 
Or, for every $z\in P\cap \Pi$ and $s\in \Co$ we should have the following  implications:
$$
 s-z\in \bc_{W_2}(w_2)+K\Rightarrow s+z\in \bc_{W_1}(w_1)+K\Rightarrow  s-z\in \bc_{W_2}(w'_2)+K.
$$
Set $\vs:=s-z-\bc_{w_2}$. The first implication then reads as:
$$
\vs\in K\Rightarrow \vs+2z+\bc_{W_2}(w_2)-\bc_{W_1}(w_1)\in K
$$
or, equivalently, $2A+\bc_{W_2}(w_2)-\bc_{W_1}(w_1)\in K$. 
Similarly, the second implication can be rewritten as
$
-2C+\bc_{W_1}(w_1)-\bc_{W_2}(w_2')\in K.
$
Adding the two conditions yields $-2\ve+\bc_{W_2}(w_2)-\bc_{W_2}(w'_2)\in K$; 
$\bc_{W_2}-\bc_{W_2}(w'_2)\in 2\ve+K$.
This implies that 
$$f_{w_1w_2}e_{w_1w_2} g_{w'_2w_1}e_{w'_2w_1}:\bZ_{A_{w'_2}}\to \bZ_{A_{w_2}}$$
 passes through $\tau_{2\ve}:\bZ_{A_{w'_2}}\to T_{2\ve!}\bZ_{A_{w_2'}}$, which implies the
statement for $fg$. Proof for $gf$ is similar. $\Box$.

\bigskip

Let us keep the assumption $W_1=W_1^+$, $W_2=W_2^-$ and consider now the case
when  $X=\Pi\cap P$ is not bounded.  Then at least one of the following is true: 

i) there is no $A\in \Co$ such that $X\subset A+K$;

ii) there is no $C\in \Co$ such that $X\subset C-K$.

\begin{Lemma} \label{obnul} Let us keep the same notation as in the previous Lemma.
 In the case i)  we have $\hom(S_{W_1};S_{W_2})=0$. In the case ii) we have 
$\hom(S_{W_2};S_{W_1})=0$.
\end{Lemma}
 \textsc{Proof.}
In Case i), given $w_1\in W_1$ and $w_2\in W_2$, it is impossible that $A_{w_2}\subset A_{w_1}$,
 And similarly for the Case ii).
$\Box$

\subsubsection{Lemma on extension}\label{exten}

We keep the same assumptions on $W_1,W_2$, namely,
$$ W_1=W_1^+, \ \ \ W_2=W_2^-.$$ 

Let $Y$ be a locally closed non-empty
connected subset of $\Co$. Let $Y+K$ (resp.  $ Y-K$) be the arithmetic sum (resp. difference) of $Y$ and $K$. 
Let $Y_+$, $Y_-$ be connected locally closed subsets
satisfying $Y\subset Y_+\subset Y+K$; $Y\subset Y_-\subset Y-K$. Let $Z$ be an arbitrary connected
 locally closed  subset $\Co$
containing $Y$.
\begin{Lemma} \label{HWNpg77} 
1) The restriction maps
$$
\hom_{Y_+}(S_{W_1^+};S_{W_2^-})\to \hom_{Y}(S_{W_1^+};S_{W_2^-});
$$
$$
\hom_{Y_-}(S_{W_2^-};S_{W_1^+})\to \hom_{Y}(S_{W_2^-};S_{W_1^+})
$$
are isomorphisms;

2) the restriction maps 
$$
\hom_{Z}(S_{W_1^+};S_{W_2^+})\to \hom_{Y}(S_{W_1^+};S_{W_2^+});
$$
$$
\hom_{Z}(S_{W_2^-};S_{W_1^-})\to \hom_{Y}(S_{W_2^-};S_{W_1^-})
$$
are isomorphisms.
\end{Lemma}
\textsc{Proof.} 1) Follows from Proposition \ref{PiPHom}: the inclusion $A_{w_2}\subset A_{w_1}$, $w_i\in W_i$
occurs on $Y_+\times\Co$ iff it occurs on $Y\times \Co$, and similar for the inclusion
$A_{w_1}\subset A_{w_2}$ on $Y_-\times \Co$.

2) Follows from Proposition \ref{PiPHom} in a similar way.

$\Box$
\subsubsection{Decomposition Lemma}\label{decomp}

Let now  $Y:=\ell:=c+(0,\infty).e^{i\alpha}$ be a ray which goes to the right. 
Let $a\in \Co$. We have natural maps
$\lambda^{+}_a:\bZ_{A^+_a}\to \bZ_{A^-_{-2c+a}}$;
$\lambda^-_a:\bZ_{A^-_a}\to \bZ_{A^+_{2c+a}}$;
 coming from  the inclusions
of the corresponding sets.
\begin{Lemma}
Let $f:\bZ_{A^+_a}\to S_{W_2}$, $g:\bZ_{A^-_a}\to S_{W_1}$ be a map of sheaves. 
Then $f$  and $g$ can be uniquely factored
as $f=f'\lambda^+_a$;  $g=g'\lambda^-_a$.
\end{Lemma}
 
\textsc{Proof.}  
Let $w\in W_2$. 
A simple analysis shows that $A_a^+\subset A_w$ is equivalent to $A_{-2c+a}^-\supset A_w$.
Proposition \ref{PiPHom} now implies the factorization of $f$. The factorization of $g$ can be proven
similarly.
$\Box$

\subsection{Restriction $\Phi^K|_\Pi$}

 As above, let $\Pi$ be a closed  $(-\alpha)$-strip.

 The goal of this subsection is to construct
an isomorphism
\begin{equation}
\phi_\Pi:(\Lambda^{K+}*S_+\oplus \Lambda^{K-}*S_-)|_{\Pi\times \Co}
\stackrel\sim\to \Phi^K|_{\Pi\times\Co}. \label{se16e556}
\end{equation}



%

Denote by 
$$
\phi_\Pi^\pm:\Lambda^{K\pm}*S_\pm|_{\Pi\times \Co}\to \Phi^K|_{\Pi\times \Co}
$$
the components.

\subsubsection{Notation}
\begin{figure}  \includegraphics{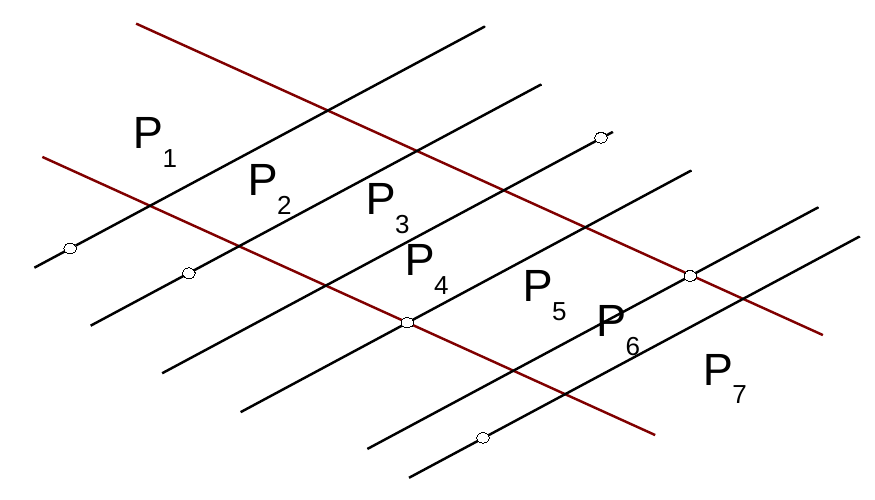} 
\captionempty 
\label{CoShHp10}
\end{figure}

Let us number all $\alpha$-strips that intersect $\Pi$ as $P_1,P_2,...,P_n$
 (there are only finitely many such stripes, Sec \ref{fas}) as shown on the picture 
\ref{CoShHp10}
so that  we number the strips from the left to the right.  The strips $P_1$ and $P_n$ are necessarily half planes.

\subsubsection{Prescription of $\phi_\Pi^+|_{(\Pi\cap P_1)\times \Co}$}
We have an identification
$$
\Phi^K|_{\Pi\cap P_1}=(\Phi^K|_{P_1})|_{(\Pi\cap P_1)\times \Co}=
(\Lambda^{K+}*S_+\oplus \Lambda^{K-}*S_-)|_{(\Pi\cap P_1)\times \Co}.
$$
This identification  gives rise to a map (embedding onto a direct summand):
$$
\Lambda^{K+}*S_+\to \Phi^K|_{(\Pi\cap P_1)\times \Co}.
$$
We assign $\phi_\Pi^+|_{(\Pi\cap P_1)\times \Co}$ to be this map.

\noindent {\bf  Remark.} {In the section \ref{jan18s2} we will inductively extend this definition to the whole $\Pi\times \C$. Construction of $\phi^-_\Pi$ will be performed in section \ref{jan18s1}. An attempt to construct $\phi^-_\Pi$  {\it starting from a prescribed map on $(\Pi\cap P_1)\times \C$} fails. }

\subsubsection{Extension of $\phi^+_\Pi$ to $\Pi{\times \C}$} \label{jan18s2}
For a subset $A\subset \Co$, set $\un{A}:=(\Pi\cap A)\times \Co\subset \Pi\times \Co$.

Let us define $\phi^+_\Pi$ by 
constructing maps
$$ j^+_k: \ \Lambda^{K+}* S_+|_{\un{P_k}} \to \Phi^{{K}}|_{\un{P_k}},$$
which agree on the intersections:
\begin{equation}\label{sklei}
j^+_{k+1}|_{\un{P_k\cap P_{k+1}}}=j^+_{k}|_{\un{P_k\cap P_{k+1}}}.
\end{equation}

We have identifications 
\begin{equation}\label{iottk}
\iota_k:\Lambda^{K+}*S_+\oplus \Lambda^{K-}*S_-|_{\un{P_k}} \to
(\Phi^K|_{P_k\times \Co})|_{\un{P_k}}=\Phi^{K}|_{\un{P_k}}
\end{equation}
coming from the gluing construction of $\Phi_K$.

We have
$$
\iota_{k}|_{\un{P_k\cap P_{k+1}}}=
\iota_{k+1}|_{\un{P_{k}\cap P_{k+1}}}\circ \Gamma^{P_{k}P_{k+1}}_{\Phi^K},
$$
where 
$
\Gamma^{P_{k}P_{k+1}}_{\Phi^K}
$
is as in (\ref{oct3e3a}).

We can now  prescribe $j_k^+$ in the following form:
$
j_k^+=\iota_k\circ i_k^+
$
where 
$$
i_k^+:\Lambda^{K+}*S_+|_{\un{ P_k}}\to
 (\Lambda^{K+}*S_+\oplus\Lambda^{K-}*S_-)|_{\un{P_k}}.
$$
The agreement conditions (\ref{sklei}) now read as:
\begin{equation}
\label{sklei2}
i_{k+1}^+|_{\un{P_k\cap P_{k+1}}}=
\Gamma^{P_{k}P_{k+1}}_{\Phi^K}i_k^+|_{\un{P_k\cap P_{k+1}}}.
\end{equation}

The assignment from the previous subsection means that $i_1^+$
is the identity embedding onto the direct summand.
Let us construct the remaining maps $i_k$ inductively.   Suppose $i_k$ has been already defined.
According to Claim (\ref{HWNpg77}),  the map
$ \Gamma^{P_{k}P_{k+1}}_{\Phi^K}i_k^+|_{\un{P_k\cap P_{k+1}}}$
extends uniquely to $\un{P_{k+1}}$ by Claim \ref{HWNpg77} \\ 
(this is where the choice of $+$ sign is crucial).
We assign $i_{k+1}^+$ to be this map.
It is clear that thus defined map $i_{k+1}^+$ satisfies (\ref{sklei2}) so that the maps $j_{k+1}^+$
give rise to a well defined map
$\phi_\Pi^+$, as we wanted.

Let us denote by $i_k^{++}:\Lambda^{K+}*S_+|_{\un{P_k}}\to \Lambda^{K+}*S_+|_{\un{P_k}};$
$i_k^{+-}:\Lambda^{K+}*S_+|_{\un{P_k}}\to \Lambda^{K-}*S_-|_{\un{P_k}};$
the components of the map $i_k^+$.

\subsubsection{Estimate} For $k=2,...,n-1$, denote by $\ve_k$ the diagonal vector of the parallellogram $P_k\cap \Pi$
such that $\ve_k\in \Int K$ (there is a unique such a diagonal vector). 
Let $\ve_\Pi\in \Int K$ be a vector such that $\ve_k\in \ve_\Pi+K$ for all $k$.

The following Claim can be now proved by a direct computation. 
\begin{Claim} \label{ocenka} 
1) $i^{++}_k \equiv 1 \mod F^{\varepsilon_{{\Pi}}}$  for all $k=1,...,n$.\\
2) Let ${\cal R}_\Pi \subset \{ 1,2,...,n{-1}\}$ consist of all $k$ s.th. $P_k\cap P_{k+1}$ goes to the right.
 We then have a transform
$$ \Gamma^{P_{k}P_{k+1}}_{+-} \ : \ \Lambda^{K+} * S_+ |_{\un{P_k\cap P_{k+1}}}
 \to \Lambda^{K-}* S_- |_{\un{P_k\cap P_{k+1}}}, $$
where $ \Gamma^{P_{k}P_{k+1}}_{+-}$ is the corresponding 
component of $\Gamma^{P_{k}P_{k+1}}_{\Phi^K}.$,
which extends uniquely to $\un{ P_{k+1} \cup ... \cup P_n}$.  $\Gamma^{P_kP_{k+1}}_{+-}$
is the same as $N_\ell^K$, where $\ell=P_k\cap P_{k+1}$ from (\ref{nell}).

We then have:
\begin{equation}\label{formocenka}
i^{+-}_k \equiv - \sum_{k'\in {\cal R}_\Pi; \ k'<k} \Gamma^{P_{k'}P_{k'+1}}_{+-} \ \mod F^{\varepsilon_\Pi}. 
\end{equation}
\end{Claim}

\subsubsection{Construction of $\phi_\Pi^-$} \label{jan18s1} The map $\phi_\Pi^-$ is constructed in a fairly similar
way (the major difference is that we need to start the construction from $\un{P_n}$ and then
continue to the left until we reach $\un{P_1}$.

Similar to above, we define  $\phi_\Pi^-$  in terms of the restrictions to $\un{P_k}$:
$$
\phi_\Pi^-|_{\underline{P_k}}=\iota_k\circ i_k^-,
$$
where $\iota_k$ is the same as above, see (\ref{iottk}), and
$$
i_k^-:\Lambda^{K-}*S_-|_{\un{P_k}}\to \Lambda^{K+}*S_+\oplus \Lambda^{K-}*S_-|_{\un{P_k}}.
$$
We have the following analogue of Claim \ref{ocenka}.
\begin{Claim} \label{ocenka1} Let $\varepsilon_\Pi\in Int \ K$ be as in Claim \ref{ocenka}. We have
1) $i^{--}_k \equiv 1 \mod F^{\varepsilon_{{\Pi}}}$  for all $k=1,...,n$.\\
2) Let ${\cal L}_\Pi \subset \{ 1,2,...,n-1\}$ consist of all $k$ s.th. $P_k\cap P_{k-1}$ goes to the left.
 We then have transform
$$ \Gamma^{P_{k-1}P_k}_{-+} \ : \ \Lambda^{K-} * S_- |_{\un{P_k\cap P_{k-1}}}
 \to \Lambda^{K+} * S_+ |_{\un{P_k\cap P_{k-1}}} $$
which extends uniquely to $\un{ P_{k-1} \cup ... \cup P_1}$. We then have:
$$ i^{-+}_k \equiv - \sum_{k'\in {\cal L}_\Pi; \ k'>k} \Gamma^{P_{k'-1}P_{k'}}_{-+} \ \mod F^{\varepsilon_\Pi}. $$
\end{Claim}

\subsubsection{The map $\phi_\Pi$ is an isomorphism}

Now that we have constructed the maps $\phi_\Pi|_{\un{P_k}}$ from \eqref{se16e556}, let us prove that 
they are isomorphisms.

We can write
\begin{equation}\label{ippk}
\phi_\Pi|_{\un{P_k}}=\iota_k\circ i_{\Pi P_k},
\end{equation}
where $i_{\Pi P_k}$ is an endomorpism of $\Lambda^{K+}*S_+\oplus \Lambda^{K-}*S_-|_{\un{P_k}}$
whose components  $i^{\pm\pm}_k$ have been constructed above.
We will abbreviate $i_{\Pi P_k}=i_k$.
The problem reduces to showing invertibility of $i_k$.

 Let us use the matrix notation
$$i_k= \left( \raisebox{2pc}{ \xymatrix{ i^{++}_k & i^{-+}_k \\ i^{+-}_k & i^{--}_k } }\right) \ \in 
\End \left(\ \left.   \raisebox{2pc}{ \xymatrix{ \Lambda^{K+} *  S_+\ar@{}[d]|
{\oplus} \\ 
{\Lambda^{K- }*  S_- }}}\right|_{\un{P_k}} \right) .$$
We have
\begin{equation}\label{inver}
\left( \begin{array}{cc} i^{++}_k & i^{-+}_k \\ i^{+-}_k & i^{--}_k \end{array} \right) \ \equiv \ 
\left( \begin{array}{cc} 1 & i^{-+}_k \\ i^{+-}_k & 1 \end{array} \right), 
\end{equation}
as follows  from Claims \ref{ocenka} and \ref{ocenka1}. 

Lemma \ref{plusminus} implies that 
$$ \left( \begin{array}{cc} 0 & i^{-+}_k \\ i^{+-}_k & 0 \end{array} \right)^2
 = \left( \begin{array}{cc}  i^{-+}_k \circ i^{+-}_k & 0 \\ 0 & i^{+-}_k \circ i^{-+}_k \end{array} \right)\equiv 0. $$

It now follows that  $X:=\left( \begin{array}{cc} i^{++}_k & i^{-+}_k \\
 i^{+-}_k & i^{--}_k \end{array} \right) $
 is invertible (Sec \ref{filr}).

We can multiply (\ref{inver}) by $X^{-1}$ so as to get:
$$
i_kX^{-1}\equiv \Id,
$$
which implies that $i_kX^{-1}$ and, thereby, $i_k$ is invertible.
Furthermore, we get:
\begin{equation}\label{obratim}
i_k^{-1}\equiv \left( \begin{array}{cc} 1 &- i^{-+}_k \\
- i^{+-}_k & 1\end{array} \right) 
\end{equation}
\subsection{The maps $\phi_{\Pi_1}$, $\phi_{\Pi_2}$ for a pair neighboring strips $\Pi_1$ and $\Pi_2$  }
\begin{figure} \includegraphics{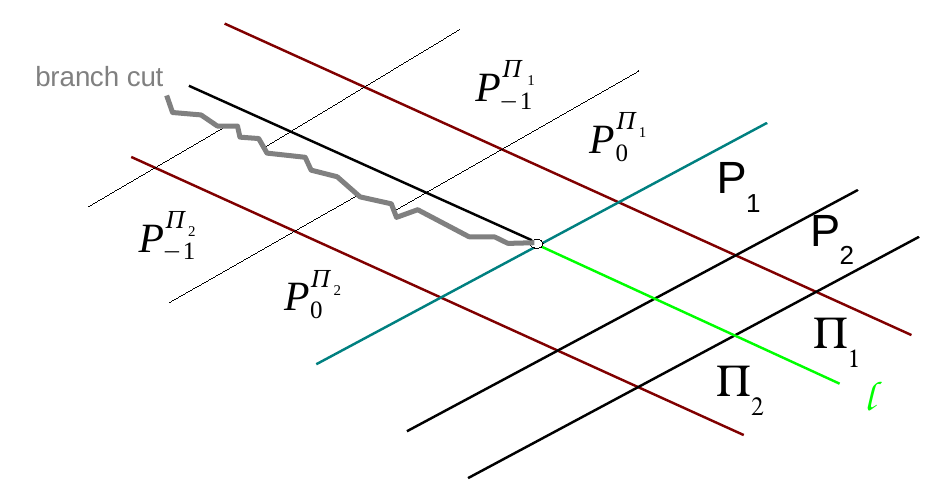}\captionempty
\label{CoShHp15} \end{figure}

Consider now the neighboring strips $\Pi_1$ and $\Pi_2$ and let $\ell=\Pi_1\cap \Pi_2$. Let us find
 the relation between $\Phi^\pm_{\Pi_1}|_\ell$ and $\Phi^\pm_{\Pi_2}|_\ell$. 
 Suppose $\ell$ goes to the right, fig. \ref{CoShHp15}.

We have  a canonical isomorphism $$
H_{\Pi_1\Pi_2}:(\Phi|_{\Pi_1\times \Co})|_{\ell} \simeq (\Phi|_{\Pi_2\times \Co})|_{\ell}. 
$$ 
Using the isomorphisms  $\phi_{\Pi_1},\phi_{\Pi_2}$ as in \eqref{se16e556},  we get
an isomorphism
$$
\tA_{\Pi_1\Pi_2}:=\phi_{\Pi_2}^{-1}|_{\ell\times \Co}\circ H_{\Pi_1\Pi_2}\circ \phi_{\Pi_1}|_{\ell\times \Co}:
$$
\begin{equation}\label{tila}
\Lambda^{K+}*S_+\oplus\Lambda^{K-}*S_-|_{\ell\times \Co}\to \Lambda^{K+}*S_+\oplus
\Lambda^{K-}*S_-|_{\ell\times \Co}.
\end{equation}
Let $P_1,P_2,...,P_n$ be all $\alpha$-strips which intersect $\ell$, fig.\ref{CoShHp15}. We then have
 commutative diagrams
$$ \xymatrix{ \left.  \Lambda^{K+} * {S}_+ \oplus \Lambda^{K- }* { S}_-  \right|_{\ell \cap P_k}  
\ar[rr]^{\tA_{\Pi_1\Pi_2}} \ar[rd]_{i_{\Pi_{{1}} P_k}|_\ell} &&
\left.  \Lambda^{K+}* { S}_+ \oplus  \Lambda^{K-} * { S}_-  \right|_{\ell \cap P_k}  
\ar[ld]^{i_{\Pi_2 P_k}|_\ell} \\
& \left.  \Lambda^{K+} * {S}_+ \oplus \Lambda^{K-} * { S}_-  \right|_{\ell \cap P_k }  
} $$  
which implies that 
$$ \tilde A_{\Pi_1 \Pi_2} |_{\ell\cap P_k} \ = \ ( i_{\Pi_2 P_k}|_{\ell\cap P_k})^{-1} \circ i_{\Pi_1 P_k}|_{\ell\cap P_k}. $$
These formulas determine $\tilde A_{\Pi_1\Pi_2}$. Let us compute:
$$ i_{\Pi_2 P_k} \circ \tilde A_{\Pi_1\Pi_2} |_{\ell\cap P_k} = i_{\Pi_1 P_k}|_{\ell\cap P_k} $$
$$ \left( \begin{array}{cc} 1 & i^{-+}_{\Pi_2 P_k} \\ i^{+-}_{\Pi_2 P_k} & 1 \end{array} \right)
 \circ \tilde A_{\Pi_1 \Pi_2} |_{\ell\cap P_k} \equiv
 \left( \begin{array}{cc} 1 & i^{-+}_{\Pi_1 P_k} \\ i^{+-}_{\Pi_1 P_k} & 1 \end{array} \right) \ .
  $$

Formula (\ref{inver})  yields
$$ \left( \begin{array}{cc} 1 & i^{-+}_{\Pi_2 P_k} \\
 i^{+-}_{\Pi_2 P_k} & 1 \end{array} \right)^{-1} \ \equiv \ \left( \begin{array}{cc} 1 & -i^{-+}_{\Pi_2 P_k} \\ 
-i^{+-}_{\Pi_2 P_k} & 1 \end{array} \right) \  . $$
Therefore,
$$ \tilde A_{\Pi_1 \Pi_2} |_{\ell\cap P_k} \equiv \left( \begin{array}{cc} 1 & -i^{-+}_{\Pi_2 P_k} \\
 -i^{+-}_{\Pi_2 P_k} & 1 \end{array} \right) \times \left( \begin{array}{cc} 1 & i^{-+}_{\Pi_1 P_k} \\
 i^{+-}_{\Pi_1 P_k} & 1 \end{array} \right) \equiv \ \ \ \ \ \  $$
\begin{equation} \ \ \ \ \ \  \equiv \left. \left( \begin{array}{cc} 1 & i^{-+}_{\Pi_1 P_k} - i^{-+}_{\Pi_2 P_k} \\
 i^{+-}_{\Pi_1 P_k} - i^{+-}_{\Pi_2 P_k} & 1 \end{array} \right) \right|_{\ell\cap P_k} \  \label{sp1e1020} \end{equation}
because $i^{+-} _{\Pi_2 P_k}\circ i^{-+}_{\Pi_1 P_k} \equiv 0$ and  
$i^{-+}_{\Pi_2 P_k}\circ i^{+-}_{\Pi_1 P_k}\equiv 0$ by Lemma \ref{plusminus}

Let us, cf. fig.\ref{CoShHp15}, number all the $\alpha$-strips that meet $\Pi_1$ or $\Pi_2$:
$$ P_{-m_1}^{\Pi_1}, P_{-m_1+1}^{\Pi_1}, ..., P_0^{\Pi_1}, P_1, P_2, ..., P_n ; $$
$$ P_{-m_2}^{\Pi_2}, P_{-m_2+1}^{\Pi_2}, ..., P_0^{\Pi_2}, P_1, P_2, ..., P_n . $$
Let us also set $P_1^{\Pi_1}:=P_1^{\Pi_2}:=P_1$.
Lemma \ref{ocenka} yields,

$$
i^{+-}_{\Pi_1P_k}\equiv -\sum'_{l<k}\Gamma^{P_{l}P_{l+1}}-
\sum'_{m\leq 0}\Gamma^{P^{\Pi_1}_{m+1}P^{\Pi_1}_m};
$$

$$
i^{+-}_{\Pi_1P_k}\equiv -\sum'_{l<k}\Gamma^{P_{l}P_{l+1}}-
\sum'_{m\leq 0}\Gamma^{P^{\Pi_2}_{m}P^{\Pi_2}_{m+1}},
$$
where only those terms are included into the sums, for which the intersection ray of the corresponding
$\alpha$-strips goes to the right.
Hence,
$$
i^{+-}_{\Pi_1P_k}-i^{+-}_{\Pi_2P_k}\equiv \sum'_{m\leq 0}\Gamma^{P^{\Pi_2}_{m}P^{\Pi_2}_{m+1}}-
\sum'_{m\leq 0}\Gamma^{P^{\Pi_1}_{m}P^{\Pi_1}_{m+1}} .
$$
Let $\ell:=\Pi_1\cap \Pi_2$ be of the form $\{\hat c(\ell)+re^{-i\alpha}\: \ r>0\}$.

It now follows that 
\begin{equation}
 i^{+-}_{\Pi_1 P_k} - i^{+-}_{\Pi_2 P_k} |_{\ell\cap P_k} 
\equiv -\Gamma^{P_0^{\Pi_1}P_1}_{+-}  \ . \ \label{au17e1144}
 \end{equation}

Thus:
$$\tilde A_{\Pi_1 \Pi_2}|_{\ell\cap P_k} \equiv
 \left( \begin{array}{cc} 1 & * \\ -\Gamma^{P_0^{\Pi_1}P_1} & 1 \end{array} \right)  \ . $$

This means that the same is true for $\tilde A_{\Pi_1\Pi_2}|_\ell$.

Let us write $\tA_{\Pi_1\Pi_2}$ in the matrix form.
$$ \tilde A_{\Pi_1\Pi_2} = 
\left( \raisebox{2pc}{\xymatrix{ \tilde A^{++}_{\Pi_1\Pi_2} & 
\tA^{-+}_{\Pi_1\Pi_2} \\ \tilde A^{+-}_{\Pi_1\Pi_2} &
 \tilde A^{--}_{\Pi_1 \Pi_2} }} \right) \ : 
\ \left. \raisebox{2pc}{\xymatrix{ \Lambda^{K+} * S_+ \ar@{}[d]|{\oplus} \\
 \Lambda^{K-} * S_-  } } \right|_\ell \to \left. 
\raisebox{2pc}{\xymatrix{ \Lambda^{K+} * S_+ \ar@{}[d]|{\oplus} \\ 
\Lambda^{K- }* S_-  }} \right|_\ell \ . $$
Lemma \ref{obnul} implies that $\tA^{-+}_{\Pi_1\Pi_2}=0$. Indeed, the corresponding map
is defined on an unbounded  set $\Pi_1\cap \Pi_2$; since the intersection ray goes to the right, we are under
the conditions of the case {\it i)} of that Lemma.

Let us summarize our findings.
\begin{Claim} \label{obatilda}  Let $\Pi_1,\Pi_2$ be neighboring strips and $\ell=\Pi_1 \cap \Pi_2$ goes
 to the right.  Assume that $\Pi_1$ is above $\Pi_2$.  Then\\
  1) the map $$ \tilde A_{\Pi_1\Pi_2}  : \ 
\left. \raisebox{2pc}{\xymatrix{ \Lambda^{K+} * S_+ \ar@{}[d]|{\oplus} \\
 \Lambda^{K-} * S_-  } } \right|_\ell \to \left.
 \raisebox{2pc}{\xymatrix{ \Lambda^{K+} * S_+ \ar@{}[d]|{\oplus} \\ \Lambda^{K-} * S_-  }} 
\right|_\ell $$ is of the form 
$$ \tilde A_{\Pi_1\Pi_2} =
 \left( \begin{array}{cc} \tilde A^{++}_{\Pi_1\Pi_2} & 0 \\
 \tilde A^{+-}_{\Pi_1\Pi_2} & \tilde A^{--}_{\Pi_1 \Pi_2} \end{array} \right) ;$$

2) $\tilde A^{++}_{\Pi_1 \Pi_2} \equiv Id \ $;
$\tilde A^{--}_{\Pi_1 \Pi_2} \equiv Id \ $;
$\tilde A^{+-}_{\Pi_1 \Pi_2} \equiv -\Gamma^{P_0^{\Pi_1}P_1}_{-+}; \ $
where $P_1$ is the leftmost $\alpha$-strip that meets both $\Pi_1$ and $\Pi_2$ and $P_0^{\Pi_1}$ is
 the rightmost $\alpha$-strip that meets $\Pi_1$ but not $\Pi_2$.
\end{Claim}

Similar result holds true in the case when the intersection ray $\Pi_1\cap \Pi_2$ goes to the left
(proof is omitted).

\begin{Claim} \label{obatilda1}  Let $\Pi_1,\Pi_2$ be neighboring strips and $\ell=\Pi_1 \cap \Pi_2$ goes to the left. 
Assume that $\Pi_1$ is below $\Pi_2$.
Then\\  1) the map $$ \tilde A_{\Pi_1\Pi_2}  : \ 
\left. \raisebox{2pc}{\xymatrix{ \Lambda^{K+} * S_+ \ar@{}[d]|{\oplus} \\ 
\Lambda^{K-} * S_-  } } \right|_\ell \to \left. 
\raisebox{2pc}{\xymatrix{ \Lambda^{K+} * S_+ \ar@{}[d]|{\oplus} \\
 \Lambda^{K-} * S_-  }} \right|_\ell $$ is of the form 
$$ \tilde A_{\Pi_1\Pi_2} = \left( \begin{array}{cc} \tilde A^{++}_{\Pi_1\Pi_2} & \tilde A^{-+}_{\Pi_1\Pi_2} \\ 
0 & \tilde A^{--}_{\Pi_1 \Pi_2} \end{array} \right); $$

2) $\tilde A^{++}_{\Pi_1 \Pi_2} \equiv Id $;
$\tilde A^{--}_{\Pi_1 \Pi_2} \equiv Id $;
$\tilde A^{-+}_{\Pi_1 \Pi_2} \equiv -\Gamma^{ P_0^{\Pi_1}P_1}_{-+} $
where $P_1$ is the rightmost $\alpha$-strip that meets both $\Pi_1$ and $\Pi_2$ and $P^{\Pi_1}_0$ 
is the leftmost $\alpha$-strip that meets $\Pi_1$ but not $\Pi_2$. 
\end{Claim}

\subsubsection{Identifications}

Let $\ell=\Pi_1\cap \Pi_2$, $\ell\in \cL^{-\alpha}$.  
%

In the notation of section \ref{Afat}, we can identify $\tS_{\ell} \stackrel\sim\to S_{{\mathbf A}^{-1}(\ell)}  $; $B_w: \tS_w \stackrel\sim\to S_{{\mathbf A}^{-1}(w)}$ for every $w\in \tW$.
For a word $w=\ell_n\cdots \ell_1L$ or $w= \ell_n\cdots \ell_1R$, set $|w|:=n$ (we set $|L|=|R|=0$).

Let $C_w:=(-1)^{|w|}B_w:\tS_w\to S_{{\mathbf A}^{-1}(w)}$.

Let us define  identifications
\begin{equation}
\bB_\pm,\bC_\pm :\tS_\pm\to S_\pm \label{se22e1059}
\end{equation}
where 
$$
\bB_\pm|_{\tS_w}=B_w;\quad \bC_\pm|_{\tS_w}=C_w.
$$

We can conclude from  2)s of Claims \ref{obatilda},  \ref{obatilda1} that 
\begin{equation}\label{agamma}
\tA_{\Pi_1\Pi_2}\equiv \bC^{-1}\Gamma^{\Pi_1\Pi_2}_{\Psi^K}\bC,
\end{equation}
where $\Gamma^{\Pi_1\Pi_2}_{\Psi^K}$ is as in (\ref{se21e844}).

\subsection{The isomorphism $I_{\Psi\Phi}:\Psi^K\to \Phi^K $}

Using the above developed results, we will construct a map 
$I_{\Psi\Phi}:\Psi^K\to \Phi^K$ which satisfies  (\ref{soglas}) (recall that such a map is unique).
Equivalently, for each $(-\alpha)$-strip $\Pi$, let us specify maps
$$
I_{\Psi\Phi,\Pi}:\Psi^K|_{\Pi\times \Co}\to \Phi^K|_{\Pi\times \Co}
$$
which agree on itersections: if $\Pi_1\cap \Pi_2=\ell\neq \emptyset$, then we should have:
\begin{equation}\label{klei1}
I_{\Psi\Phi,\Pi_1}|_{\ell\times \Co}=I_{\Psi\Phi,\Pi_2}|_{\ell\times \Co}.
\end{equation}
Let us now reformulate condition  (\ref{soglas}).

Let $\bP_0$ be an  $\alpha$ strip  and $\bPi_0$ be a $-\alpha$-strip such that
$\bx_0\in \bP_0\cap \bPi_0$ (these strips are unique).

{Denote $\cF_0^K:=\Z_{{\mathbf x}_0\times K}$, cf.\eqref{eq144}.}

Let $$i^0_\Phi:\cF_0^{{K}}\to \Phi^K|_{(\bPi_0\cap \bP_0)\times \Co};$$
$$i^0_\Psi:\cF_0^{{K}}\to \Psi^K|_{(\bPi_0\cap \bP_0)\times \Co}$$
be the restrictions of $i_\Phi,i_\Psi$.  Since $\cF_0^K$ is supported on $(\bPi_0\cap \bP_0)\times \Co$,
the condition (\ref{soglas}) is equivalent to:
\begin{equation}\label{soglas1}
I_{\Psi\Phi}|_{(\bPi_0\cap \bP_0)\times \Co}i^0_\Psi=i^0_\Phi.
\end{equation}

We have identifications
$$
\tiota_\Pi:\Lambda^{K+}*\tS_+\oplus\Lambda^{K-}*\tS_-|_{\Pi\times \Co}\to \Psi^K|_{\Pi\times\Co}
$$
$$
\phi_\Pi:\Lambda^{K+}*S_+\oplus\Lambda^{K-}*S_-|_{\Pi\times\Co}\to \Phi^K|_{\Pi\times \Co}.
$$

Here $\tilde \iota_\Pi$ is defined similarly to \eqref{iottk} but for $\tilde S_\pm$, $\Psi^K$ and $(-\alpha)$-strips
 instead of $S_\pm$, $\Phi^K$ and $\alpha$-strips; and $\phi_\Pi$ is as in \eqref{se16e556}.

One can now equivalently look for $I_{\Psi\Phi,\Pi}$ in the form:
\begin{equation}
I_{\Psi\Phi,\Pi}=   \phi_\Pi U_\Pi\tilde \iota_\Pi^{-1}, \label{UPiDef}
\end{equation}
where
$$
U_\Pi:\Lambda^{K+}*\tS_+\oplus\Lambda^{K-}*\tS_-|_{\Pi\times \Co}
\to
\Lambda^{K+}*S_+\oplus\Lambda^{K-}*S_-|_{\Pi\times \Co}
$$
is to be calculated.

Since $\Pi$ satisfies both i) and ii) in Lemma \ref{obnul},
we have
$$\hom_{\Pi\times \Co}(\Lambda^{K\pm}*\tS_{\pm};\Lambda^{K\mp}*\tS_{\mp})=0.
$$
Thus, we must have:
\begin{equation}\label{nesmesh}
U_\Pi(\Lambda^{K\pm}*\tS_{\pm})\subset \Lambda^{K\pm}*S_{\pm}.
\end{equation}

Using (\ref{tila}) and (\ref{au22e3}), we rewrite
the gluing condition 
 (\ref{klei1})  as follows:
\begin{equation}\label{klei2}
U_{\Pi_2}|_{\ell\times \Co}=\tA_{\Pi_1\Pi_2}U_{\Pi_1}|_{\ell\times \Co}{\Gamma}^{\Pi_2\Pi_1}_{\Psi^K}.
\end{equation}

Let us now rewrite the condition (\ref{soglas1}) (from now on all our maps are restricted
onto $(\bPi_0\cap \bP_0)\times \Co$, unless otherwise specified).
Let 
$$
\nu:\cF_0^K\to \Lambda^{K+}*S_L\oplus \Lambda^{K-}*S_R
$$
 be the map given by the left arrow in (\ref{i11eq128}).  Let $\nu^+:\cF_0^K\to \Lambda^{K+}*S_L$;
 $\nu^-:\cF_0^K\to \Lambda^{K-}*S_L$ be the components of $\nu$.

We have the following obvious embeddings:
$$I_L: \Lambda^{K+}*S_L\to \Lambda^{K+}*S_+\oplus \Lambda^{K-}*S_-;\quad I_R:
 \Lambda^{K-}*S_R\to \Lambda^{K+}*S_+\oplus \Lambda^{K-}*S_-;$$
$$\tI_L: \Lambda^{K+}*S_L\to \Lambda^{K+}*\tS_+\oplus \Lambda^{K-}*\tS_-;\quad 
\tI_R: \Lambda^{K-}*S_R\to \Lambda^{K+}*\tS_+\oplus \Lambda^{K-}*\tS_-.$$

The formula (\ref{ippk}) can  now be rewritten as
$$
\phi_{\bPi_0}=\iota_{\bP_0}i_{\bPi_0\bP_0}.
$$
We, therefore, can split
\begin{equation}
i^0_\Phi=\iota_{\bP_0} (I_L\oplus I_R)\nu
=\phi_{\bPi_0} i_{\bPi_0\bP_0}^{-1} (I_L\oplus I_R)\nu. \label{se17e1116}
\end{equation}
Next, we have
$$
i^0_\Psi=\tiota_{\bPi_0} (\tI_L\oplus \tI_R)\nu.
$$

Combining \eqref{UPiDef} and \eqref{se17e1116}, we have 
$$
I_{\Psi\Phi,\bPi_0}i^0_\Psi
=\phi_{\bPi_0}^{-1}U_{\bPi_0}(\tI_L\oplus \tI_R)\nu;
$$
so that the condition (\ref{soglas1}) is equivalent to the condition
\begin{equation}\label{soglas2}
U_{\bPi_0}(\tI_L\oplus \tI_R)\nu=  i_{\bPi_0\bP_0}^{-1} (I_L\oplus I_R)\nu \ \ \text{as maps}
 \ \ {\cal F}_0^K \ \to \
 \Lambda^{K+} * S_+ \oplus \Lambda^{K- }* S_- |_{\bPi_0\times \Co} \ . 
\end{equation}
Denote 
$$ i_{\bPi_0\bP_0}^{-1} (I_L\oplus I_R)\nu=:\cI_0. $$

Let us make this condition \eqref{soglas2} more specific. 
\begin{Lemma}\label{razkladec} Let $\cJ:\cF^K_0\to (\Lambda^{K+}*S_+\oplus
\Lambda^{K-}*S_-)[2]$ be an arbitrary map in $\DerCat( (\Pi_0\cap P_0)\times \C)$.
There exist unique maps
$$
\cJ^+:\Lambda^{K+}*S_L\to \Lambda^{K+}*S_+;
$$
$$
\cJ^-:\Lambda^{K-}*S_R\to \Lambda^{K-}*S_-
$$
such that
$$
\cJ=(\cJ^+\oplus \cJ^-)\nu.
$$
\end{Lemma}

\textsc{Proof}
We have   identifications:
$$
\beta  \ : \ R\hom_{\Co}(\bZ_{K};i_{\bx_0}^{-1}(\Lambda^{K+}*S_+\oplus \Lambda^{K-}*{S_-}))
 \stackrel{\sim}{\to} \ \ \ \  $$ $$ \stackrel{\sim}{\to}
R\hom_{\Co}(\bZ_{K};i_{\bx_0}^!(\Lambda^{K+}*S_+\oplus \Lambda^{K-}*{S_-})[2]) \stackrel{\sim}{\to} 
R\hom(\cF^K_0; (\Lambda^{K+}*S_+\oplus \Lambda^{K-}*{S_-})[2]),
$$
where $i_{\bx_0}:\Co\to (\Pi_0\cap P_0)\times \Co$ is the inclusion
$s\mapsto (\bx_0,s)$. 
Consider two more identification 
$$
\alpha^+ \ : \ R\hom(\Lambda^{K+}*S_L;\Lambda^{K+}*S_+)\stackrel\sim\to
 R\hom(i_{\bx_0}^{-1}\Lambda^{K+}*S_L;i_{\bx_0}^{-1}\Lambda^{K+}*S_+)=
R\hom(\bZ_{K};i_{\bx_0}^{-1}\Lambda^{K+}*S_+);
$$
$$
\alpha^- \ : \ R\hom(\Lambda^{K-}*S_R;\Lambda^{K-}*S_R)\stackrel\sim\to
 R\hom(i_{\bx_0}^{-1}\Lambda^{K-}*S_L;i_{\bx_0}^{-1}\Lambda^{K-}*S_-)=
R\hom(\bZ_{K};i_{\bx_0}^{-1}\Lambda^{K-}*S_-);
$$
and let $\alpha=\alpha^+ \oplus \alpha^-$. Then  we have a chain of identifications
$$
R\hom(\Lambda^{K+}*S_L;\Lambda^{K+}*S_+)\oplus R\hom(\Lambda^{K-}*S_R;\Lambda^{K-}*S_-)
$$
$$
\stackrel\alpha\to R\hom_{\Co}(\bZ_{K};i_{\bx_0}^{-1}(\Lambda^{K+}*S_+\oplus \Lambda^{K-} *{S_-}) )
$$
$$
\stackrel\beta\rightarrow R\hom(\cF^K_0;(\Lambda^{K+}*S_+\oplus \Lambda^{K- }* {S_-})[2]).
$$
Let 
$$
\gamma:R\hom(\Lambda^{K+}*S_L;\Lambda^{K+}*S_+)\oplus R\hom(\Lambda^{K-}*S_R;\Lambda^{K-}*S_-)
\to 
 R\hom(\cF^K_0;\Lambda^{K+}*S_+\oplus \Lambda^{K-}*{S_-})[2])
$$
be given by the pre-composition with $\nu$. One can check that $\gamma\beta=\alpha$
so that $\gamma$ is an isomorphism.

The statement now follows.
$\Box$.

\bigskip

Let  $\cI_0^\pm$ denote the maps obtained from $\cI_0$ by means of lemma \ref{razkladec}. Observe 
 that the maps $\cI_0^\pm$ uniquely extend from $(\bPi_0\cap \bP_0)\times \Co$ onto 
$\bPi_0\times \Co$.
Denote the resulting extensions by the symbol 
$\cI^\pm:\Lambda^{K\pm}*S_{L/R}|_{\bPi_0\times \Co}\to \Lambda^{K+}*S_+\oplus 
\Lambda^{K-}*S_-|_{\bPi_0\times \Co}$.

Rewrite the condition \eqref{soglas2} in the form:
$$ U_{\Pi_0} (\tilde I_L \oplus \tilde I_R) \nu = (\cI^+_0 \oplus \cI^-_0)\nu. $$
It now follows that the condition (\ref{soglas2}) (and thus also \eqref{soglas}) will be satisfied
iff
\begin{equation}
\label{nachus}
U_{\bPi_0}|_{\Lambda^{K+}*S_L}=\cI^+;\   U_{\bPi_0}|_{\Lambda^{K-}*S_R}=\cI^-.
\end{equation}
Indeed, the implicaton $\eqref{nachus}\Rightarrow \eqref{soglas2}$ is 
obvious, and $\eqref{soglas2}\Rightarrow \eqref{nachus}$ follows from \eqref{nesmesh}.

\subsubsection{Estimate}
Let us prove the following estimates:
\begin{Claim}
\begin{equation}
\cI^{+}\equiv I_L;\quad \cI^{-}\equiv I_R. \label{PutinByaka} 
\end{equation}
\end{Claim}


Let us bring the current notation into correspondence with that in Lemmas \ref{ocenka},\ref{ocenka1}.
Set $\Pi:=\bPi_0$. Let us denote all the $\alpha$-strips intersecting $\Pi$
by $P_0,P_1,\ldots,P_n$ in the order from the left to the right, in the same way as in Lemmas
\ref{ocenka}, \ref{ocenka1}. Suppose that $\bP_0=P_k$ so that $i_{\bPi_0 \bP_0}=i_k$ in
the notation of Lemmas \ref{ocenka},\ref{ocenka1}. 

Let us now write $i_{\bPi_0  \bP_0}^{-1}=i_k^{-1}=\Id+a_0$, where $a_0$ is an endomorhipsm
of $\Lambda^{K+}*\tS_+\oplus\Lambda^{K-}*\tS_-$.  Let $\ba:=a_0(I_L\oplus I_R)\nu$.
Our statement now reads as $\ba^+\equiv 0$; $\ba^-\equiv 0$.

According to $(\ref{obratim})$, we have
$$
a_0\equiv 
 \left( \begin{array}{cc} 0 &- i^{-+}_k \\
- i^{+-}_k & 0\end{array} \right) 
$$
so that
\begin{equation}\label{baa}
\ba=-(i^{+-}_kI_L\oplus i^{-+}_kI_R)\nu.
\end{equation}
Let us now examine the map $i^{+-}_kI_L\nu$.  We have
$$i^{+-}_kI_L:\Lambda^{K+}*S_L|_{\bPi_0\cap\bP_0\times \Co}\to
 \Lambda^{K-}*S_-|_{\bPi_0\cap\bP_0\times \Co}=
\bigoplus\limits_{w\in \wa_\pravo} \bZ_{{\zA}(K,w)},
$$
where, as in {\eqref{oc21e7}, \eqref{oc21e8}}, ${\zA}(K,w):=\{(z,s)|s-z\in K+\hat c(w)\}\subset (\bPi_0\cap \bP_0)\times \Co$.

As above, let $\waprime_\pravo\subset \wa_\pravo$ consists of all $w$ such that ${\zA}(K,w)\subset {\zA}(K,L)$,
where
$$
{\zA}(K,L)=\{(z,s)|s+z-\bx_0\in K\}\subset (\bPi_0\cap \bP_0)\times \Co.
$$
Let $E_w:\bZ_{{\zA}(K,L)}\to \bZ_{{\zA}(K,w)}$ be the corresponding map of sheaves.
We then have
$$
i^{+-}I_L=\sum\limits_{w\in \waprime_\pravo} n_wE_w,
$$
where for each $(z,s)\in {\zA}(K,L)$ there are only finitely many $w$ such that
$n_w\neq 0$ and $(z,s)\in {\zA}(K,w)$.

Let $A$ be a unique  vertex of the parallelogram $\bPi_0\cap \bP_0$ such that $\bPi_0\cap \bP_0\subset A+ K$.
The condition ${\zA}(K,w)\subset {\zA}(K,L)$ is then equivalent to
$
2A-\bx_0+\hat c(w)\in K,
$
or 
$
\hat c(w)+\bx_0=-2(A-\bx_0)+\ve_w
$
where $\ve_w\in K$. Observe that $\bx_0-A\in \Int K$ because $\bx_0\in \Int \bPi_0\cap\bP_0$.
It now follows that for each $w\in \waprime_\pravo$, the map $E_w\nu^+:\cF_0\to \bZ_{{\zA}(K,w)}$ factors as
$$
\cF_0\stackrel{\nu^-}\to \Lambda^-*S_R=\bZ_{{\zA}(K,R)}\to \bZ_{\{(z,s)|s-z+\bx_0+2(A-\bx_0)\in K\}}
\stackrel{F_w}\to 
\bZ_{\{(z,s)|s-z+\bx_0+2(A-\bx_0)-\ve_w\in K\}}=\bZ_{{\zA}(K,w)},
$$
where all the arrows except the leftmost one are induced by the closed embeddings of the corresponding 
closed sets. It is easy to check that the sum $\sum n_wF_w$ gives rise to a well-defined map
$$
J:\bZ_{\{(z,s)|s-z+\bx_0+2(A-\bx_0)\in K\}}\to \bigoplus\limits_{w\in \wa_\pravo} {\zA}(K,w)
$$

Let $\delta:=2(A-\bx_0)$. We have $bZ_{\{(z,s)|s-z+\bx_0+2(A-\bx_0)\in K}=T_{\delta*}\bZ_{{\zA}(K,R)}.$
Let $\tau_\delta:\bZ_{{\zA}(K,R)}\to T_{\delta*}\bZ_{{\zA}(K,R)}$ be the map induced by the closed embedding
of the corresponding closed sets.
We then have a factorization
$$
i^{+-}_kI_L\nu=J\tau_\delta\nu^{-},
$$
which implies that $(i^{+-}_kI_L\nu)^+=J\tau_\delta\equiv 0$. Similarly, one can check that
$(i^{-+}_kI_R\nu)^-\equiv 0$, which, by virtue of (\ref{baa}), that $\ba=0$.
$\Box$

\subsection{Inductive construction of the maps $U_\Pi$.}
We will now construct the maps $U_\Pi$ satisfying (\ref{klei2}) and (\ref{nachus}).
Taking into account \eqref{nesmesh}, it is possible to construct $U_\Pi$ in terms of its components 
$$ 
U_\Pi^w:\Lambda^{K+}*\tS_w\to \Lambda^{K+}*S_+, \text {  for all  } w\in \wma_\levo;
$$
$$ 
U_\Pi^w:\Lambda^{K-}*\tS_w\to \Lambda^{K-}*S_-, \text {  for all  } w\in \wma_\pravo.
$$

\subsubsection{Rewriting the gluing condition}

Let us rewrite the conditions (\ref{klei2}).

{\it Case 1}: $\ell$ goes to the left and $w\in \wa_\levo$ (set $\pm=+$ on both sides of \eqref{se22e849})
or  $\ell$ goes to the right and $w\in \wa_\pravo$ (set $\pm=-$ on both sides of \eqref{se22e849})
 Let us rewrite (\ref{klei2}):
\begin{equation}
U^w_{\Pi_2}|_{{\ell\times \Co}}=\tA_{\Pi_1\Pi_2}U_{\Pi_1}^w|_{{\ell\times \Co}}:
\Lambda^{K\pm}*S_w|_\ell\to
 \Lambda^{K\pm}*S_\pm|_{{\ell\times \Co}}. \label{se22e849}
\end{equation}
Every map  as on the RHS extends uniquely to $\Pi_2$ (Lemma \ref{HWNpg77}) 

so that we can equivalently rewrite
\begin{equation}\label{okonch1}
U^w_{\Pi_2}=(\Gamma^{\Pi_1\Pi_2}_{\Psi^K}U_{\Pi_1}^w|_{\ell})_\ext,
\end{equation}
where  $\ext$  means the extension onto $\Pi_2$.

 Case 2:
\begin{equation}\label{case2}
\text{$\ell$ goes to the left and $w\in \wa_\pravo$ (set $\pm=-$) or 
$\ell$ goes to the right and $w\in \wa_\levo$ (set $\pm=+$):}
\end{equation}

$$
U^w_{\Pi_2}|_{{\ell\times \Co}}=
\Gamma^{\Pi_1\Pi_2}_{\Psi^K}(U_{\Pi_1}^w|_{{\ell\times \Co}}\oplus
  \vt(\Pi_2,\Pi_1)U_{\Pi_1}^{\ell w}|_{{\ell\times \Co}}N_\ell^w),
$$
where
$
N_\ell^w:\Lambda^-_\ell*S_w\to \Lambda^+_\ell*S_{\ell w}
$
is as in (\ref{nellw}).

Recall that $\tA^{\mp\pm}_{\Pi_1\Pi_2}=0$ by Claims \ref{obatilda}, \ref{obatilda1}, 
so that we can rewrite  the RHS as (using notation from Sec \ref{oboznvt})
$$
\tA^{\pm\pm}_{\Pi_1\Pi_2}U_{\Pi_1}^w|_{{\ell\times \Co}}+
(\tA^{\pm\mp}_{\Pi_1\Pi_2}U_{\Pi_1}^w|_{{\ell\times \Co}}+
\tA^{\mp\mp}_{\Pi_1\Pi_2}\vt(\Pi_2,\Pi_1)U_{\Pi_1}^{\ell w}|_{{\ell\times \Co}} N_\ell^w).
$$
So that we  have  (by separating $+$ and $-$ components):
\begin{equation}\label{predv1}
U^w_{\Pi_2}|_{{\ell\times \Co}}=\tA^{\pm\pm}U_{\Pi_1}^w|_{{\ell\times \Co}}.
\end{equation}
\begin{equation}\label{predv2}
\tA^{\pm\mp}_{\Pi_1\Pi_2}U_{\Pi_1}^w|_{{\ell\times \Co}}+
\tA^{\mp\mp}_{\Pi_1\Pi_2}\vt(\Pi_2,\Pi_1)U_{\Pi_1}^{\ell w}|_{{\ell\times \Co}} N_\ell^w)=0.
\end{equation}
As above, (\ref{predv1}) can be equivalently rewritten in the same way as (\ref{okonch1}).

Let us rewrite  (\ref{predv2}):
$$
U_{\Pi_1}^{\ell w}|_{{\ell\times \Co}} N_\ell^w=
-\vt(\Pi_2,\Pi_1)\tA^{\mp\mp}_{\Pi_2\Pi_1}\tA^{\pm\mp}_{\Pi_1\Pi_2}U_{\Pi_1}^w|_{{\ell\times \Co}}.
$$
Given a map
$\cK:\Lambda^{K\pm}*S_w|_\ell\to \Lambda^{K\mp}*S_{\mp}|_\ell$, one can uniquely factor
it as  
$$
\cK=\cK' N_\ell^w,
$$
where $\cK':\Lambda^{K\mp}*S_{lw}|_\ell\to \Lambda^{K\mp}*S_\mp|_\ell$ (Sec \ref{exten})
which extends uniquely
to a map
$$
\cK'_{\ext}:\Lambda^{K\mp}*S_{lw}|_{\Pi_2}\to \Lambda^{K\mp}*S_\mp|_{\Pi_2}
$$
by Lemma \ref{HWNpg77}.  
In view of this remark, we finally write
\begin{equation}
\label{okonch3}
U_{\Pi_1}^{\ell w}=
\left(-\vt(\Pi_2,\Pi_1)\tA^{\mp\mp}_{\Pi_2\Pi_1}\tA^{\pm\mp}_{\Pi_1\Pi_2}U_{\Pi_1}^w|_{\ell} \right)_{\ext} .
\end{equation}

Let us summarize. Gluing conditions (\ref{klei2}) can be equivalently formulated as follows:

For every pair of neighboring strips $\Pi_1,\Pi_2$, $\ell=\Pi_1\cap \Pi_2$, we have (\ref{okonch1}).
In the case (\ref{case2}) we also have (\ref{okonch3}).

Condition (\ref{okonch1}) implies that 
\begin{equation}
\label{oc1}
U_{\Pi_2}^w|_\ell \equiv U_{\Pi_1}^w|_{\ell}.
\end{equation}

\subsubsection{Constructing $U^w_\Pi$}
Let us proceed by the induction in the length of $w$. In the case $\Pi=\Pi_0$ and $w=L$ or $w=R$, 
$U^w_{\Pi_0}$ is determined by
 (\ref{nachus}).

Given an arbitrary strip $\Pi$, there is a unique sequence 
\begin{equation}\label{posledovat}
 \Pi_0,\Pi_1,\ldots,\Pi_n=\Pi
\end{equation}
 where all $\Pi_i$ are different and $\Pi_i\cap\Pi_{i+1}\neq \emptyset$
(because the graph formed by the strips is a tree).
Formulas (\ref{okonch1}) (applied for all pairs $\Pi_i,\Pi_{i+1}$)  determine $U^L_\Pi,U^R_\Pi$
for all $\Pi$.

Suppose that $U^w_\Pi$ for all words $w$ of length $\leq N$.  Let $w=\ell w'$ be a word 
of length $N+1$ (so that the length of $w'$ is $N$).   Let $\ell=\Pi_1\cap \Pi_2$. The formulas 
(\ref{okonch3})   determine $U^w_{\Pi_1}$.
Given an arbitrary strip $\Pi$ we can join it with $\Pi_1$ by a path and define $U^w_{\Pi}$
using (\ref{okonch1})
in the same way as above.

\subsubsection{Estimate} We are going to prove the following estimate.
Let $\Pi$ be a strip. Consider a map $\bC=\bC_+ \sqcup \bC_-$, cf. \eqref{se22e1059}. We will prove

\begin{Claim}\label{oc}
$$
U_\Pi^w\equiv \bC I_w=(-1)^{|w|}I_w.
$$
\end{Claim}
\textsc{Proof.}
Let us use induction in $|w|$.
If $w=L$ or $w=R$ and $\Pi$ is arbitrary, the estimate follows from  (\ref{oc1}).
Suppose that the estimate is the case for all $w$ with $|w|\leq N$. Let
now $|w'|=N+1$ and $w'=lw$, $|w|=N$. Let $\ell=\Pi_1\cap \Pi_2$.

Combining \ref{okonch3} and the inductive assumption, we have:

$$
\bC^{-1}U_{\Pi_1}^{\ell w}  \equiv \left(-\vt(\Pi_2,\Pi_1)\bC^{-1}\tA^{\mp\mp}_{\Pi_2\Pi_1}\tA^{\pm\mp}_{\Pi_1\Pi_2}\bC I_w|_{\ell} \right)_{\ext} 
\left(-\vt(\Pi_2,\Pi_1)\bC^{-1}\tA_{\Pi_1\Pi_2}\bC I_w|_{\ell} \right)_{\ext}
$$
$$
\stackrel{\text{Claims \ref{obatilda},\ref{obatilda1}}}{\equiv} \ \  \left(-\vt(\Pi_2,\Pi_1)\bC^{-1}\tA^{\pm\mp}_{\Pi_1\Pi_2}\bC I_w|_{\ell} \right)_{\ext} 
\left(-\vt(\Pi_2,\Pi_1)\bC^{-1}\tA_{\Pi_1\Pi_2}\bC I_w|_{\ell} \right)_{\ext}
$$ 
$$
\stackrel{(\ref{agamma})}\equiv( -\vt(\Pi_2,\Pi_1)\tGamma_{\Pi_1\Pi_2}^w|_{\ell})_{\ext}
$$
$$
\equiv(N_\ell^w)|_{\ext}=I_{\ell w},
$$
and \eqref{oc1} allows us to extend this equality to other strips. $\Box$ 

\subsubsection{Proof of Proposition (\ref{ipip})} \label{proofipip}
Let us first find an expression for the maps 
$
J_{\Pi P}
$ as in (\ref{ippk1}).
We have
\begin{equation}
I_{\Psi\Phi,\Pi}|_{\Pi\cap P\times \Co}\stackrel{\eqref{UPiDef}}=   \phi_\Pi|_{\Pi\cap P\times \Co} U_\Pi|_{\Pi\cap P\times\Co}
\tilde \iota_\Pi^{-1}|_{\Pi\cap P\times \Co}
\stackrel{\eqref{ippk}}=\iota_{\Phi P}|_{\Pi\cap P\times \Co}i_{\Pi P} U_\Pi|_{\Pi\cap P\times \Co}
\tilde \iota_{\Psi \Pi}^{-1}|_{\Pi\cap P\times \Co}.
\end{equation}
Comparison with $\eqref{ippk1}$ yields:

$$
J_{\Pi P}=i_{\Pi P}U_\Pi|_{\Pi\cap P}.
$$

We then have (for every $w\in \wa$)
$$
J_{\Pi P}I_w\equiv i_{\Pi P}I_w(-1)^{|w|},
$$
by Claim \ref{oc}.

Let us write $$i_{\Pi P}I_w:\bZ_{{\zA}(K,w)}\to \bigoplus\limits_{w'\in \wa} \bZ_{{\zA}(K,w')}
$$
as 
$$
i_{\Pi P}I_w=\sum\limits_{w'\in \waprime} m^{\Pi P}_{w w'} e_{ww'},
$$
where the sum is taken over all $w'$ such that ${\zA}(K,w')\subset {\zA}(K,w)$ and 
$e_{ww'}:\bZ_{{\zA}(K,w)}\to \bZ_{{\zA}(K,w')}$ is induced by this embedding.
We are to show that $m^{\Pi P}_{w w'}\neq 0$ implies that ${\zA}(K,w)\neq {\zA}(K,w')$.
Assume, on the contrary that ${\zA}(K,w)={\zA}(K,w')$ for $w,w'\in \wa$. Since $P\cap \Pi\neq \emptyset$,
this is only possible when $w,w'\in \wa_\pravo$ or $w,w'\in \wa_\levo$.  Suppose $w,w'\in \wa_\pravo$. Lemma
\ref{ocenka} then implies that either $w'=w$, or $\hat c(w')-\hat c(w)\in \Int K$, i.e. $w\neq w'$, as we wanted.
The case $w,w'\in \wa_\levo$ is treated in the same way by means of Lemma \ref{ocenka1}. $\Box$


\vskip3pc

{\bf \large Acknowledgements}

A.G.'s work was supported by World Premier International Research Center Initiative (WPI Initiative), MEXT, Japan, and the travel expenses were partially covered by NSF.

D.T.'s work was partially supported by NSF.

Both authors are profoundly grateful to Prof. Boris Tsygan for discussions and encouragement of this project. {We thank Professors T.Kawai and Y.Takei for their valuable feedback on this manuscript.}


\end{document}